\documentclass[12pt]{elsart}
\makeatletter\def\ps@copyright{}\makeatother

\usepackage[a4paper, height=220mm, width=155mm]{geometry}
\usepackage{amsmath, amssymb}
\usepackage{graphicx}

\newcommand{\eVq}{\ensuremath{\text{eV}^2}}
\newcommand{\Dmq}{\Delta m^2}
\newcommand{\Sol}{\text{sol}}
\newcommand{\Atm}{\text{atm}}
\newcommand{\Sbl}{\textsc{sbl}}

\def\d{\delta}

\def\t{\theta}

\newcommand{\be}{\begin{equation}}

\newcommand{\ee}{\end{equation}}

\newcommand{\bea}{\begin{eqnarray}}

\newcommand{\eea}{\end{eqnarray}}

\newcommand{\ba}{\begin{array}}

\newcommand{\ea}{\end{array}}

\newcommand{\NO}{\nonumber}

\newcommand{\mrm}{\mathrm}

\begin{document}

\begin{frontmatter}

\begin{flushright}
    IFT-UAM/CSIC-08-91 \\
    RM3-TH/08-21 \\
    EUROnu-WP6-08-01 \\
\end{flushright}

\title{The discovery channel at the Neutrino Factory:
$\nu_\mu\to\nu_\tau$
pointing to sterile neutrinos} 

\author[ad:Madrid]{Andrea Donini,}
\author[ad:Tokyo]{Ken-ichi Fuki,}
\author[ad:Madrid]{J. L\'opez-Pav\'on,}
\author[ad:Roma]{Davide Meloni}
and
\author[ad:Tokyo]{Osamu Yasuda}

\address[ad:Madrid]{%
  Instituto F\'{\i}sica Te\'orica UAM/CSIC,
  Cantoblanco, E-28049 Madrid, Spain}

\address[ad:Roma]{%
  I.N.F.N., Sezione di Roma III and Universit\`a degli Studi di "Roma Tre",
 Dipartimento di Fisica Via Vasca Navale 84,
 I-00146 Rome, Italy}

\address[ad:Tokyo]{%
Tokyo Metropolitan University, Department of Physics, Minami-Osawa 1-1
 Hachioji, Tokyo 192-0397, Japan}

\begin{abstract}
We study the potential of a Neutrino Factory in constraining the
parameter space of a scheme with one sterile neutrino separated from
three active ones by an $O(1)$ \eVq\, mass-squared difference.  
We present approximated analytic expressions for the oscillation probabilities, 
showing that  the greatest sensitivity to sterile neutrinos at a Neutrino Factory can be achieved using 
the $\nu_\mu \to\nu_\mu$ and the $\nu_\mu \to\nu_\tau$ oscillations. 
We have studied two setups: a Neutrino Factory with 50 GeV (20 GeV) stored muons, 
with two detectors of the Hybrid-MIND type (a magnetized ECC next to a 
magnetized iron calorimeter),  located at $L=3000, 7500$ km ($L=4000, 7500$ km) from the source. 
Four channels have been used: $\nu_e \to \nu_\mu,\nu_\tau$; 
$\nu_\mu \to \nu_\mu,\nu_\tau$.
The relevant backgrounds, efficiencies and systematic errors have been taken into account,
and we have discussed dependence of the sensitivities on the systematic errors.
We have found that the 50 GeV (20 GeV) setup can constrain
$\sin^2 2 \theta^{\rm (4fam)}_{13} \leq 7 \times 10^{-5} (2 \times 10^{-4})$;
$\theta_{34} \leq 12^\circ (14^\circ)$; and $\theta_{24}\leq 7.5^\circ (8^\circ)$.  
Our results hold for any value of $\Dmq_\Sbl \gtrsim 0.1~\eVq$.
Eventually we have shown that, if a positive signal is found, the proposed setup is able to measure
simultaneously $\theta_{34}$ and $\delta_3$ with a precision of few degrees and few tens of degrees, respectively, 
solving the so-called "intrinsic" and "sign degeneracies".
Combination of $\nu_\mu$ disappearance and of the $\nu_\mu \to\nu_\tau$ channel, that will be called  "{\it the discovery channel}'', at the two baselines is able to measure at 99\% CL a new CP-violating phase $\delta_3$
for $\sin^2 2 \theta_{34} \geq 0.06$.
\end{abstract}

\vspace*{\stretch{2}}
\begin{flushleft}
    \vskip 2cm
    \small
    {PACS: 14.60.Pq, 14.60.Lm}
\end{flushleft}

\end{frontmatter}

\section{Introduction}

From the results of solar~\cite{Cleveland:1998nv,
Abdurashitov:1999zd,Hampel:1998xg, Fukuda:2001nj,:2008zn,
Ahmad:2001an, Ahmed:2003kj,Aharmim:2008kc},
atmospheric~\cite{Fukuda:1998mi, Ambrosio:2001je},
reactor~\cite{Apollonio:1999ae, Apollonio:2002gd,
Boehm:2001ik,Eguchi:2002dm} and accelerator~\cite{Ahn:2002up,
Aliu:2004sq,Michael:2006rx,Adamson:2008zt} neutrino experiments, we now know that 
neutrinos have masses and mixings.  In the framework of three flavor oscillations,
neutrino oscillations are described by three mixing angles,
$\theta_{12}$, $\theta_{13}$, $\theta_{23}$, and one CP phase $\delta$,
as well as two independent mass-squared differences, $\Dmq_{21}$ and
$\Dmq_{31}$.  In the standard parametrization \cite{Amsler:2008zz} for
the three flavor mixing matrix
$U_\text{PMNS}$~\cite{Pontecorvo:1957cp, Maki:1962mu,
Pontecorvo:1967fh, Gribov:1968kq}, when $\theta_{13}$ is small,
($\Dmq_{21}$, $\theta_{12}$) and ($\Dmq_{31}$, $\theta_{23}$)
correspond to the mass-squared difference and the mixing angle of the
solar and atmospheric oscillations, respectively.  From the solar
neutrino experiments we have $\Dmq_\Sol \simeq 7.7 \times
10^{-5}~\eVq$, $\sin^2\theta_{12}\simeq 0.30-0.31$
\cite{Schwetz:2008er,Fogli:2008cx}, and from the atmospheric neutrino
experiments $|\Dmq_\Atm| \simeq 2.4 \times 10^{-3}~\eVq$,
$\sin^2 2 \theta_{23}\simeq 0.47-0.50$
\cite{Schwetz:2008er,Fogli:2008cx}.  As for $\theta_{13}$, the reactor
data~\cite{Apollonio:1999ae, Apollonio:2002gd, Boehm:2001ik} and
three-family global analysis of the experimental data give an upper
bound\footnote{ In Refs.~\cite{Fogli:2008jx,Fogli:2008cx,Ge:2008sj}, a
global analysis of the neutrino oscillation data has been shown, in
which a non-vanishing value for $\theta_{13}$ is found. This result,
however, is compatible with $\theta_{13} = 0$ at less than $2 \sigma$,
and it has not been confirmed by the other groups performing global
fits \cite{Schwetz:2008er}. }, $\sin^2\theta_{13} \leq 0.04$, while we
have no information on $\delta$ at present (see, however,
Ref.~\cite{GonzalezGarcia:2007ib}).

In order to determine precisely the remaining two parameters
$\theta_{13}$ and $\delta$, long baseline experiments with intense
neutrino beams have been proposed
\cite{Itow:2001ee,Ayres:2004js,Ishitsuka:2005qi,Hagiwara:2005pe,Diwan:2003bp,Geer:1997iz,Zucchelli:2002sa}.
As in the case of the B factories~\cite{belle,babar}, these precision
measurements will allow us to look for deviation from the standard three flavor oscillations scenario.  
Possible scenarios for such
deviations include non-standard interactions which affects the
neutrino productions and detections~\cite{Grossman:1995wx}, those
which modify the neutrino
propagations~\cite{Guzzo:1991hi,Roulet:1991sm}, light sterile
neutrinos~\cite{Pontecorvo:1967fh}, unitarity violation due to the
effect of heavy fermions~\cite{Antusch:2006vwa,Abada:2007ux}, etc.  These
scenarios (except the non-standard interactions which change the neutrino
propagation, only), in general break unitarity of the PMNS matrix.  As in the B physics, test of
unitarity is one of the important problems which should be
investigated in the future long baseline experiments (see
Ref.~\cite{Group:2007kx} for a review).  Among the proposed long
baseline experiments with high intense neutrino beams, a Neutrino
Factory~\cite{Geer:1997iz}, which uses a muon storage ring to produce
neutrino beams from muon decays, is expected to have excellent
sensitivity to $\theta_{13}$ and $\delta$. 

One of the advantages of a Neutrino Factory is that the flux is flavor-rich, well under control and with no $\nu_\tau$ contamination. This last point is of particular relevance for new physics searches in neutrino oscillations, 
since the $\nu_\mu\to\nu_\tau$ oscillations provide one of the most promising signal of non-standard physics in oscillations (as it will be shown in Sec.~\ref{sec:probs} for sterile neutrinos; see also Refs.~\cite{Ota:2001pw,Ota:2002na,FernandezMartinez:2007ms, Altarelli:2008yr}) and $\nu_\tau$ detection is important to check unitarity of the PMNS matrix (although, clearly, 
unitarity violations of the PMNS matrix are best studied in weak decay processes, \cite{Antusch:2006vwa}).
A Neutrino Factory with multi-GeV muons is a powerful facility to look for $\tau$'s signals, if detectors
dedicated to $\tau$-detection are provided.  Notice that oscillations into $\nu_\tau$ cannot be measured
by so-called $\beta$ beams~\cite{Zucchelli:2002sa} and that high energy conventional superbeams
are affected by  $\nu_\tau$ contamination of the flux (through $B$-mesons decay).

Four-neutrino mass schemes have attracted much attention
since the announcement by the LSND group on
evidence for neutrino oscillations $\bar{\nu}_\mu\to\bar{\nu}_e$
with a mass squared difference 
$\Delta m^2\sim O(1)$ eV$^2$ \cite{Athanassopoulos:1996jb,Athanassopoulos:1997pv,Aguilar:2001ty}.
Because the mass squared difference suggested by the LSND result
is much larger than those for the solar and atmospheric neutrino
oscillations, in order to explain all these data in terms of
neutrino oscillations, it is necessary to introduce 
\emph{at least} a fourth light neutrino state.
From the LEP data \cite{Amsler:2008zz,LEPfinal}, which indicate that
the number of weakly interacting light neutrinos is three, the fourth
state has to be a sterile neutrino.  For this reason, the LSND signal
could be considered as an evidence for the existence of a sterile
neutrino.  Recently the MiniBooNE experiment
\cite{AguilarArevalo:2007it} gave a negative result for neutrino
oscillations with the mass squared difference $\Delta
m^2\sim O(1)$ eV$^2$ which was suggested by the LSND data, and it
has become difficult for four-neutrino models to explain the LSND
data.  The so-called (3+2)-scheme with two sterile neutrinos has also
been proposed \cite{Sorel:2003hf} to account for LSND, but also in this case,
tension with the disappearance experiments remains, as long as we take into
account the LSND data.  Adding a third sterile neutrino does not seem
to help~\cite{Maltoni:2007zf}, and in general global analyses seem to
indicate that sterile neutrinos alone are not enough to account for
all the data in terms of neutrino oscillations.  Models with sterile
neutrinos and exotic physics have been therefore
proposed~\cite{Barger:2005mh, PalomaresRuiz:2005vf,deGouvea:2006qd,
Schwetz:2007cd, Nelson:2007yq}.

While the efforts to account for all the data including the LSND in
terms of neutrino oscillations have been unsuccessful, sterile
neutrino scenarios which satisfy all the experimental constraints
\emph{except} LSND are still possible.  
Even if the inclusion of light sterile neutrinos is not needed to explain 
the present experimental data, it is certainly
worth investigating scenarios where sterile neutrinos do appear and constrain their parameter space. 
Light singlet fermions are indeed present in the low-energy spectrum of many theories and models including them 
represent, for example, phenomenologically natural frameworks to break three-flavor unitarity.

In Ref.~\cite{Donini:2007yf} the (3+1)-scheme
without imposing the LSND constraint was studied in the context of the
CNGS experiments~\cite{cngs}, finding that if the OPERA detector is
exposed to the nominal CNGS beam intensity, a null result can improve
a bit the present bound on $\theta_{13}$, but not those on the
active-sterile mixing angles, $\theta_{14},\theta_{24}$ and
$\theta_{34}$.

In this paper, we have extended the analysis in
Ref.~\cite{Donini:2007yf} to the case of a Neutrino Factory
experiment.  We have first of all extended the analytic computation of
the oscillation probabilities for the ($3+1$)-model at long baseline
experiments in matter using the formalism by
Kimura-Takamura-Yokomakura (KTY)~\cite{Kimura:2002hb,Kimura:2002wd}.  
Approximated formul\ae~in powers of $\theta_{13}$, of the deviations from maximality of $\theta_{23}$ ($\delta \theta_{23}$)
and of the active-sterile mixing angles, $\theta_{i4}$, have been obtained.  
On the basis of this analysis, we have found that the greatest sensitivity 
to the active-sterile mixing angles is achieved using the $\nu_\mu \to \nu_\mu$ and $\nu_\mu \to
\nu_\tau$ channels (as it was noticed, for example, in
Refs.~\cite{Donini:2001xy, Donini:2001xp} and refs. therein).
To take full advantage of these signals, detectors capable of both $\nu_\mu$ and $\nu_\tau$
identification are needed. In our numerical analysis we have, thus, assumed 
a detector of the Hybrid-MIND type \cite{Abe:2007bi}: a 50 kton magnetized iron
calorimeter next to a 4 kton Emulsion Cloud Chamber with magnetized
iron plates.  This detector has a greater efficiency to $\nu_\mu \to
\nu_\tau$ than the standard OPERA-type ECC, with lead plates acting as
target.

We have then extensively analyzed the physics reach of a 50 GeV Neutrino Factory which has 
$2 \times 10^{20}$ useful muon decays per year aimed at two detectors
of the Hybrid-MIND type located at $L=3000$ km and $L=7500$ km from the source, 
with both polarities running for 4 years each. 
As a consistency check, we have also studied the case of a 20 GeV Neutrino Factory
which has $5 \times 10^{20}$ useful muon decays per year aimed at the same
two detectors of the Hybrid-MIND type located at $L=4000$ km and
$L=7500$ km, again with both polarities running for 4 years each.  
The latter option was the scenario which was suggested in the {\em
International Scoping Study for a future Neutrino Factory and
Super-Beam facility}~\cite{Group:2007kx}. 

Four signals have been considered: the "standard" Neutrino Factory channels, the golden channel $\nu_e \to \nu_\mu$ 
\cite{Cervera:2000kp} and the silver channel $\nu_e \to \nu_\tau$ \cite{Donini:2002rm};  the $\nu_\mu$ disappearance channel; and the novel signal $\nu_\mu \to \nu_\tau$, that will be named in this paper the "{\it  discovery channel}". 
We think that this name is appropriate due to its high-sensitivity to new physics signals in neutrino oscillations,
as found from the theoretical analysis of oscillation probabilities. At the same time, however, we are forced to remind that
the effective potential of this channel is tightly linked to the performances of the proposed detectors
able to identify $\tau$'s, that are still in a very preliminary design phase and that should be studied further.

Using the first two channels at the 50 GeV Neutrino Factory, we can extend the three-family $\theta_{13}$-sensitivity plots to the four-family ($\theta^{(\rm 4fam)}_{13},\theta_{14}$)-plane, putting a stringent upper bound on $\theta^{(\rm 4fam)}_{13}$ mixing angle, $\sin^22\theta^{(\rm 4fam)}_{13}\lesssim O(10^{-4})$ at 90\% CL.
Using the combination of the $\nu_\mu$ disappearance channel and of the "{\it discovery channel }" $\nu_\mu \to \nu_\tau$
at the 50 GeV Neutrino Factory, we are able to constrain the active-sterile mixing angles $\theta_{34}$ and $\theta_{24}$,
$|\theta_{34}|\lesssim 12^\circ$ and $|\theta_{24}|\lesssim 7.5^\circ$ at 90\% CL. 
We have found that the combination of the shortest baseline data with the longest baseline ones
significantly improves the sensitivity in both planes, ($\theta^{(\rm 4fam)}_{13},\theta_{14}$) and ($\theta_{24},\theta_{34}$).

We have, then, compared the results for the 50 GeV Neutrino Factory with those that can be obtained at the 20 GeV
ISS-inspired one (with $2 \times 10^{20}$ and $5\times 10^{20}$ useful muon decays per year per baseline, respectively).
We have found that the former setup has a greater potential than the latter for sterile neutrino searches, in particular
for the simultaneous measurement of $\theta_{24}$ and $\theta_{34}$. We interpret this result as  consequence of
the larger $\tau$ statistics that can be collected using higher energy muons.

We have also studied the region of the four-family parameter space for which a four-neutrino signal cannot be 
confused with the three-family model, determining the "discovery potential" of the Neutrino Factory in the 
$(\theta_{13}^{(\rm 4fam)},\theta_{14})$- and $(\theta_{24},\theta_{34})$-planes.

We have, then,  performed a preliminary study of the potential of the 50 GeV Neutrino Factory 
to measure four-family parameters, showing 99\% CL contours in the ($\theta_{34},\delta_3$)-
and ($\theta_{13}^{(\rm 4fam)},\delta_2$)-planes. We have found that the shortest baseline data  
are affected by "intrinsic" \cite{BurguetCastell:2001ez} and "sign" degeneracies \cite{Minakata:2001qm}. 
On the other hand, the longest baseline data have a rather good precision in $\theta_{34}$ and are not affected by those degeneracies. The combination of the two baselines, thus, is able to measure simultaneously $\theta_{34}$ and 
$\delta_3$ with a precision of a few degrees and a few tens of degrees, respectively.
In this context, we have also studied the "$\delta_3$-discovery potential". We have found that the $\nu_\mu$ disappearance
channel can measure a CP-violating $\delta_3$ for $\sin^2 2 \theta_{34} \geq 0.4$ with a 50\% CP-coverage; 
the combination of the $\nu_\mu$ disappearance and of the $\nu_\mu \to \nu_\tau$ "discovery" channels, however, 
is able to measure a CP-violating $\delta_3$ for for $\sin^2 2 \theta_{34} \geq 0.06$ with a 80\% CP-coverage. 
The combination of the two channels is extremely effective in solving correlations and degeneracies, 
thus improving the $\delta_3$-discovery potential by an order of magnitude in $\sin^2 2 \theta_{34}$.
We have also shown how the measurement of the three-family like
CP-violating phase $\delta_2$ is modified by the presence of
non-vanishing active-sterile mixing angles.

Eventually, we have studied the impact of statistic and systematic errors on the performance of the detector proposed
to identify $\tau$'s (the magnetized ECC component of the Hybrid-MIND detector, MECC). We have found that, to improve significantly the potential of the MECC, systematic errors should be kept at the level of a few percents. 
On the other hand, increasing the mass of the MECC from 4 kton to 8 kton is found to be of marginal impact. 
It is clear from this analysis that the $\tau$-detector section of the Hybrid-MIND should be studied further to understand
its ultimate potential. To this purpose, we will  take great advantage of the understanding of the ECC technology and systematics after the first years of OPERA data taking. 
 
An analysis such as this is not completely new: first studies of
sterile neutrinos at a Neutrino Factory were presented at the first
NuFact workshop in Lyon in 1999 \cite{Donini:1999jc,Donini:1999he} in the
framework of the so-called (2+2)-schemes and subsequently extended to
the case of (3+1)-schemes in Ref.~\cite{Donini:2001xy}.  The
possibility to use the Neutrino Factory detectors, optimized to look
for three-family oscillations, to disentangle three- from
four-neutrino signals was considered in
Ref.~\cite{Kalliomaki:1999ii,Donini:2001xp}. Recently, in
Ref.~\cite{Dighe:2007uf} a four-family neutrino analysis in the spirit
of Ref.~\cite{Donini:2007yf} has been performed. The main differences
between this paper and Ref.~\cite{Dighe:2007uf} are (i) that
careful numerical analyses are carried out here by taking
into account backgrounds, efficiencies and systematic errors specific to the
considered signals and setup, and (ii)
that the four channels $\nu_\mu \to\nu_\tau$, $\nu_\mu \to\nu_\mu$,
$\nu_e \to\nu_\mu$, $\nu_e \to\nu_\tau$ at the Neutrino Factory are
considered and their contributions are clarified in the present paper.

The paper is organized as follows. In Sec.~\ref{sec:schemes} the main
features of four-neutrino schemes and the present bounds on the mixing
angles in these scenarios are briefly summarized.  Furthermore
we compute approximated oscillation probabilities
in matter in the atmospheric regime using the KTY formalism~\cite{Kimura:2002hb,Kimura:2002wd}
(details of our computations are given in App. A). In
Sec.~\ref{sec:NUFACT} we remind the details of the considered Neutrino
Factory setup. In Sec.~\ref{sec:results} we present our results using
various channels at the Neutrino Factory. Finally, in
Sec.~\ref{sec:concl} we draw our conclusions.

\section{Four neutrino schemes}
\label{sec:schemes}

\begin{figure}[t] \centering
    \includegraphics[width=0.3\linewidth]{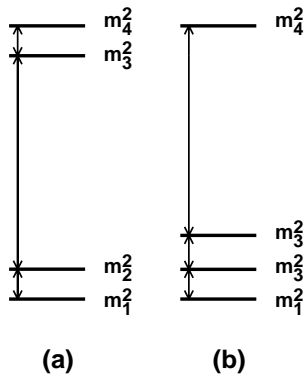}
    \caption{\label{fig:schemes}\sl%
      The two classes of four--neutrino mass spectra, (a): (2+2) and
      (b): (3+1).}
\end{figure}

Four-neutrino schemes consist of one extra sterile state in addition
to the three weakly interacting ones.
Depending on whether one or two
mass eigenstate(s) are separated from the others by the largest
mass-squared gap\footnote{The only assumption
for the largest mass-squared difference is that oscillations
caused by this mass-squared difference are averaged.
So the results hold for any value of $\Dmq_\Sbl \gtrsim 0.1~\eVq$.
Interesting models
with "sterile" neutrino with masses  $m \sim O(1)$ KeV can be
found, for example, in Ref.~\cite{Boyarsky:2005us}.},
the schemes are
called (3+1)- and (2+2)-schemes, as is shown in Fig.~\ref{fig:schemes}.
In the (3+1) schemes, there is a group of three
close-by neutrino masses that is separated from the fourth one by the
larger gap.  In (2+2) schemes, there are two pairs of close masses
separated by the large gap.
These two classes lead to very different phenomenological consequences.

\subsection{(2+2)-schemes}
A characteristic feature of (2+2) schemes is that the extra sterile
state cannot simultaneously decouple from \emph{both} solar and
atmospheric oscillations. 
The fraction of sterile
neutrino contributions to solar and atmospheric oscillations
is given by $\eta_s \equiv |U_{s1}|^2+|U_{s2}|^2$ and $1 - \eta_s \equiv |U_{s3}|^2+|U_{s4}|^2$, respectively,
where the mass squared differences $\Delta m^2_{21}$ and $|\Delta m^2_{43}|$
are assumed to be those of the solar and atmospheric oscillations.
The experimental results show that mixing among active neutrinos give dominant contributions
to both the solar and atmospheric oscillations (see, e.g., Ref.~\cite{Maltoni:2004ei}).
In particular, in Fig.~19 of Ref.~\cite{Maltoni:2004ei} we can see that at the 99\% level
$\eta_s \le 0.25$ and $1 - \eta_s\le 0.25$,
which contradicts the unitarity condition $\sum_{j=1}^4|U_{sj}|^2= 1$.
In fact the (2+2)-schemes are excluded at 5.1$\sigma$ CL \cite{Maltoni:2004ei}.
This conclusion is independent of whether we take the LSND data
into consideration or not and we will not consider (2+2)-schemes
in the rest of this paper.

\subsection{(3+1)-schemes with the LSND constraint}
On the other hand, (3+1)-schemes are not affected by the tension
between the solar and atmospheric constraints on sterile neutrino
oscillations, because as long as the mixing of sterile neutrino is
small, then phenomenology of solar and atmospheric oscillations is
approximately the same as that of the three flavor framework.  The
(3+1) schemes start having a problem only when one tries to account
for LSND and all other negative results of the short baseline
experiments.  To explain the LSND data while satisfying the
constraints from other disappearance experiments, the oscillation
probabilities of the appearance and disappearance channels have to
satisfy the following relation \cite{Okada:1996kw,Bilenky:1996rw}:
\begin{eqnarray}
\sin^22\theta_{\mbox{\rm\tiny LSND}}(\Delta m^2)
<\frac{1}{4}\,\sin^22\theta_{\mbox{\rm\scriptsize Bugey}}(\Delta m^2)
\cdot
\sin^22\theta_{\mbox{\rm\tiny CDHSW}}(\Delta m^2)
\label{relation31}
\end{eqnarray}
where $\theta_{\mbox{\rm\tiny LSND}}(\Delta m^2)$,
$\theta_{\mbox{\rm\tiny CDHSW}}(\Delta m^2)$,
$\theta_{\mbox{\rm\scriptsize Bugey}}(\Delta m^2)$ are the value of
the effective two-flavor mixing angle as a function of the mass
squared difference $\Delta m^2$ in the allowed region for LSND
($\bar{\nu}_\mu\rightarrow\bar{\nu}_e$), the CDHSW experiment
\cite{Dydak:1983zq} ($\nu_\mu\rightarrow\nu_\mu$), and the Bugey
experiment \cite{Declais:1994su}
($\bar{\nu}_e\rightarrow\bar{\nu}_e$), respectively.  The reason that
the (3+1)-scheme to explain LSND is disfavored is basically because
eq.~(\ref{relation31}) is not satisfied for any value of $\Delta m^2$.
A (3+2)-scheme with two sterile neutrino has also been proposed
\cite{Sorel:2003hf} to account for LSND, and it may be
possible to reconcile the LSND and MiniBooNE data by introducing a CP
phase \cite{Karagiorgi:2006jf,Maltoni:2007zf}.  Also in this case,
however, tension with CDHSW~\cite{Dydak:1983zq} and
Bugey~\cite{Declais:1994su} remains, as
in the case of the (3+1)-scheme.

\subsection{(3+1)-schemes without the LSND constraint}
\label{sec:bounds}

If we give up our effort to account for the LSND data,
on the other hand,
we no longer have the constraint (\ref{relation31}).
In this case we have only the upper bound on the extra mixing angles
and this scenario satisfies all the experimental constraints
(except that of LSND).
Throughout this work, therefore,
we will consider a (3+1)-scheme
without taking the LSND data into account while satisfying all the
negative constraints, as it was done in Ref. \cite{Donini:2007yf}.

It has been discussed that the mixing angles of four neutrino schemes
may be constrained by big-bang nucleosynthesis
(see Refs.~\cite{Okada:1996kw,Bilenky:1998ne} and references therein),
and if such arguments are applied,
then the mixing angles of sterile neutrinos would have to be very
small.  However, it is known that in some model
\cite{Foot:1996qc} neutrino oscillations themselves create large
lepton asymmetries which prevent sterile neutrinos from being in
thermal equilibrium, so it is not so clear whether the arguments in
\cite{Okada:1996kw,Bilenky:1998ne} hold.  At present, therefore, it is
fare to say that there is not yet general consensus on this issue (see
Ref.~\cite{Cirelli:2004cz} and references therein).
In this paper we will not impose cosmological constraints
on our scheme.

The mixing matrix $U$ can be conveniently parametrized in terms of six
independent rotation angles $\theta_{ij}$ and three (if neutrinos are
Dirac fermions) or six (if neutrinos are Majorana fermions) phases
$\delta_i$.  In oscillation experiments, only the so-called ``Dirac
phases'' can be measured, since the ``Majorana phases'' 
appear only as an overall phase of the oscillation amplitude
and disappear in the oscillation probability.
The Majorana or Dirac nature
of neutrinos can thus be tested only in $\Delta L = 2$ transitions
such as neutrino-less double $\beta$-decay or
lepton number violating decays~\cite{Amsler:2008zz}. In the following
analysis, with no loss in generality, we will restrict ourselves to
the case of 4 Dirac-type neutrinos only.

A generic rotation in a four-dimensional space can be obtained by
performing six different rotations along the Euler axes. Since the
ordering of the rotation matrices $R_{ij}$ (where $ij$ refers to the
plane in which the rotation takes place) is arbitrary, plenty of
different parametrizations of the mixing matrix $U$ are allowed. In
this paper we are interested in the so-called ``atmospheric regime'',
with oscillations driven by the atmospheric mass difference, $ \Delta_\Atm= \Dmq_\Atm L/2E \sim O(1)$.  We will then make use of the following
parametrization, adopted in Ref.~\cite{Maltoni:2007zf}:
\begin{equation}
    \label{eq:3+1param2}
    U =
    R_{34}(\theta_{34} ,\, 0) \; R_{24}(\theta_{24} ,\, 0) \;
    R_{23}(\theta_{23} ,\, \delta_3) \;
    R_{14}(\theta_{14} ,\, 0) \; R_{13}(\theta_{13} ,\, \delta_2) \; 
    R_{12}(\theta_{12} ,\, \delta_1) \,.
\end{equation}
In eq.~(\ref{eq:3+1param2}), $R_{ij}(\t_{ij},\ \d_l)$ are the complex
rotation matrices in the $ij$-plane defined as:
\be
[R_{ij}(\t_{ij},\ \d_{l})]_{pq} = \left\{ \ba{ll} \cos \t_{ij} & p=q=i,j \\
1 & p=q \not= i,j \\
\sin \t_{ij} \ e^{-i\d_{l}} &	p=i;q=j \\
-\sin \t_{ij} \ e^{i\d_{l}} & p=j;q=i \\
0 & \mrm{otherwise.}\ea \right.
\ee

It is convenient to put phases in $R_{12}$ (so that it automatically
drops in the limit $\Delta_\Sol = \Dmq_\Sol L/2E \to 0$) and $R_{13}$ (so that it
reduces to the ``standard'' three-family Dirac phase when sterile
neutrinos are decoupled). The third phase can be put anywhere; we will
place it in $R_{23}$.  Note that in the one-mass dominance regime \cite{DeRujula:1979yy}
(i.e. for $\Delta_\Atm, \Delta_\Sol  \to 0$)
all the phases automatically disappear from the oscillation probabilities for some choices 
of the four-family PMNS matrix parametrization.

The mixing matrix elements in the parametrization (\ref{eq:3+1param2}) are
given in the Appendix \ref{app:uaj}.

This parametrization has been used in Ref. ~\cite{Donini:2007yf} to put bounds
on the active-sterile mixing angles $\theta_{i4}$ using existing data (including MiniBooNE but
neglecting LSND). These bounds can be summarized as follows: 
\begin{enumerate}
\item Bounds from $\nu_e$ disappearance reactor experiments \par
Reactor experiments such as Bugey and Chooz can put stringent bounds on $\theta_{13}$ and
$\theta_{14}$ in this parametrization: $\theta_{13} \leq 13^\circ$ and $\theta_{14} \leq 10^\circ$ at 99\% CL,
for any value of $\Dmq_\Sbl > 0.1~\eVq$, with some correlation between the two (in particular the four-family 
Chooz bound on $\theta_{13}$ is slightly modulated by $\theta_{14}$). 
\item Bounds from $\nu_\mu$ disappearance experiments \par
A ``negative'' result in a  $\nu_\mu$ disappearance experiment at
"atmospheric" $L/E$ (such as K2K or MINOS), in which $\nu_\mu$
oscillations can be very well fitted in terms of three-family
oscillations, puts a stringent bound on the mixing angle $\theta_{24}$.  
The bound from such experiments found in Ref.~\cite{Donini:2007yf} is: $\theta_{24} \leq 14^\circ$
at 99\% CL, for  $\Dmq_\Sbl \geq 0.1~\eVq$.
\item Bounds on $\theta_{34}$ \par
Neither $\nu_e$ nor $\nu_\mu$ disappearance probabilities in vacuum depend strongly on $\theta_{34}$
(as it can be seen in Ref.~\cite{Donini:2007yf}).
An upper bound on $\theta_{34}$, however, can be drawn as
the result of indirect searches for $\nu_\mu \to \nu_s$ conversion in atmospheric neutrino experiments, 
that take advantage of the different interaction
with matter of active and sterile neutrinos.
Present bounds on $\theta_{34}$ arise, thus, from a measurement of spectral distortion. 
On the other hand, bounds on $\theta_{13},\theta_{14}$ and $\theta_{24}$ are
mainly drawn by a flux normalization measurement. As a consequence,
the bound on $\theta_{34}$ that we can draw by non-observation of
$\nu_\mu \to \nu_s$ oscillation in atmospheric experiments is less
stringent than those we have shown before.  For this reason,
$\theta_{34}$ can be somewhat larger than $\theta_{13},\theta_{14}$
and $\theta_{24}$: $\theta_{34} \leq 32^\circ$ at 99\% CL. 
\end{enumerate}

These bounds are depicted in Fig.~2 of Ref.~\cite{Donini:2007yf}, 
where 90\%, 95\%, 99\% and 3$\sigma$ CL contours in the ($\theta_{13}-\theta_{14}$)-
and ($\theta_{24}-\theta_{34}$)-planes are shown for
$\Delta_\Sol \to 0$ and $\Dmq_\Atm = 2.4 \times 10^{-3}~\eVq$ and $\theta_{23} = 45^\circ$.

\subsection{Oscillation probabilities at the Neutrino Factory in the (3+1)-scheme}
\label{sec:probs}

To understand the details of the different channels with the greatest sensitivity to the
four-family neutrino schemes, it is useful to obtain simple analytical expressions
for the different channels in matter. Hereafter, we will assume a constant Earth density along the neutrino path, 
computed using the PREM \cite{Dziewonski:1981xy}. 
Notice that, in the framework of Neutrino Factory experiments simulations, 
the impact of non-constant matter density has been thoroughly studied, showing that for the baselines
and the muon energy considered in this paper the details of the density profile crossed by neutrinos does not modify the results.\footnote{This is totally different, for example, in the case of $\beta$-Beams with Li/B decaying ions. 
In those setups, a resonance is crossed for $O$(5 GeV) neutrinos for $L = O(10000)$ km baselines \cite{Agarwalla:2006vf}.}

To derive the oscillation probabilities, we adopt the KTY
formalism~\cite{Kimura:2002hb,Kimura:2002wd} (the details are given in
the Appendix \ref{app:formulae}).
Furthermore, to get simplified forms of the formul\ae,
it is convenient to obtain the probabilities expanding with respect to the following small parameters:
\bea
\epsilon \equiv \theta_{34}\sim \sqrt{ \theta_{13}} \sim \sqrt{ \theta_{14}}\sim\;
\sqrt{ \theta_{24}}\sim\sqrt{\delta \theta_{23}}\;\lesssim 4\times 10^{-1} \, ,  \NO 
\eea
with $\delta \theta_{23} = \theta_{23} - \pi/4$.

The relevant oscillation probabilities in matter, expanded to third order in $\epsilon$, are given by
\bea
\label{eq:pee}        
P_{ee}&\sim& 1+O\left( \epsilon^4 \right), \\
\label{eq:pemu}    
P_{e\mu}&\sim& P_{e\tau}\;\sim\; P_{e s}\; \sim\;  O\left(\epsilon^4 \right), \\
\label{eq:pmumu} 
P_{\mu\mu}&=&1- \sin^2\frac{\Delta_{31} L}{2}
-2\left( A_nL\right)s_{24}\,s_{34}\cos\delta_3\sin \Delta_{31} L
+ O\left( \epsilon^4 \right) \; ,\\
P_{\mu\tau}&=&\left( 1-s_{34}^2 \right) \sin^2\frac{\Delta_{31} L}{2}
+\left\lbrace s_{24}\,s_{34}\sin\delta_3+2\left( A_nL\right)s_{24}\,
s_{34}\cos\delta_3\right\rbrace \sin\Delta_{31}L \NO \\
&& 
\label{eq:pmutau}
+ O\left( \epsilon^4 \right) , \\
\label{eq:pmus} 
P_{\mu s}&=&s_{34}^2\sin^2\frac{\Delta_{31} L}{2}
-s_{24}\,s_{34}\sin\delta_3\sin\Delta_{31}L 
+O\left( \epsilon^4 \right)\;,
\eea

where $\Delta_{31}=\Delta m_{31}^2/2E$, and we take the convention
that the central value of $|\Delta m^2_{31}|$ is $\Dmq_\Atm$, which is determined by
the two flavor analysis of the atmospheric neutrino data.
The matter density parameter $A_n$ is $A_n = \sqrt{2} G_F n_n/2$.
Notice that at $O(\epsilon^3)$ the expansion parameter $\delta \theta_{23}$ is not present in the
oscillation probabilities (it only arises at the next order in $\epsilon$).

From eqs.~(\ref{eq:pee})-(\ref{eq:pmus}), it can be easily verified that
unitarity of the PMNS matrix is satisfied to this order in $\epsilon$. As it can be seen,
moreover, the $\nu_e$ decouples within this approximation. We can thus
conclude that the "classic" Neutrino Factory channels, such as the "golden channel"  $\nu_e \to \nu_\mu$ 
and the "silver channel" $\nu_e \to\nu_\tau$, are of limited interest to study sterile neutrinos,
as we will see later in the numerical analysis.\footnote{
This has been known since long.  See for example Refs.~\cite{Donini:2001xy,Donini:2001xp,Donini:2007yf}.}  
Leading sensitivity to $\theta_{34}$  is provided by the first term in $P_{\mu\tau}$, eq.~(\ref{eq:pmutau}), that is 
proportional to $(1-s^2_{34})$. Sensitivity to $\theta_{24}$ is best achieved using the $\nu_\mu \to \nu_\mu$
oscillations, though the relevant term appears at $O(\epsilon^4)$, as it will be shown in Sec.~\ref{sec:th24th34}. 
In eqs.~(\ref{eq:pmumu})-(\ref{eq:pmutau}) we can also see that combined sensitivity to $\theta_{24}$ and $\theta_{34}$ is achievable through the $O(\epsilon^3)$ matter-dependent $s_{24} s_{34} \cos \delta_3$ term in $P_{\mu\mu},P_{\mu\tau}$. 
The bounds on these two angles are those that can be improved the most by the Neutrino
Factory experiments. We can thus safely say that the $\nu_\mu$ disappearance
channel and the $\nu_\mu\rightarrow\nu_\tau$ appearance channel are the most relevant signals
to look for sterile neutrinos.
This will be confirmed by the numerical analysis later.

Notice that  $\theta_{34}$ is considerably less constrained than the rest of the new angles 
(which are as constrained as $\theta_{13}$). Therefore,  the $\nu_\mu \to \nu_\tau$ channel would have at the theoretical level the strongest sensitivity to sterile neutrino parameter space:  $P_{\mu\tau}$ is the only one (with 
the exception of  $P_{\mu s}$, that cannot be directly measured) that is of $O(\epsilon^2)$ in the expansion parameters.
Moreover, $\nu_\mu \to \nu_\tau$ can also look for CP-violating signals in four-family scenarios through the $O(\epsilon^3)$ matter-independent $s_{24} s_{34} \sin \delta_3$ term, something out of the reach of the $\nu_\mu$ disappearance channel
(that can only measure $\delta_3$ through CP-conserving signals).

As it is well known,  $\tau$-detection experiments are extremely difficult at the experimental level. If the experimental problems could be overcome, $\nu_\mu \to \nu_\tau$ would be the important  channel to study sterile neutrinos as well as other kinds of new physics,
such as the non-standard interactions~\cite{Ota:2001pw,Ota:2002na},
or unitarity violations~\cite{FernandezMartinez:2007ms, Altarelli:2008yr}.
In particular, if CP violation occurs due to these new physics, then
the $\nu_\mu \to \nu_\tau$ channel is quite powerful in measuring
the new CP-violating phases.
For this reason, we will name it throughout this paper as "{\it the discovery channel}".

Notice that the $\nu_\mu \to \nu_\tau$ channel has not been studied in detail yet in the framework of neutrino factories
studies (with the possible exceptions of Refs.~\cite{Bueno:2000fg,Campanelli:2000we}). 

\section{The Neutrino Factory and the Hybrid-MIND detector}
\label{sec:NUFACT}

In a Neutrino Factory \cite{Geer:1997iz,DeRujula:1998hd} muons are first produced with a multi-MW proton source, 
accelerated up to energies of several GeV's, and finally injected into a storage ring with long straight sections aiming to 
one ore more detectors.
The muon decays $ \mu^+ \rightarrow e^+ \, \nu_e \, \bar \nu_\mu$ and $\mu^- \rightarrow e^- \, \bar \nu_e \, \nu_\mu $
provide a very well known two-flavor neutrino flux \cite{Broncano:2002hs} with energies in the range $E_\nu \in [0,E_\mu]$.
Neutrino Factory designs have been proposed in Europe \cite{Autin:1999ci,Gruber:2002tn}, the US
\cite{Finley:2000cn,Ozaki:2001bb,Alsharoa:2002wu,Zisman:2003bh}, and Japan \cite{Kuno:2001tb}.  
The dedicated {\em International Scoping Study for a future Neutrino Factory and Super-Beam facility}~\cite{Group:2007kx} 
showed that, provided sufficient resources, an accelerator complex capable of providing about $10^{21}$ muon decays 
of a given polarity per year can be built.  

The Neutrino Factory setup that we propose for sterile neutrinos searches and that we examine in detail in the rest of the paper is defined as follows: 
muons of both polarities are accelerated up to $E_\mu = 50$ GeV and injected into one storage 
ring with a geometry that allows to aim at two far detectors, the first located at 3000 km and the second
at 7500 km from the source. An alternative option, considered in the final ISS Accelerator Report \cite{Zisman:2008zz}, 
is to inject the muon beam into different storage rings, each of them aimed to a single far detector.
The number of useful muon decays per year aimed at each detector has been fixed
to $2 \times 10^{20}$. This number is rather conservative since, in the final ISS Physics Report \cite{Group:2007kx},
with a similar storage ring(s) geometry, $5 \times 10^{20}$ useful muon decays per year aimed at each detector are considered (i.e., $10^{21}$ total useful muon decays per year). Four years of data taking for each muon polarity are envisaged.

\begin{figure}[t]
\begin{center}
\hspace{-0.5cm}
\begin{tabular}{cc}
\includegraphics[width=7.5cm]{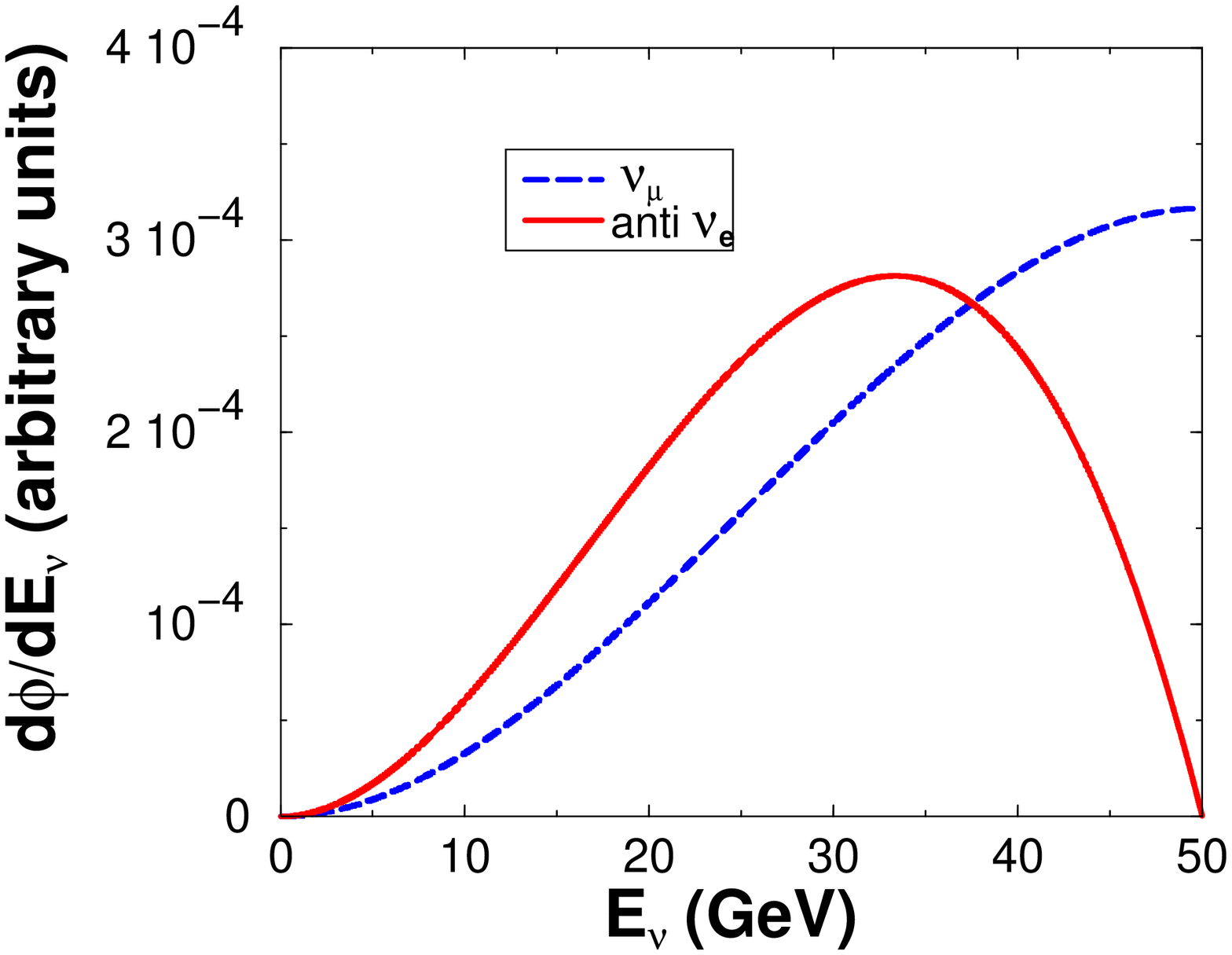} &
\includegraphics[width=7.5cm]{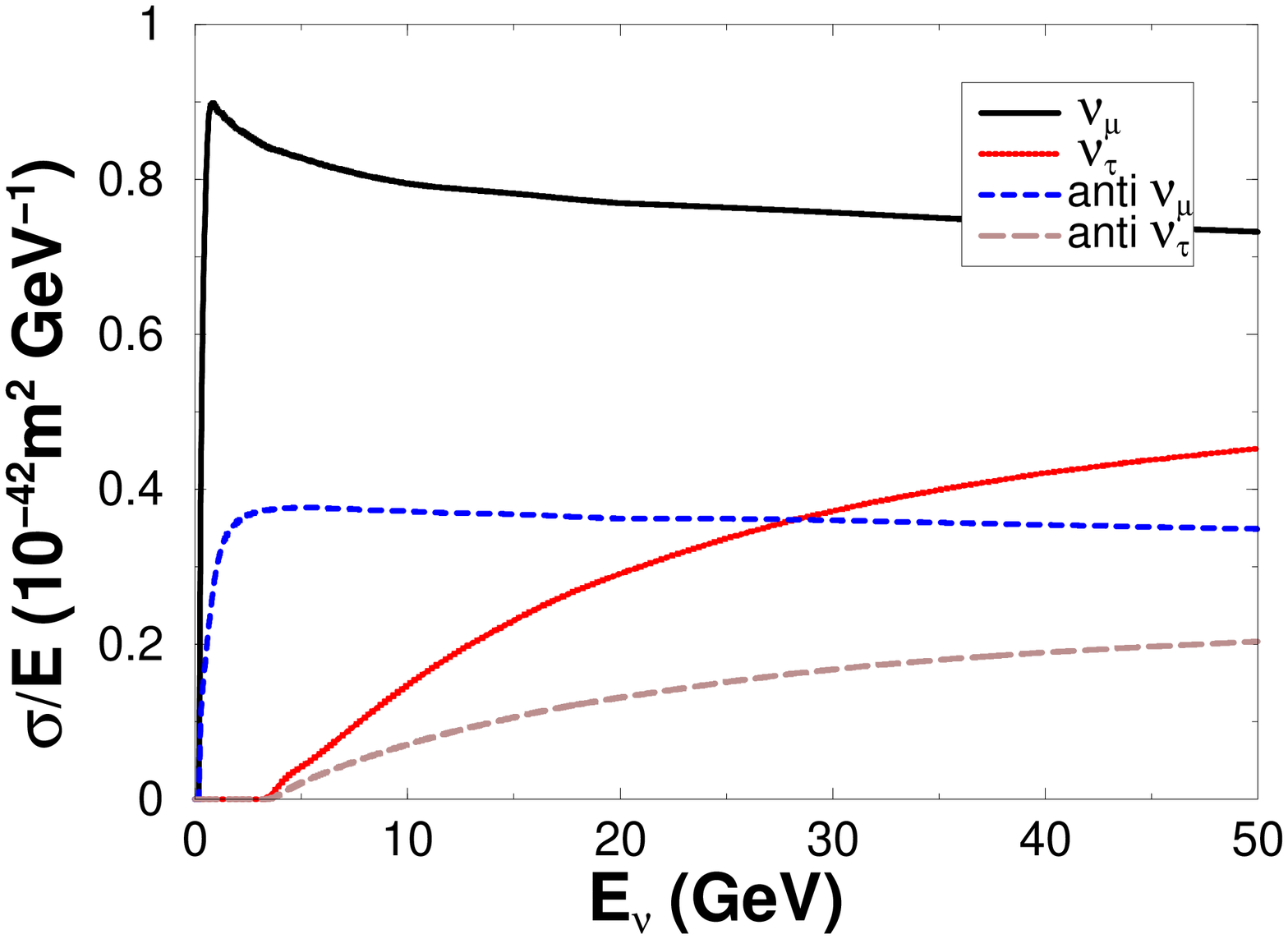}
\end{tabular}
\caption{\it 
Left: 50 GeV Neutrino Factory fluxes at $L=3000$ km;
Right: the $ \nu_\mu N$ and $\nu_\tau N$ cross-sections on iron \cite{Lipari,Lipari:1994pz}.}
\label{fig:fluxcross}
\end{center}
\end{figure}

The Neutrino Factory fluxes (for $\mu^-$ accumulated in the storage ring)  at $L=3000$ km as a function of the neutrino energy for $E_\mu = 50$ GeV are shown in Fig.~\ref{fig:fluxcross}(left). 

Two detectors of different technologies have been considered to detect the $\nu_\mu$ and 
$\nu_\tau$ signals.  The first one is a magnetized iron calorimeter \cite{CerveraVillanueva:2008zz}, that was proposed with a slightly different design in Ref.~\cite{Cervera:2000vy} to measure the "golden" $\nu_e \to \nu_\mu$ wrong-sign muons signal. 
The second detector is a Magnetized Emulsion Cloud Chamber (MECC) \cite{Abe:2007bi}, an evolution of the ECC
modeled after OPERA that was first considered for Neutrino Factory studies in Refs.~\cite{Donini:2002rm,Autiero:2003fu}
to look for the "silver channel" $\nu_e \to \nu_\tau$ (a channel that can be looked at only in a Neutrino Factory setup).

The $\nu_\mu N$ and $\nu_\tau N$ cross-sections on iron as a function of the neutrino energy
for both neutrinos and anti-neutrinos are shown in Fig. ~\ref{fig:fluxcross}(right). 

Notice that the adopted setup, similar to those proposed in the first Neutrino Factory studies (see, for example, Refs.~\cite{Cervera:2000kp,BurguetCastell:2001ez}), slightly differs from the setup that was suggested in the final 
ISS Physics Report. The latter consists in stored muons with energies in the range $E_\mu \in [20,30]$ GeV, aiming to
two detectors located at $L = 4000$ km and $L=7500$ km from the source. 
The longest baseline corresponds to the so-called "magic baseline" \cite{Huber:2003ak}, 
where three-family CP-violating effects vanish. We have chosen the same baseline for the far detector in our
setup, although in the (3+1)-neutrino model not all of the CP-violating phases dependence decouple at this distance
(as it can be seen from eqs.~(\ref{eq:pmumu},\ref{eq:pmutau}) and in Sec.~\ref{sec:results}).
 The shortest baseline was optimized in the ISS Physics Report to look for CP-violating three-family signals, finding 
that a detector with a baseline of $L = 4000$ km performed slightly better than at $L=3000$ km.  
This optimization, however, is no longer valid when looking for the (3+1)-neutrino model signals. 
We decided, therefore, to adopt the $L=3000$ km baseline used in previous studies, for which possible 
sites have already been explored.\footnote{For example, the island of La Palma in the Canary Islands, the island
of Longyearbyen in Norway or the Oulu mine in Pyh\"osalmi in Finland are possible options for the location 
of a detector at approximately 3000 km from the source for a CERN-based Neutrino Factory. }
Eventually, we use a stored muon energy that is larger than the optimal value adopted in the ISS Physics Report, 
again chosen to maximize the sensitivity to three-family observables such as $\theta_{13}$, the sign of the 
atmospheric mass difference and the three-family CP-violating phase $\delta$. 
It is indeed well known from previous studies (see, for example, 
Ref.~\cite{Kopp:2008ds} for an optimization of the muon energy to look for NSI signals at the Neutrino Factory) that to look 
for new physics the higher the muon energy the better. An evident motivation for this  is that
the $\nu_\mu \to\nu_\tau$ channel is very important
to look for new physics in neutrino oscillations, 
and high neutrino energies are required to circumvent the extremely low $\nu_\tau N$ cross-section in the 
tens of GeV energy range (see Fig.~\ref{fig:fluxcross}, right).

To show that the 50 GeV Neutrino Factory setup proposed above is more suited to look for sterile neutrino signals,
we will compare our results with an ISS-inspired Neutrino Factory, defined as follows: 
muons of both polarities are accelerated up to $E_\mu = 20$ GeV and injected into storage 
ring(s) with a geometry that allows to aim at two far detectors, the first located at 4000 km and the second
at 7500 km from the source. The number of useful muon decays per year aimed at each detector has been fixed in this case
to $5 \times 10^{20}$, following the final ISS Physics Report \cite{Group:2007kx}. Four years of data taking for each muon polarity are envisaged. When studying the performance of the ISS-inspired 20 GeV setup, we consider the same detectors as in the case of the 50 GeV Neutrino Factory.  

\subsection{The MIND detector: $\nu_e \to \nu_\mu$ and $\nu_\mu \to \nu_\mu$}
\label{sec:MIND}

The baseline detector, a 50 kton magnetized iron calorimeter of the MINOS type, 
was originally optimized to reduce the background to the wrong-sing muon signal for
$\nu_e \to \nu_\mu$ oscillations (represented dominantly by right-sign muons with a wrong charge
assignment and charmed meson decays) to the $10^{-6}$ level. To achieve this extremely ambitious
signal-to-background ratio,  tight kinematical cuts were applied. Such cuts,
although strongly reducing the background, have the disadvantage of an
important suppression of the signal below 10 GeV, an energy region that, on the other hand, 
has been shown to be extremely important. A non-negligible signal below and above
the first oscillation peak (that, for $L \sim 3000$ km, lies precisely in this energy range) 
is crucial to solve many of the parametric degeneracies \cite{BurguetCastell:2001ez,Minakata:2001qm,Fogli:1996pv,Barger:2001yr}
 that bother the three-family ($\theta_{13},\delta$)-measurement. 
To reduce such problems, a modification of the original detector proposal called MIND (Magnetized Iron Neutrino
Detector) was presented in Ref.~\cite{CerveraVillanueva:2008zz}. 

Studies of the four-family $\nu_e \to \nu_\mu$ oscillation (still the best channel to measure $\theta_{13}$) 
are not different from those performed in the framework of the three-family model. In particular, 
no new sources of background are expected. Quite the contrary, those backgrounds induced by wrongly
reconstructed $\bar \nu_\mu$ are expected to decrease for large values of $\theta_{34}$, due to the increased
oscillation into sterile neutrinos, see eq.~(\ref{eq:pmus}).
When looking at $\nu_e \to \nu_\mu$ oscillations, we will therefore take advantage of the wrong-sign muon identification efficiency presented in the ISS Detector Report \cite{Abe:2007bi}:
$\epsilon_{e\mu} = 0.7$ above 10 GeV, with the efficiency increasing linearly from $\epsilon_{e\mu} = 0.1$ at 1 GeV. 

The MIND detector can also be used to look for $\nu_\mu \to \nu_\mu$ disappearance (one of the channels of interest
to study sterile neutrino models), providing a very good measurement of the atmospheric parameters $\theta_{23}$ and $\Dmq_\Atm$ and giving some handle to solve the ``octant degeneracy'' (see, e.g., 
Ref.~\cite{Bueno:2000fg,Donini:2005db}) .
For the right-sign muon sample there is no need to accurately tell the charge of the muon, since the background 
induced by misidentified wrong-signs muons is negligible with respect to the signal, \cite{Huber:2008yx}. 
We can safely use for this signal the muon identification efficiency of  the MINOS experiment~\cite{Ables:1995wq}: 
$\epsilon_{\mu\mu} = 0.9$ above 1 GeV.  Notice that at the MIND detector it is not possible  to single out $\tau$'s decaying to muons. We cannot thus use MIND to study the leading  $\nu_\mu \to \nu_\tau$ oscillation and the "silver" $\nu_e \to \nu_\tau$ channel. 

We have considered as background for the $\nu_\mu$ disappearance channel $10^{-5}$ of all
neutral current events, \emph{all} wrong-sign muon events and the right-sign muons coming from $\nu_\mu \to \nu_\tau$
oscillation with $\tau$ decaying into muons. The inclusion of this background has no effect on our results for this channel, 
that are remarkably systematic-dominated. 

Different treatments of the energy response of the detector can be found.  
For example, in Ref.~\cite{Cervera:2000kp} a constant energy resolution $\Delta E_\nu = 0.2 E_\mu$ 
was considered, by grouping events in five bins of energy width $\Delta E_\nu = 10$ GeV.  
On the other hand,  in Ref.~\cite{Huber:2006wb} a finer binning was adopted, with a more refined treatment
of the energy resolution: 43 bins of variable $\Delta E_\nu$ were considered
in the energy range $E_\nu \in [1 {\rm GeV}, E_\mu]$, folding the event distribution with a Gau\ss ian resolution kernel
of variable width, $\sigma_E=0.15 \times E_\nu$.  In this paper  we have followed the first approach, 
grouping events into 10 constant energy resolution bins with  $\Delta E_\nu = 0.1 E_\mu$, 
leaving possible improvements of the detector simulation following Ref.~\cite{Huber:2006wb} 
(if needed) for future publications. 

Throughout our numerical simulations we have assumed 2\% and 5\% for
the bin-to-bin uncorrelated systematic errors on the golden channel
and on the $\nu_\mu$ disappearance signal.  We have also assumed
1\% and 5\% for normalization and energy spectrum distorsion as
the correlated systematic errors for all the channels.\footnote{
In principle, some of the correlated systematic errors
could be common among the different channels (such as the detector
volume uncertainty) or different baselines (such as the cross
section uncertainty), but for simplicity we assume here
that all the correlated systematic errors are independent
among the different channels or different baselines.
The results in the present paper may be, therefore,
somewhat conservative.}
The dependence of the sensitivities on the systematic errors
will be discussed in section \ref{sec:systematics}.

Notice that a different proposal for a magnetized iron calorimeter that can be used 
for a Neutrino Factory experiment has been advanced in Ref.~\cite{Indumathi:2004pn}.

\subsection{The MECC detector:  $\nu_e \to \nu_\tau$ and  $\nu_\mu \to \nu_\tau$}
\label{sec:OPERA}

Two technologies were considered in the literature to study neutrino oscillations into $\tau$'s: 
Liquid Argon detectors \cite{Bueno:2000fg,Campanelli:2000we} and Emulsion Cloud Chamber techniques.  
In both cases, the $\nu_e,\nu_\mu \to \nu_\tau$ signal can be tagged looking for right-sign muons in coincidence
with a $\tau$ decay vertex, to distinguish them from $\nu_\mu$ disappearance muons.
Therefore, a detector with muon charge identification and vertex reconstruction is needed. 

A dedicated analysis to use an ECC modeled after OPERA at the Neutrino Factory to look for the "silver channel" 
$\nu_e \to \nu_\tau$  \cite{Donini:2002rm} has been published in Ref.~\cite{Autiero:2003fu}. In that reference, 
a 5 kton ECC was considered, with a detailed study of the main sources of background.

The result of that analysis was that $\tau$'s can be identified with a very low background, at the price of a very
low efficiency, $O$(5\%). One of the main motivation of such low efficiency was that only the $\tau\to \mu$
decay channel was used, i.e.  only 17\% of the total amount of produced $\tau$'s. 
In the OPERA detector used at the CNGS and in the analysis of Ref.~\cite{Autiero:2003fu}, the $\tau$ charge identification is achieved using two large spectrometers located at the end of two thick active sections, with bricks made of repeated layers of lead (acting as the target for $\nu_\tau N$ interactions) and emulsions. 
In the Magnetized ECC (MECC) proposal~\cite{Abe:2007bi} the lead plates are replaced 
by iron plates, again interleaved with emulsions layers. 
The ECC can be, thus, directly magnetized through the iron plates. Emulsion spectrometers (currently in their test phase, \cite{Fukushima:2008zz}) are located at the end of the ECC section. Eventually, the MECC is placed in front of the MIND detector, to form the so-called Hybrid-MIND setup. 
The efficiency of this detector to $\tau$'s is much higher than in the case of the standard ECC, since
$\tau$ decay into electrons and into hadrons can be used in addition to $\tau \to \mu$ decays. 
The expected efficiency is, thus, approximately  five times larger than in the case of the ECC 
(see, for example, Refs.~\cite{Abe:2007bi,Meloni:2008bd} and refs. therein). 

The MECC bricks are bigger than the corresponding ECC ones, due to the replacement of the lead target
with iron. The huge volume to be magnetized put, thus,  a tight limit on the maximum foreseeable detector mass. 
We will therefore consider throughout the paper a  4 kton MECC \cite{Migliozzi} 
located in front of the 50 kton MIND, and use this detector to study both the silver channel $\nu_e \to \nu_\tau$
and the novel "discovery channel" $\nu_\mu \to \nu_\tau$. 
For the silver channel signal, we will use an energy dependent efficiency $\epsilon_{e\tau}$ taken from 
Ref.~\cite{Autiero:2003fu}, multiplying it by a factor five to take into account that all $\tau$'s decay channel
can be used at the MECC. 
A detailed study of the efficiency  to the $\nu_\mu \to \nu_\tau$ channel at the MECC,  on the other hand, 
is lacking. We will therefore assume a constant efficiency $\epsilon_{\mu\tau} = 0.65$ above 5 GeV,
increasing by a factor five the efficiency considered in Ref.~\cite{Donini:2007yf}.
This assumption must be checked in further studies of the MECC-type detectors exposed to a Neutrino Factory beam.

The backgrounds for the silver and the discovery channels should be also correspondingly increased at the MECC
with respect to the ECC ones.  At the ECC, the expected signal-to-background ratio (after some kinematical cuts) for $\nu_\mu \to \nu_\tau$
(using the $\tau \to \mu$ decay channel, only) is 50:1 or larger \cite{Donini:2007yf},  the dominant source of background
for the process $\nu_\mu \to \nu_\tau \to \tau^- \to \mu^-$ being represented by non-oscillated muons
that produce charmed mesons eventually decaying into $\mu^-$ either through NC or CC in which the muon
is not observed. 
No detailed study of the expected background for the $\nu_e \to \nu_\tau$ or the $\nu_\mu \to \nu_\tau$
signals at the MECC exposed to a Neutrino Factory beam has been performed yet, though. 
We have thus decided to make the assumption that, using MECC, all $\tau$ decay channels
should be affected by similar backgrounds. We have therefore consistently multiplied the backgrounds for 
$\nu_e \to \nu_\tau$ and $\nu_\mu \to \nu_\tau$ computed in Refs.~\cite{Autiero:2003fu,Donini:2007yf} 
by a factor five. This is possibly a conservative assumption\footnote{
Notice that, for the silver channel, at least the background induced by 
right-sign $\tau$'s with wrong charge assignment should be depleted in a four-family neutrino scenario with
respect to the standard three-family one. This particular background is strongly affected by active-sterile mixing angles, 
since, in the allowed region of the parameter space, $\nu_\mu \to \nu_\tau$ oscillations are significantly depleted with
respect to the standard three-neutrino ones. }, since
the MECC is expected to have a signal-to-background ratio for this signal slightly better than the ECC \cite{ScottoLavina:2008}. We consider, however, this to be the only 
reasonable choice that we can take at this preliminary stage of four-family neutrino detailed study at the Neutrino Factory 
(not to be compared with the order-of-magnitude estimations made in previous papers). 

Also in this case, we have grouped events into 10 bins with $\Delta E_\nu = 5$ GeV constant energy resolution. 
We have assumed 10\% for the bin-to-bin uncorrelated systematic
errors, 1\% and 5\% for normalization and energy spectrum distorsion
as the correlated systematic errors throughout the numerical
simulations for both the $\nu_e \to \nu_\tau$ and the $\nu_\mu \to
\nu_\tau$ signals.  Once again, the results here may be slightly
conservative because we may be overcounting the systematic errors.

\begin{table}[t] \centering
    \begin{tabular}{|c|c|c|c|c|c|}
	\hline\hline
	$(\theta_{13}; \theta_{14}; \theta_{24}; \theta_{34})$ & 
	$N_{\tau^-}^{CNGS}$ & $N_{\tau^-}^{3000}$ &  $N_{\tau^+}^{3000}$ & $N_{\tau^-}^{7500}$ & $N_{\tau^+}^{7500}$ \\
	\hline
	$(5^\circ;  5^\circ;  5^\circ; 20^\circ)$    &     8.9    & 559   &  10  &   544  & 2 \\
	$(5^\circ;  10^\circ; 5^\circ; 20^\circ)$   &              & 557  &   29   &  544  & 5  \\
	$(5^\circ;  5^\circ; 10^\circ; 20^\circ)$   &     8.3    & 474  &   11   &  529   & 2  \\
 	$(5^\circ;  5^\circ; 10^\circ; 30^\circ)$   &    10.5   & 384  &  18    &  454  & 3 \\
        $(5^\circ;  10^\circ;  5^\circ; 30^\circ)$   &            & 424  &  59    & 441   &  11   \\
	$(5^\circ;  5^\circ; 10^\circ; 30^\circ)$   &    10.5   & 384  &  18    &  454  & 3 \\
	$(10^\circ; 5^\circ;  5^\circ; 20^\circ)$   &     8.5    & 522  &   22    &  512  &  2 \\
	$(10^\circ; 10^\circ;  5^\circ; 20^\circ)$   &            & 517  &   42    &  510  &  6 \\
	$(10^\circ; 5^\circ; 10^\circ; 20^\circ)$  &     7.9    & 443  &   22    &  498  & 2  \\
	$(10^\circ; 5^\circ;  5^\circ; 30^\circ)$   &     6.5    & 397  &   30   &  413  & 4  \\
	$(10^\circ; 10^\circ;  5^\circ; 30^\circ)$   &            & 389  &   74   &  412  & 11  \\
	$(10^\circ; 5^\circ; 10^\circ; 30^\circ)$  &    10.3   & 361  &   30  &  428  & 4  \\
	3 families, $\theta_{13} = 5^\circ$          &    15.1   & 797  &    3     & 666  &   0     \\
	3 families, $\theta_{13} = 10^\circ$        &    14.4   & 755  &    12  &  632  & 1    \\
	\hline \hline
    \end{tabular}
    \caption{\label{tab:rates3}\sl%
      Event rates for the $\nu_\mu \to \nu_\tau$ and $\bar \nu_e \to \bar \nu_\tau$
      channels for 1 kton MECC detector, exposed to a $2 \times 10^{20}$ ($\nu_\mu,\bar \nu_e$) flux for one year,
      for different values of
      $\theta_{14}, \theta_{24}$ and $\theta_{34}$ in the (3+1)
      scheme. The other unknown angle, $\theta_{13}$ has been fixed
      to: $\theta_{13} = 5^\circ, 10^\circ$. The CP-violating phases
      are: $\delta_1 = \delta_2 = 0$; $\delta_3 = 90^\circ$.  As a
      reference, rates at the 1.8 kton OPERA detector (exposed to the nominal CNGS beam intensity) 
      and the expected event rates for 1 kton MECC detector
       in the case of the three-family model (i.e., for $\theta_{i4} = 0$ and maximal CP-violating 
      phase $\delta$) are also shown. In all cases, perfect efficiency is assumed.
      }
\end{table}

In Tab.~\ref{tab:rates3} we show the expected number of $\tau^-$ from $\nu_\mu \to \nu_\tau$
and  $\tau^+$ from $\bar \nu_e \to \bar \nu_\tau$ for a 1 kton MECC detector with perfect efficiency, exposed to a $2 \times 10^{20}$ ($\nu_\mu,\bar \nu_e$) flux for one year, for different values of $\theta_{13}, \theta_{14}, \theta_{24}$ 
and $\theta_{34}$. The other parameters are: $\theta_{12} = 34^\circ; \theta_{23} = 45^\circ;
\Dmq_\Sol = 7.9 \times 10^{-5}~\eVq$; $\Dmq_\Atm = 2.4 \times
10^{-3}~\eVq$ and $\Dmq_\Sbl = 1~\eVq$ (all mass differences are taken
to be positive).  Eventually, phases have been fixed to: $\delta_1 =
\delta_2 = 0; \delta_3 = 90^\circ$.  For comparison, the rates at the CNGS (for the nominal CNGS flux, 
of $4.5 \times 10^{19}$ pot/year, an active lead target
mass of 1.8 kton and 5 years of data taking) and the expected number of events
in the three-family model for a 1 kton MECC detector with perfect efficiency are also shown.
We can see that the number of expected $\tau^-$ events at the 1 Kton MECC is $O(500)$ at both baselines, 
with some dependence on the different mixing angles. The fact that at both baselines we expect a similar number
of events is a consequence of the convolution of the $\nu_\mu \to \nu_\tau$ oscillation probability 
with the $\nu_\tau N$ cross-section and the $\nu_\mu$ neutrino flux: at the shortest baseline, the probability is maximal below 10 GeV; at the longest baseline, the maximum is located in the 30 GeV bin. The higher cross-section for this energy
bin compensates for the decrease in the $\nu_\mu$ neutrino flux, thus giving a similar number of $\tau$'s in the detector.

Notice that we will not use the MECC section of the Hybrid-MIND detector to look for the golden $\nu_e \to \nu_\mu$
and the $\nu_\mu \to \nu_\mu$ disappearance signals.  Both channels can be studied, though, at the price of an increase
in the scanning load. 

\section{Sensitivity to $(3+1)$ sterile neutrinos at the Neutrino Factory}
\label{sec:results}

In this section we study the physics reach of the 50 GeV Neutrino Factory setup discussed in the previous section
to (3+1)-model sterile neutrinos. We will make use of all possible oscillation channels available at this setup, namely
the golden channel $\nu_e \to \nu_\mu$ and the disappearance channel $\nu_\mu \to \nu_\mu$ using the MIND
section of the detector; the silver channel $\nu_e \to \nu_\tau$ and the "discovery channel" $\nu_\mu \to \nu_\tau$
using the MECC section of the detector.

In Sec.~\ref{sec:th13th14} and \ref{sec:th24th34} we study how a negative result at the Neutrino Factory can 
be used to improve present bounds on the four-family model $\theta^{(\rm 4fam)}_{13}$, extending the standard 
three-family $\theta_{13}$-sensitivity analysis, and on the three active-sterile mixing angles $\theta_{14},\theta_{24}$ 
and $\theta_{34}$. 
In particular, the golden and silver channels ($\nu_e \to \nu_\mu,\nu_\tau$) cooperate to put a strong bound
on $\theta_{13}^{(\rm 4fam)}$ (see Sec.~\ref{sec:th13th14}).
On the other hand,  the $\nu_\mu$ disappearance channel and the "discovery channel" 
$\nu_\mu \to \nu_\tau$ improve significantly the exclusion bounds for (3+1)-sterile neutrinos in the ($\theta_{24},\theta_{34}$) plane (see Sec.~\ref{sec:th24th34}). Preliminary results for the combination of these two channels at a 50 GeV Neutrino Factory setup have been presented in Refs.~\cite{Fuki:2008nufact,Meloni:2008ti}.

In Sec.~\ref{sec:discrimination} we consider a different situation: in the case of a positive signal at the Neutrino Factory, 
can we discriminate between the (3+1)-sterile neutrino model and the "standard" three-family one? We will show
that, using the combination of the four available channels, we can indeed discriminate between the two models
in a significant region of the active-sterile mixing angles parameter space. 
In Secs.~\ref{sec:th13th14}, \ref{sec:th24th34}  and \ref{sec:discrimination} we will also compare our results with those that could be obtained using the 20 GeV ISS-inspired Neutrino Factory defined in Sec.~\ref{sec:NUFACT}.

In Sec.~\ref{sec:systematics} we discuss the dependence of the four channels on the systematic errors.  
It is known that the golden and silver channels are dominated by the statistical errors. 
We will show, on the other hand, that both the disappearance and discovery channels
are dominated by the systematic errors,  emphasizing that improvement on systematic errors is 
extremely important to take full advantage of the discovery channel.

In case a positive signal is found, 
it is interesting to study the various CP-violating signals. This is done  in Sec.~\ref{sec:measurement}, where we show 
the attainable precision in the simultaneous measurement of $\theta_{34}$ and of the CP-violating phase $\delta_3$
and the $\delta_3$-discovery potential, when both active-sterile mixing angles $\theta_{24}$ and $\theta_{34}$ are not exceedingly small. We also show how the measurement of the three-family--like CP-violating phase $\delta_2$ is modified by the presence of non-vanishing active-sterile mixing angles.

\subsection{Sensitivity to $(\theta_{13},\theta_{14})$}
\label{sec:th13th14}

In this subsection first we discuss sensitivity to
$(\theta_{13},\theta_{14})$ at the Neutrino Factory.  As we will see
below, sensitivity to $\theta_{14}$ turns out to be poor.
So we will consider sensitivity to other combinations of
the mixing matrix elements later.

\subsubsection{Sensitivity to $(\theta_{13},\theta_{14})$}
\label{sec:subth13th14}

The sensitivity is defined as follows: we first compute the expected number of events  for 
$\nu_e \to \nu_\mu$ and $\nu_e \to \nu_\tau$ oscillations for the input values $\theta^{\rm (4fam)}_{13} = 0$,
and $\theta_{14} = \theta_{24} = \theta_{34} = 0$,
where
$\theta_{jk}^{\rm (4fam)}$ ($(j,k)=(1,2),(1,3),(2,3)$)
and $\theta_{j4}\equiv\theta_{j4}^{\rm (4fam)}$ stand for the mixing angles in the four-family scheme denoted 
by {\rm (4fam)}.\footnote{
Using the four-family expressions for the oscillation probabilities in vacuum shown in Ref.~\cite{Donini:2007yf}, it can be seen that, for vanishing $\theta_{14}$ and $\theta_{24}$, the four-family mixing angle $\theta_{13}^{\rm (4fam)}$ maps into the three-family one, $\theta_{13}^{\rm (3fam)}$. On the other hand, for non-vanishing $\theta_{14}$, the four-family parameter is expected to be slightly smaller than the three-family one.}
This number,  that is identical in the three- and four-family models, is labeled as $N^0$.
We then compute the expected number of events in the ($\theta^{\rm (4fam)}_{13},\theta_{14}$)-plane for the same
oscillation channels in the four-family model.
The $\Delta \chi^2$, computed with respect to the "true" value $N^0$,  is eventually evaluated. 
The contour for which the 2 d.o.f.'s $\Delta \chi^2$ is $\Delta \chi^2 = 4.61$ defines, then, the region of the parameter space
of the (3+1)-sterile neutrino model that is non-compatible at 90\% CL with the input data corresponding to vanishing
$(\theta^{\rm (4fam)}_{13}, \theta_{i4})$  (to the right of the contour line) and the region that it is still allowed (to the left of the line) at this CL. 

The minimum of the $\chi^2$ is computed as follows: 
\begin{eqnarray}
\Delta \chi^2 =  
&
\min_{\rm marg \, par}
&
 \left [
\sum_{{\rm pol., (chan.)}, (L)}
\min_{\alpha's, \,\beta's}
\left \{\sum_j \frac{1}{\sigma^2_j}
\left ( (1+\alpha_s+ x_j\beta_s) N_j ({\rm 4fam}) 
 \right.\right.\right.\nonumber\\
&{\ }&
+ (1+\alpha_b+ x_j\beta_b) B_j ({ \rm 4fam}) - 
N^0_j - \left. B^0_j
\right )^2
 + \left(\frac{\alpha_s}{\sigma_{\alpha_s}}\right)^2
 + \left(\frac{\alpha_b}{\sigma_{\alpha_b}}\right)^2\nonumber\\
&{\ }&
+\left.\left(\frac{\beta_s}{\sigma_{\beta_s}}\right)^2
+\left(\frac{\beta_b}{\sigma_{\beta_b}}\right)^2\right \}
+ \Delta \chi^2_{\rm atm+re}({\rm 4fam})
\left. \right ] \, ,
\label{chi2-13-14}
\end{eqnarray}
where we have introduced the prior that comes from the four-family analysis
of the atmospheric and reactor data:\footnote{
Since $\Delta\chi^2$ is expected to depend
little on the solar neutrino oscillation parameters,
we will not vary the solar oscillation
parameters throughout this paper, so
we omit the terms on the solar parameters here.}
\begin{eqnarray}
\Delta \chi^2_{\rm atm+re} ({\rm 4fam}) &=& 
\frac{(s^{2~\rm (4fam)}_{23}-0.50)^2}{\sigma^2(s^2_{23})}
+\frac{(|\Delta m^{2~\rm (4fam)}_{32}|-2.4\times10^{-3}\mbox{\rm eV}^2)^2}
{\sigma^2(|\Delta m^{2~\rm (4fam)}_{32}|)}\nonumber\\
&{\ }&+\frac{(s^{2~\rm (4fam)}_{13}-0.01)^2}{\sigma^2(s^2_{13})}
+\frac{(s^2_{14})^2}{\sigma^2(s^2_{14})}
+\frac{(s^2_{24})^2}{\sigma^2(s^2_{24})}
+\frac{(s^2_{34})^2}{\sigma^2(s^2_{34})}.
\label{chi2-atmre4}
\end{eqnarray}
where $|\Delta m^{2~\rm (4fam)}_{32}|$
stands for the atmospheric mass squared difference in the four-family scheme,
and the errors of the oscillation parameters
in the four-flavor scheme in eq. (\ref{chi2-atmre4})
are deduced from Refs. \cite{Schwetz:2008er}
and \cite{Donini:2007yf} as follows:
\begin{eqnarray}
\sigma(s^2_{23})&=&0.07,~~
\sigma(|\Delta m^2_{32}|)=0.12\times10^{-3}\mbox{\rm eV}^2,~~
\sigma(s^2_{13})=0.016,\nonumber\\
\sigma(s^2_{14})&=&0.013,~~
\sigma(s^2_{24})=0.02,~~
\sigma(s^2_{34})=0.12.
\label{sigmas}
\end{eqnarray}
In the minimization procedure in eq. (\ref{chi2-13-14}),
``marg par'' stands for the oscillation parameters to
be marginalized over (that can be different for different plots),
and $\alpha_s$, $\alpha_b$, $\beta_s$ and $\beta_b$ are
the variables for
the correlated systematic errors, which stand for the uncertainties
in the overall normalization and in the linear distortion in the
spectral shape in the magnitude of signal (s) or background (b)
\cite{Huber:2002mx}, where we have defined $x_j\equiv
E_j/(E_{\text{max}}-E_{\text{min}})$ for neutrino energy $E_j$ for the
$j$-th bin.  Following Ref.~\cite{Huber:2002mx}, we assume the
correlated systematic errors $\sigma_{\alpha_s}=\sigma_{\alpha_b}=0.01$ 
for the normalization and $\sigma_{\beta_s}=\sigma_{\beta_b}=0.05$ for the spectrum distorsion.
In the analysis of the case with single baseline length,
we have minimized the $\chi^2$ for each baseline separately,
i.e., no sum is performed over $L$,
and in the analysis combining the two baselines, we have
minimized the sum of $\chi^2$ for each baseline.
Similarly, in the analysis of a single channel,
no sum is performed over the channels (``chan.''),
while in the analysis combining the different channels
we have summed over the different channels,
i.e., golden and silver in this Section.
In all cases  we sum up $\chi^2$
for the two possible stored muon polarities (``pol.'').
The index $j$ runs over 10 energy bins.
$B_j$ is the background correspondent to the $j$-th bin
($B^0_j$ stands for the expected background in the four-family model for vanishing $\theta^{\rm (4fam)}_{13}$ and $\theta_{i4}$
in the $j$-th bin).
Within this procedure, for any minimization that we perform the best-fit for the variables on which we marginalize over can be different. However, when we project onto the two-dimensional ($\theta_{13}^{(\rm 4fam)}, \theta_{14}$)-plane (or onto the $(\theta_{24},\theta_{34})$-plane in the next section) this information is lost. 
Notice that this approach is analogous to that used in the three-flavor
analysis in Ref.~\cite{GonzalezGarcia:2004cu}.
A final comment is in order: in the case of a real experiment, the minimum of the $\chi^2$ distribution will in general not 
correspond exactly to the "true" values of the parameters to be fitted. The $\Delta \chi^2$, computed with respect
to the best-fit point, will be therefore slightly different from our definition. 
This procedure, however, is commonly adopted to compute the expected sensitivity of an experiment that is still 
under development, so as to compare different setups on equal footing. 

The variance is defined as:
\begin{eqnarray}
\sigma_j^2 = N^0_j + B^0_j  +  [f_j N^0_j ]^2 +  [f_j B^0_j ]^2 \, ,
\label{variance}
\end{eqnarray}
where $f_j$ is the uncorrelated bin-to-bin systematic error in the $j$-th bin: 2\% for the golden channel and 10\% for the 
silver channel, irrespectively of the energy bin,  of the baseline and of the stored muon polarity.
Notice that we can use the Gau\ss ian expression for the $\chi^2$  throughout our numerical simulation, 
 with the possible exception of the silver channel data at the longest baseline, for which a Poissonian expression
 could be more appropriate due to the extremely low statistics (as it can be seen from Tab.~\ref{tab:rates3}).
 However, as it will be shown in the following discussion on the results, the impact of the silver channel at that baseline
 is negligible. 

In the numerical analysis of this section, the following parameters in common between three- and four-family models 
have been kept fixed to their central values:\footnote{We will not introduce any label to distinguish these parameters between three- and four-family models.}
$\theta_{12} = 34^\circ$, $\theta_{23} = 45^\circ$; $\Dmq_\Sol = 7.9 \times 10^{-5}~\eVq$, 
$\Dmq_{31} = 2.4 \times 10^{-3}~\eVq$. 
The following two parameters specific to the four-family model have been also kept fixed: 
$\Dmq_\Sbl = 1~\eVq$ and $\delta_1 = 0$.
Eventually, we have marginalized over $\theta_{24} \in [0,12^\circ]$, $\theta_{34} \in [0,35^\circ]$ and $\delta_2,\delta_3 \in [0,360^\circ]$.
Matter effects have been included considering a constant matter density $\rho = 3.4$ g/cm$^3$
for the shortest baseline and $\rho = 4.3$ g/cm$^3$ for the longest one, 
computed averaging over the density profile in the PREM \cite{Dziewonski:1981xy} along the neutrino path.
We have checked that marginalization over a 10\% matter density uncertainty does not modify our results.

It is useful to show here the analytic expressions for the golden and silver channels oscillation probabilities in vacuum. For both channels it is mandatory to go beyond the $O(\epsilon^3)$ of the expansion in power of small parameters that was shown in eqs.~(\ref{eq:pee}-\ref{eq:pmus}). At high orders in powers of $\epsilon$, it is no longer possible to neglect terms proportional to the 
solar mass difference. In particular, for the short baseline $L = 3000$ km (the only case in which vacuum formul\ae \
give some insight on the numerical results), the parameter $\Delta_{21} L$ ranges from $O(\epsilon^3)$ 
to $O(\epsilon^4)$, depending on the neutrino energy. We will thus show the oscillation probabilities
$P_{e\mu}$ and $P_{e\tau}$ to order $\epsilon^8$ in powers of $\sqrt{\theta_{13}},\sqrt{\theta_{14}},\sqrt{\theta_{24}},\sqrt{\delta \theta_{23}},\theta_{34}$ and $(\Delta_{21} L)^{1/4}$.
For completeness, we have also computed the oscillation probability $P_{es}$ and $P_{ee}$ to the same order in $\epsilon$ ($P_{es}$ can be found in the Appendix), checking unitarity of the mixing matrix. 
Oscillation probabilities in matter at this order in $\epsilon$ are difficult to obtain and they will not  be presented here. 

The golden channel oscillation probability expanded to order $\epsilon^8$ in vacuum is:
\begin{eqnarray}
\label{eq:emuprobeps6}
P_{e\mu} &=& 2\, \theta_{14}^2 \theta_{24}^2 + 2 \left\lbrace 
\theta_{13}^2  (1 + 2 \delta \theta_{23}- \theta_{13}^2 - \theta_{14}^2 - \theta_{24}^2 )
     \right\rbrace
    \sin^2 \frac{\Delta_{31} L}{2} 
\nonumber\\
   &+& 2 \sqrt{2}\, \theta_{13} \theta_{14} \theta_{24} (1 + \delta \theta_{23}) \sin \left (\delta_2 - \delta_3 + \frac{\Delta_{31} L }{2} \right ) 
   \sin\frac{\Delta_{31} L }{2}
\nonumber\\
&+& \sin2\theta_{12}\,  \theta_{13}  (\Delta_{21} L) 
\cos \left ( \delta_1-\delta_2+\delta_3 - \frac{\Delta_{31} L }{2} \right ) \sin \frac{\Delta_{31} L}{2}
\nonumber\\
&+& \frac{1}{\sqrt{2}} \sin 2 \theta_{12}\, \theta_{14} \theta_{24} (\Delta_{21} L)  \sin\delta_1
- s_{12}^2\, \theta_{13}^2  (\Delta_{21} L) \sin\Delta_{31} L + \frac{1}{2} \sin^2 2 \theta_{12} (\Delta_{21} L)^2 \, ,
\nonumber  \\
\end{eqnarray}
whereas the silver channel oscillation probability at the same order is: 
\begin{eqnarray}
\label{eq:etauprobeps6}
P_{e\tau} &=& 2\, \theta_{14}^2 \theta_{34}^2
\nonumber \\
 &+& 2 \left\lbrace \theta_{13}^2 ( 1 - 2 \delta \theta_{23} - \theta_{13}^2 - \theta_{14}^2
  - \theta_{34}^2 + 2 \delta \theta_{23} \theta_{34}^2) 
  - \theta_{13}^2 \theta_{24} \theta_{34}\cos\delta_3 \right\rbrace
    \sin^2 \frac{\Delta_{31} L}{2}
    \nonumber \\
 &-& 2  \sqrt{2}\, \theta_{13} \theta_{14} \theta_{24}\theta_{34}^2\sin \left (\delta_2 - \delta_3 + \frac{\Delta_{31} L }{2} \right ) \sin \frac{\Delta_{31} L}{2}
\nonumber\\
&+& 2 \sqrt{2}\, \theta_{13} \theta_{14} \theta_{34} \left (1- \delta \theta_{23} - \frac{\theta_{34}^2}{2} \right)
\sin \left ( \delta_2  + \frac{\Delta_{31} L}{2} \right ) \sin \frac{\Delta_{31} L}{2}
\nonumber\\
&-&  \sin2\theta_{12}\,  \theta_{13} (1 - \theta_{34}^2 ) (\Delta_{21} L) 
\cos \left ( \delta_1-\delta_2+\delta_3 - \frac{\Delta_{31} L}{2} \right ) \sin \frac{\Delta_{31} L}{2}
\nonumber\\
&-&\frac{1}{\sqrt{2}} \sin 2\theta_{12}\, \theta_{14} \theta_{34} (\Delta_{21}L ) \sin(\delta_1+\delta_3)
-  s_{12}^2\, \theta_{13}^2   (\Delta_{21} L) \sin \Delta_{31} L + \frac{1}{2} \sin^2 2 \theta_{12} (\Delta_{21} L)^2 \, .
\nonumber \\
\end{eqnarray}

Notice that the golden and silver channel expressions have leading terms proportional to $\theta_{13}^2$, i.e. they are $O(\epsilon^4)$ (as it was anticipated in Sec.~\ref{sec:probs}). This is the same order at which leading 
terms arise in the three-family model. We expect, thus, a similar sensitivity to $\theta_{13}$ in both three- and four-model analyses at the Neutrino Factory using these channels. 
On the other hand, we expect poor sensitivity to $\theta_{14}$ using these channels.
First of all,
dependence on $\theta_{14}$ in the golden and silver channel arise only at higher orders in 
$\epsilon$ (at $O(\epsilon^6)$ for the golden channel and at $O(\epsilon^5)$ for the silver channel, respectively). 
Secondly, all the terms proportional to $\theta_{14}$
in the above probabilities appear in combination with
either $\theta_{34}$ or $\theta_{24}$, so when we evaluate
the value of $\Delta \chi^2$ by marginalizing over
$\theta_{34}$ and $\theta_{24}$, both of which have
tendency to be small because of the priors,
these two probability can become almost independent of $\theta_{14}$.
For this reason, we expect poor sensitivity to $\theta_{14}$ using these channels. 
Nevertheless we discuss the sensitivities to both $\theta_{13}$ and $\theta_{14}$
because the sensitivity to $\theta_{13}$ can have nontrivial
dependence on the true value of $\theta_{14}$
in the four neutrino case, as we will see below.

Eventually, the $\nu_e$ disappearance probability is: 
\begin{eqnarray}
\label{eq:eeprobeps6}
P_{ee} &=& 1 - 2\,\theta_{14}^2 (1 - \theta_{14}^2) - 4 \theta_{13}^2 (1 - \theta_{13}^2 - 2\, \theta_{14}^2) \sin^2 \frac{\Delta_{31} L }{2} 
\nonumber \\
&+& 2\, s^2_{12} \theta_{13}^2 (\Delta_{21} L) \sin \Delta_{31} L -  \sin^2 2 \theta_{12} (\Delta_{21} L)^2
\end{eqnarray}
Notice that this channel has leading $O(\epsilon^4)$ dependence on both $\theta_{13}$ and $\theta_{14}$. 
For this reason the most stringent bound on this angle has been obtained by reactor experiments, 
as we have shown in Sec.~\ref{sec:bounds}. A detector with an extremely  good electron identification efficiency 
is needed to perform this task, however, something beyond the reach of the MIND or MECC detectors. We are, thus, not 
considering this channel in the present paper. 

\begin{figure}[t]
\begin{center}
\begin{tabular}{cc}
\hspace{-0.5cm}
\includegraphics[width=7.5cm]{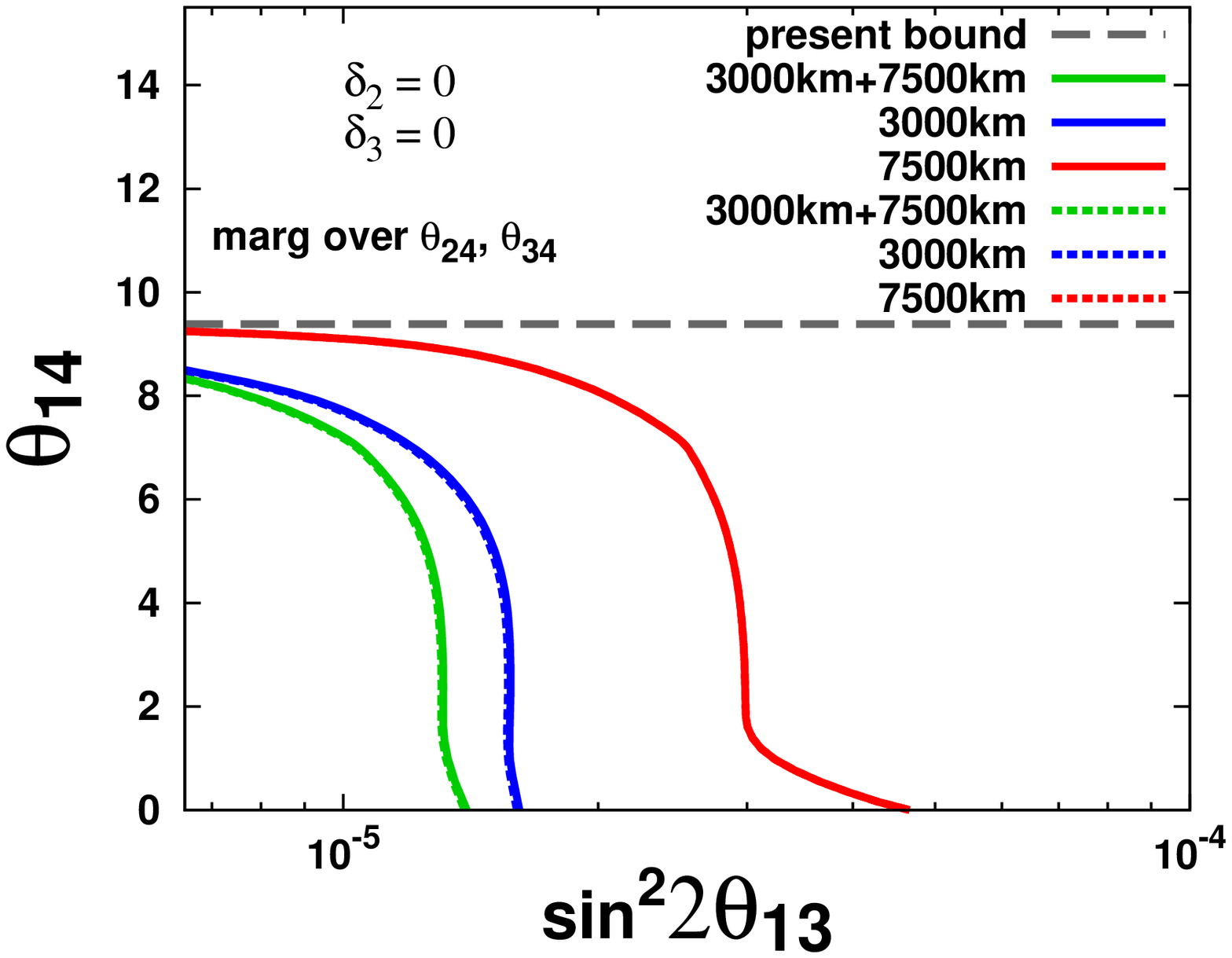} &
\includegraphics[width=7.5cm]{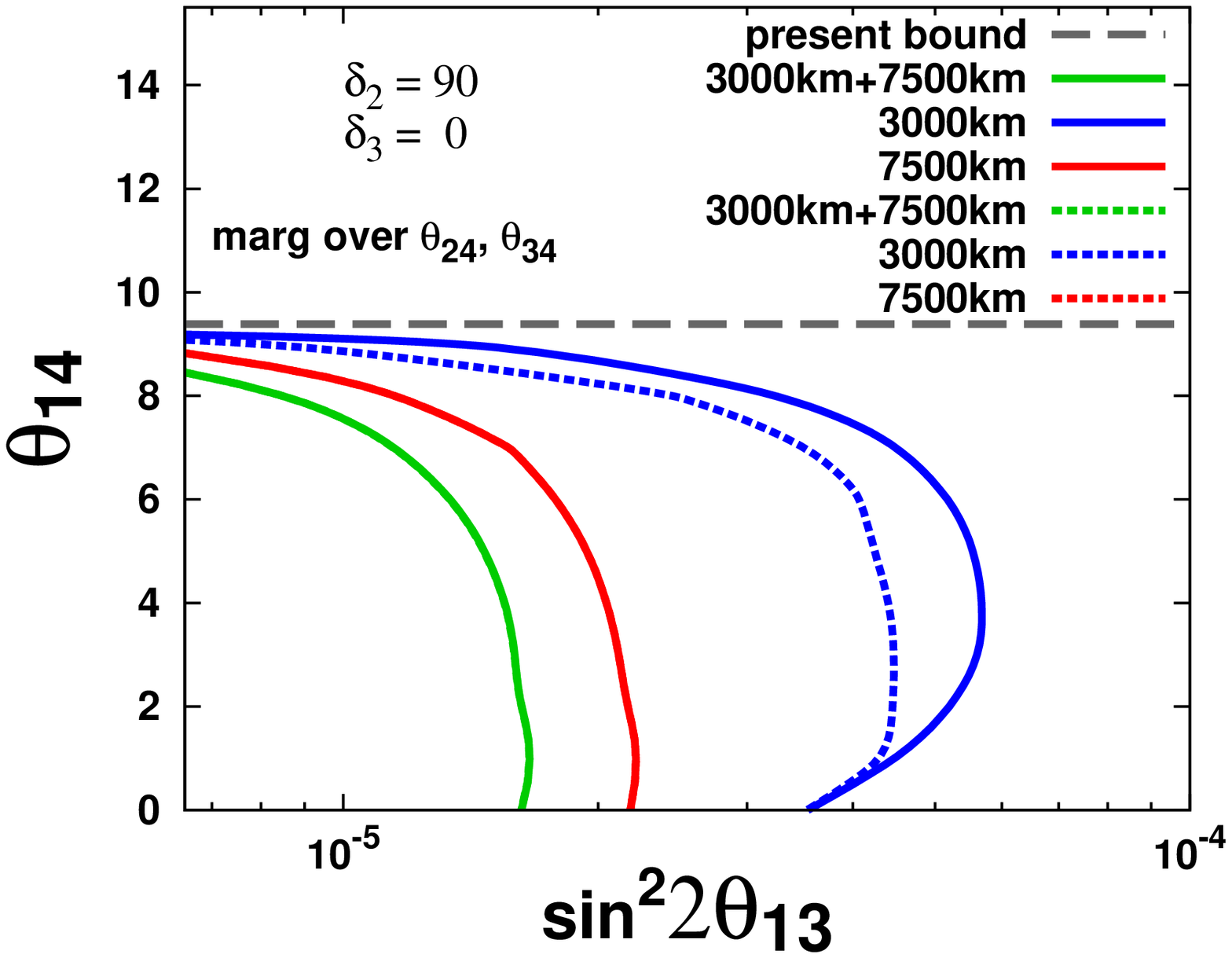} \\
\\
\includegraphics[width=7.5cm]{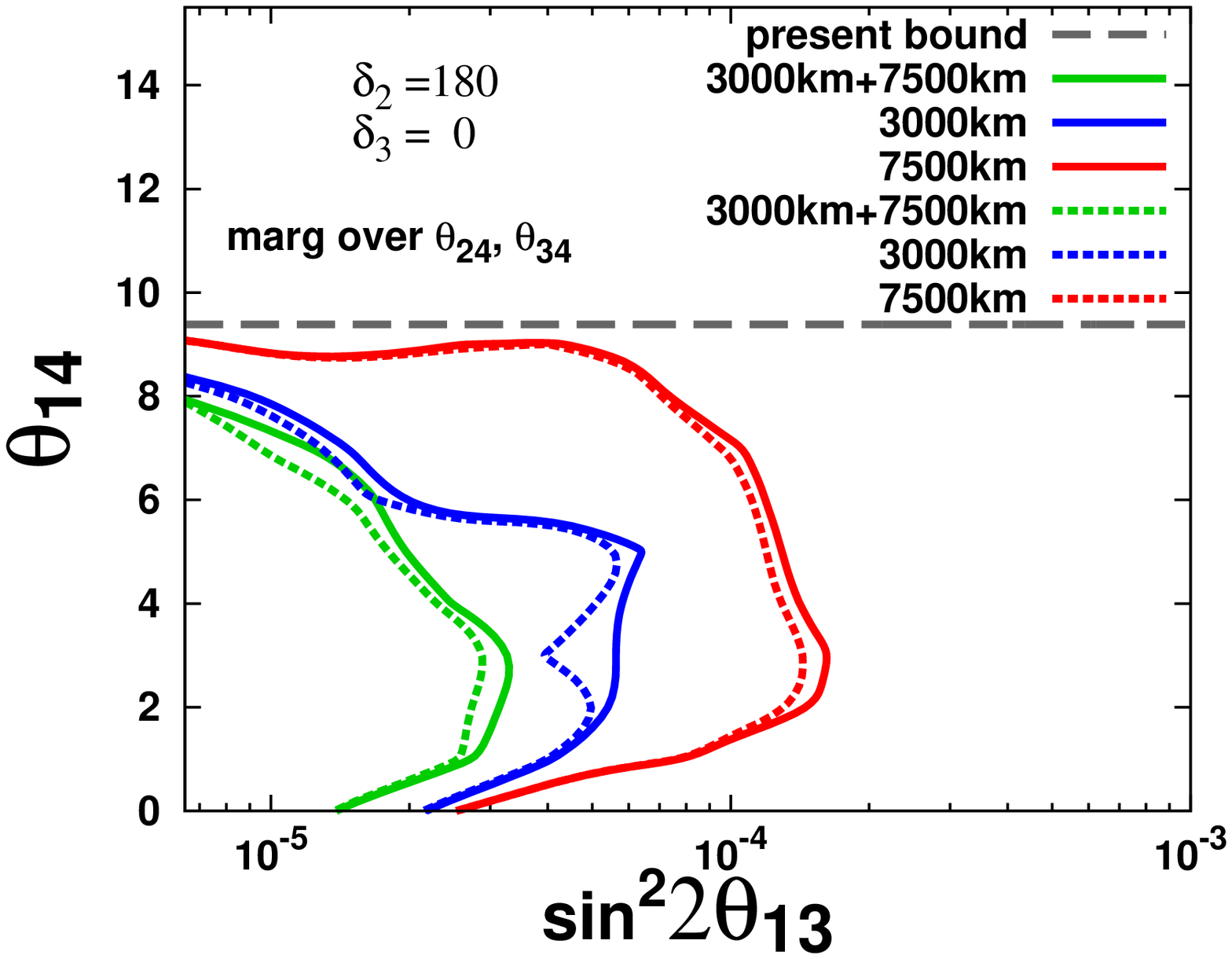} &
\includegraphics[width=7.5cm]{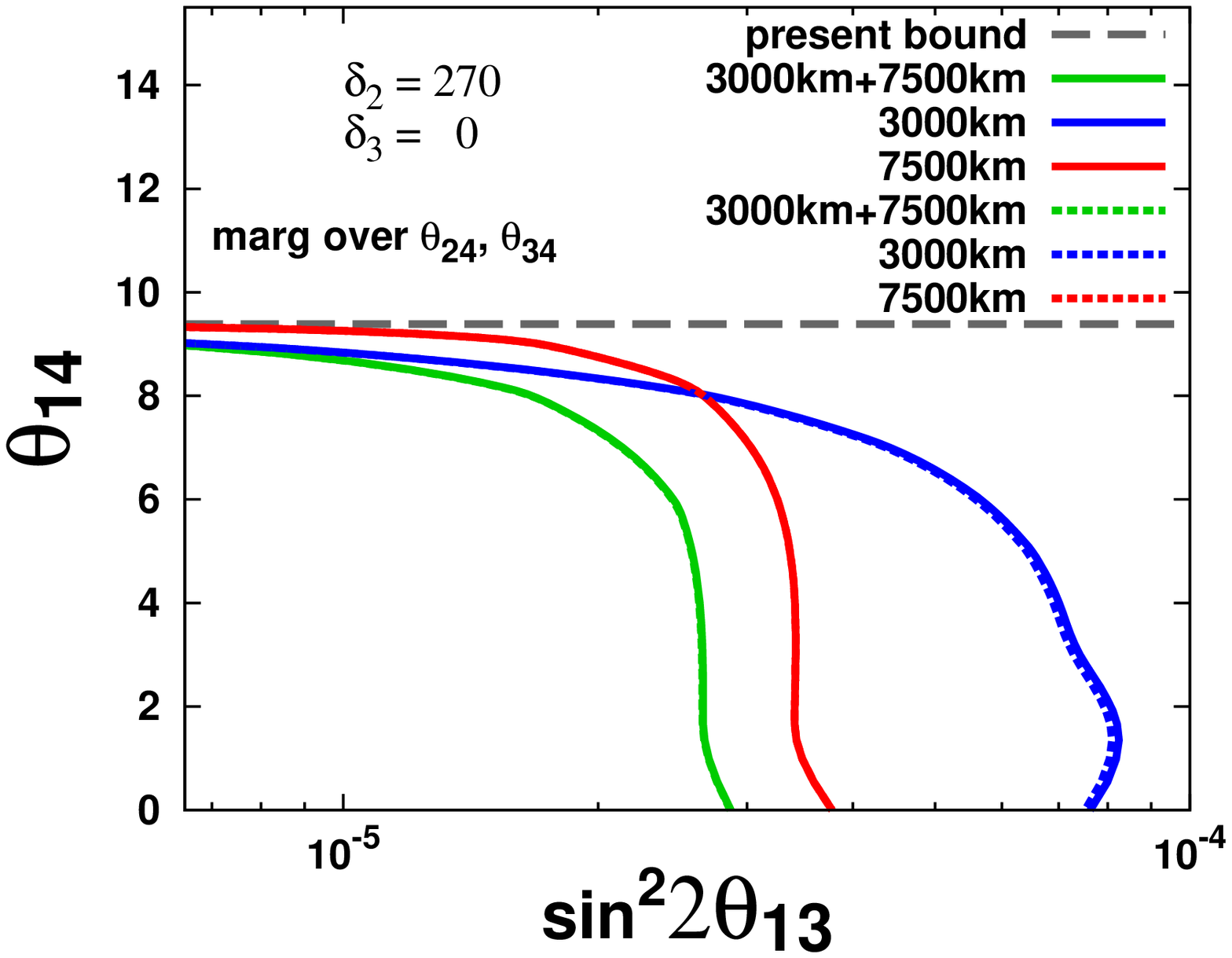}\\
\end{tabular}  
\caption{\label{fig:goldensilver1}\sl%
      Sensitivity limit at 90\% CL in the ($\sin^2 2\theta_{13}$, $\theta_{14}$) plane for $\delta_3 = 0$ and different values of $\delta_2$.
      The solid lines refer to the golden channel results, only.  Dashed lines stand for the 
      sum of golden and silver channel results. The colors are: blue for $L = 3000$ km; red for 
      $L = 7500$ km; green for the combination of the two baselines; the horizontal dashed grey line represents the present     
      bound on $\theta_{14}$.
      The four panels represent our results for: $\delta_2 = 0$ (top left); $\delta_2 = 90^\circ$ (top right);
      $\delta_2 = 180^\circ;$ (bottom left); $\delta_2 = 270^\circ$ (bottom right).
      }
\end{center}
\end{figure}

\begin{figure}[t]
\begin{center}
\begin{tabular}{cc}
\hspace{-0.5cm}
\includegraphics[width=7.5cm]{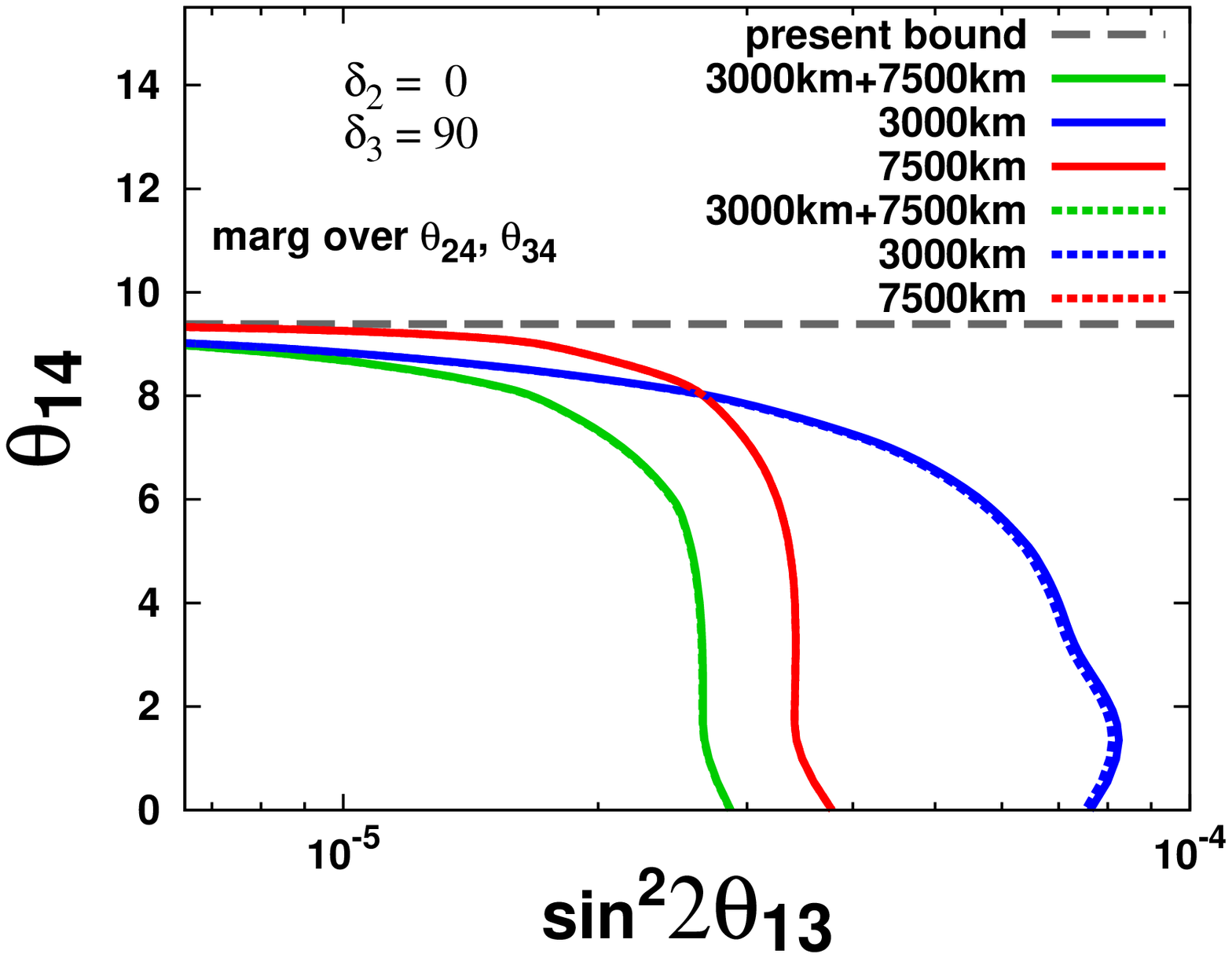} &
\includegraphics[width=7.5cm]{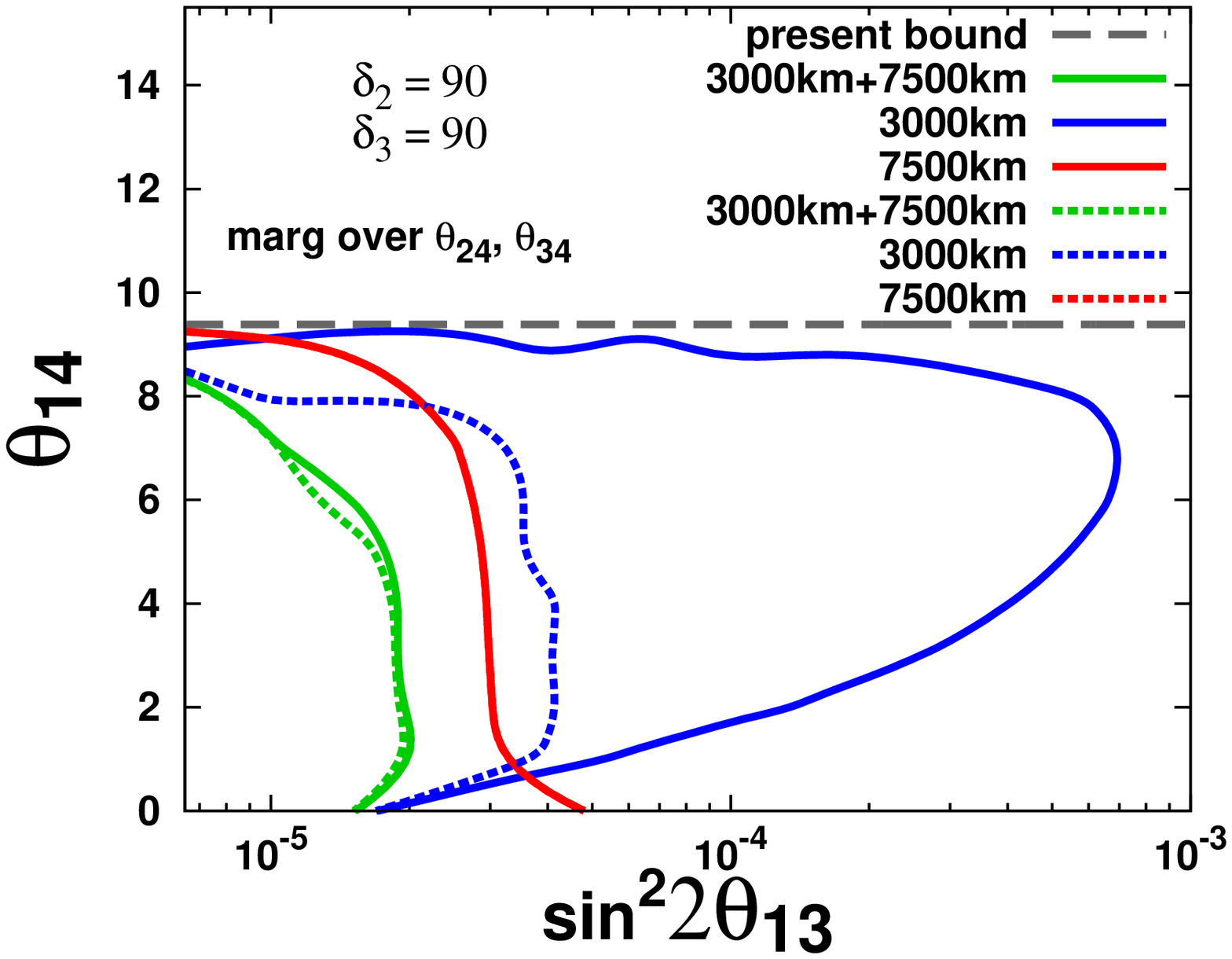} \\
\\
\includegraphics[width=7.5cm]{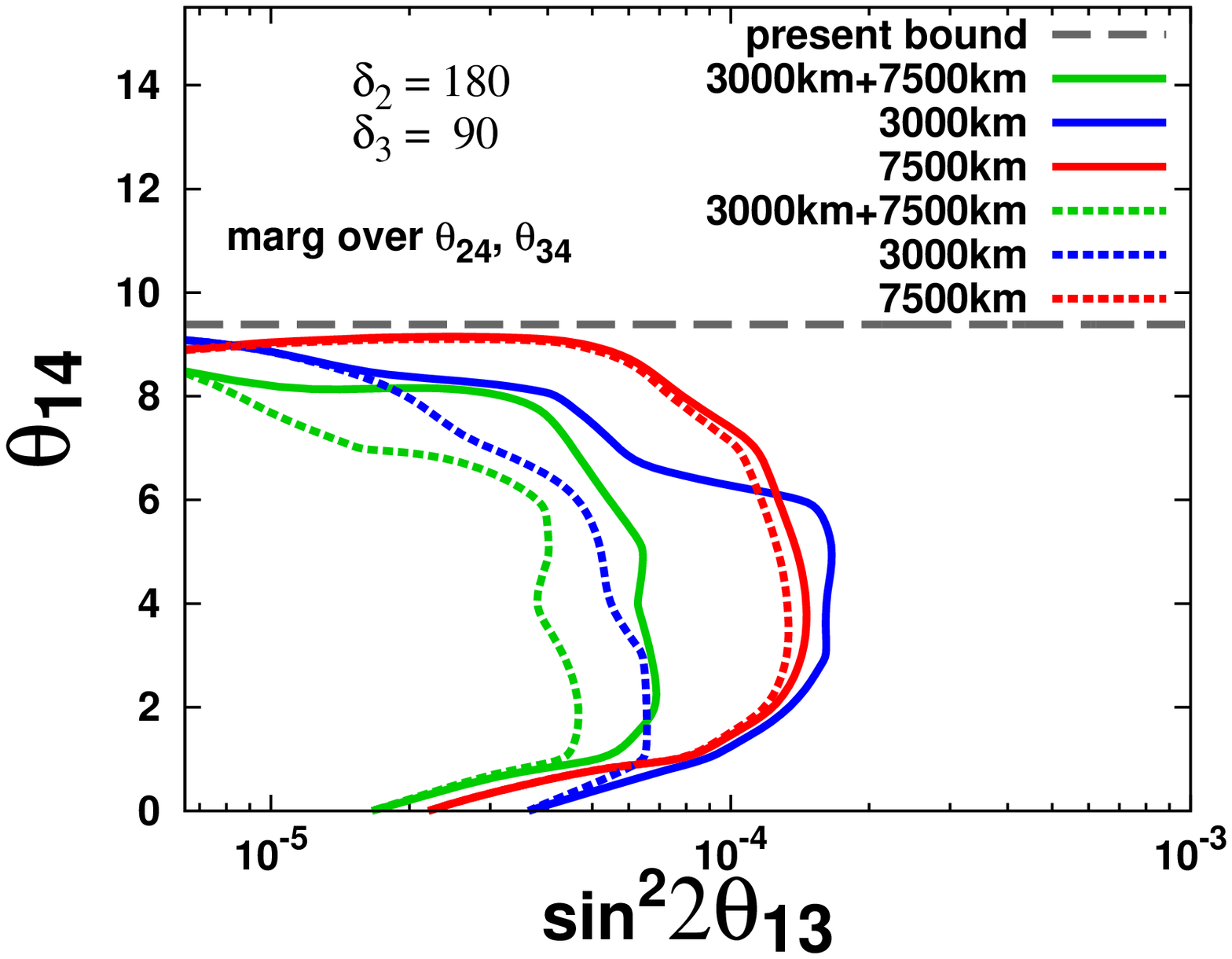} &
\includegraphics[width=7.5cm]{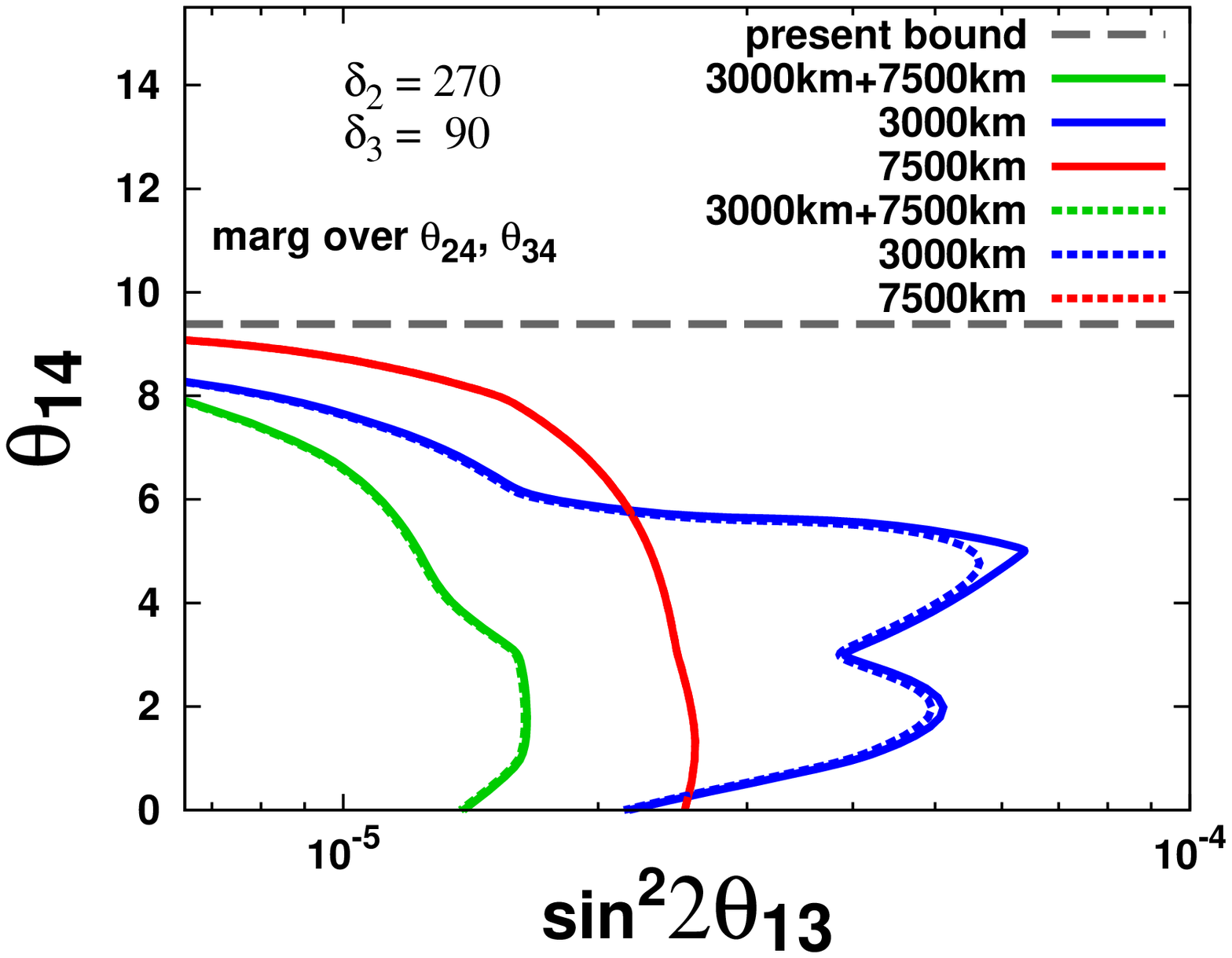}\\
\end{tabular}  
\caption{\label{fig:goldensilver2}\sl%
      Sensitivity limit at 90\% CL in the ($\sin^2 2\theta_{13}$, $\theta_{14}$) plane for $\delta_3 = 90^\circ$ and different values of $\delta_2$.
      The solid lines refer to the golden channel results, only.  Dashed lines stand for the 
      sum of golden and silver channel results. The colors are: blue for $L = 3000$ km; red for 
      $L = 7500$ km; green for the combination of the two baselines; the horizontal dashed grey line represents the present     
      bound on $\theta_{14}$.
            The four panels represent our results for: $\delta_2 = 0$ (top left); $\delta_2 = 90^\circ$ (top right);
      $\delta_2 = 180^\circ;$ (bottom left); $\delta_2 = 270^\circ$ (bottom right).
      }
\end{center}
\end{figure}

Figs.~\ref{fig:goldensilver1} and \ref{fig:goldensilver2} present  the sensitivity to (3+1)-sterile neutrinos in the 
$(\theta^{\rm (4fam)}_{13},\,\theta_{14})$-plane at 90\% CL 
at the 50 GeV Neutrino Factory setup for several representative choices of $\delta_2$ and $\delta_3$. 
In particular, $\delta_3$ has been fixed to $\delta_3 = 0$ and $90^\circ$ in Fig.~\ref{fig:goldensilver1}  and Fig.~\ref{fig:goldensilver2}, respectively. In both cases, the phase $\delta_2$ (that reduces to the three-family CP-violating phase $\delta$ in the limit  $\theta_{i4} \to 0$) has been fixed to $\delta_2 = 0, 90^\circ, 180^\circ$ and $270^\circ$. Results for the two baselines are shown both separately and summed:
blue lines stand for the shortest baseline data; red lines stand for the longest baseline data; green lines
stand for the sum of the two baselines. Solid lines stand for  golden channel data, only, whereas
dashed lines stand for the sum of golden and silver channels data. Eventually,  the horizontal dashed grey line represents 
the present bound on $\theta_{14}$ that takes into account the four-family analysis of the atmospheric and reactor data. 

First of all, notice that the golden and silver channel data show a very limited sensitivity to $\theta_{14}$ when marginalizing over
$\theta_{24}$ and $\theta_{34}$, as it was expected
by inspection of the oscillation probabilities $P_{e\mu}$ and $P_{e \tau}$, eqs.~(\ref{eq:emuprobeps6},\ref{eq:etauprobeps6}).

Secondly, when we focus on the golden channel results (solid lines), we notice a rather different behavior at the short and long baseline. 
At the short baseline, the sensitivity to $\theta_{13}^{\rm (4fam)}$ is significantly phase-dependent: the maximal sensitivity
to $\theta_{13}^{\rm (4fam)}$ ranges from $\sin^2 2 \theta_{13}^{\rm (4fam)} = 1.5 \times 10^{-5}$ for $\delta_2 = \delta_3 = 0$
to $6 \times 10^{-3}$ for $\delta_2 = \delta_3 = 90^\circ$, with a strong dependence on the $\theta_{14}$ value. 
This is a consequence of cancellations between the term proportional to $\theta_{13} (\Delta_{21} L)$ in the third line of eq.~(\ref{eq:emuprobeps6}), that in three-family represents the interference between solar and atmospheric oscillations, and the term proportional to $\theta_{13} \theta_{14} \theta_{24}$ in the second line of the same equation.

The behavior is extremely different when we consider the golden channel at the long baseline. The location of the 
far detector corresponds to the three-family  "magic baseline"  \cite{Huber:2003ak}, where the three-family dependence on 
the CP-violating phase $\delta$ vanishes. Notice that this happens as a consequence of the vanishing of the term in the 
third line of eq.~(\ref{eq:emuprobeps6}) when computed in the Earth matter. This is, indeed, the only term 
through which a $\delta$-dependence enters into the three-family golden channel oscillation probability. 
The four-family term in the third line of eq.~(\ref{eq:emuprobeps6}), that reduces to the three-family one for 
vanishing $\theta_{i4}$ and $\delta_1,\delta_3$, vanishes also. Therefore, no cancellations between different 
terms occur at this baseline and sensitivity to $\theta_{13}^{\rm (4fam)}$ depends much less than for the short baseline
on the values of $\delta_2,\delta_3$ and $\theta_{14}$. 
This is precisely what we can see in all panels of Fig.~\ref{fig:goldensilver1} and \ref{fig:goldensilver2}.

A similar effect can be observed when we add the silver channel data to the golden channel ones at the short baseline. 
In this case, the term in the third line of eq.~(\ref{eq:emuprobeps6}) cancels
 with the term in the fifth line of eq.~(\ref{eq:etauprobeps6}) at the leading order, leaving an $O(\epsilon^8) $ term
suppressed by $\theta_{13} \theta_{34}^2 (\Delta_{21} L)$. Also in this case, we see indeed in all panels of both figures
that a reduced dependence of the sensitivity to $\theta_{13}^{\rm (4fam)}$ from $\delta_2, \delta_3$
and $\theta_{14}$ is achieved. Particularly striking is the case of $\delta_2 = \delta_3 = 90^\circ$, Fig.~\ref{fig:goldensilver2}
(top right), where we can see that the sensitivity to $\theta_{13}^{\rm (4fam)}$ goes from 
$\sin^2 2 \theta_{13}^{\rm (4fam)} = 6 \times 10^{-3}$ for the golden channel alone to $\sin^2 2 \theta_{13}^{\rm (4fam)} = 3 \times 10^{-5}$ for the combination of golden and silver channels. 
On the other hand, the silver channel statistics at $L = 7500$ km from the source is extremely poor
(see Tab.~\ref{tab:rates3}). For this reason we have no impact whatsoever of this channel on the $\theta_{13}^{\rm (4fam)}$ sensitivity at the long baseline.

The performance of the detectors located at the two baselines differ significantly depending on the specific
choices of $\delta_2$ and $\delta_3$. For most of the choices of $(\delta_2,\delta_3)$, the longest baseline outperforms the shortest one, with the only exceptions of $\delta_2 = 0,180^\circ; \delta_3 = 0$ (Fig.~\ref{fig:goldensilver1}, left panels) and 
$\delta_2 = 180^\circ; \delta_3 = 90^\circ$ (Fig.~\ref{fig:goldensilver2}, bottom left panel). 

The combination of the two channels and the two baselines at the 50 GeV Neutrino Factory has a rather good sensitivity to 
$\theta_{13}^{\rm (4fam)}$: in the absence of a signal, we can exclude values of $\theta_{13}^{\rm (4fam)}$ larger than
$\sin^2 2 \theta_{13}^{\rm (4fam)} \leq 4 \times 10^{-5}$ for any of the choices of $(\delta_2,\delta_3)$ shown.
The sensitivity to the active-sterile mixing angle $\theta_{14}$, on the other hand, is $\theta_{14} \leq 9^\circ$ when marginalizing over
$\theta_{24}$ and $\theta_{34}$.

\begin{figure}[t]
\begin{center}
\hspace{-0.5cm}
\begin{tabular}{cc}
\includegraphics[width=7.5cm]{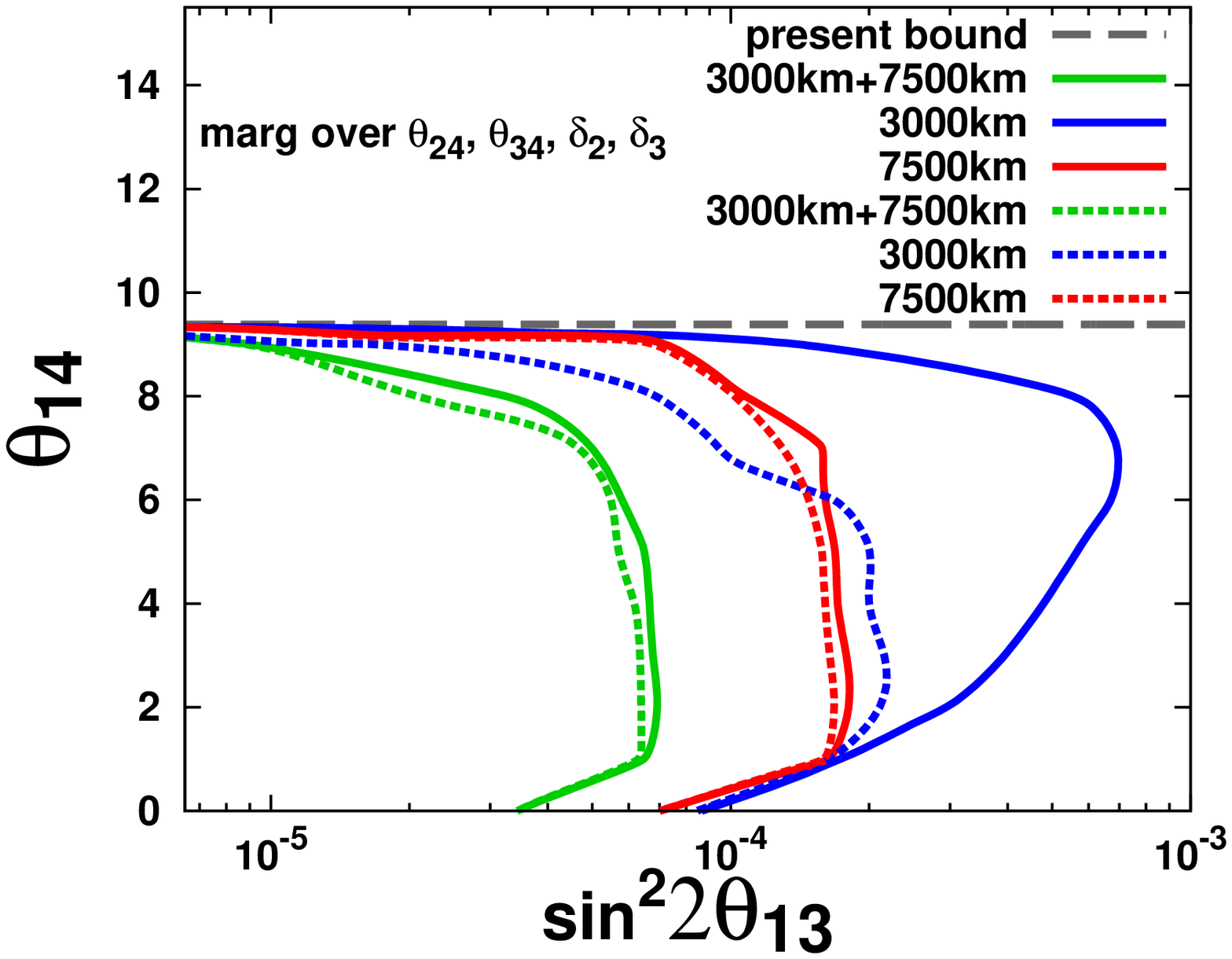} &
\includegraphics[width=7.5cm]{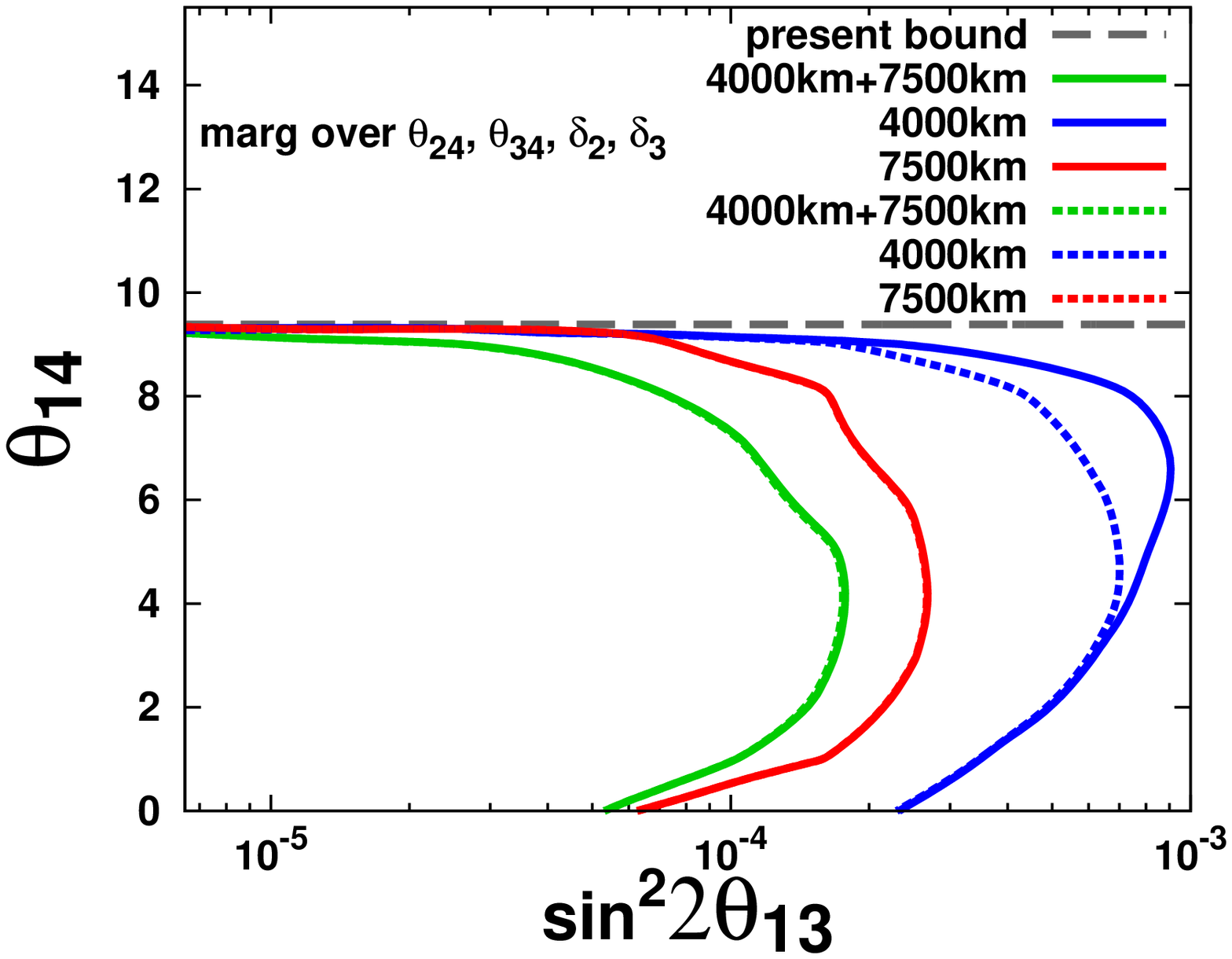} 
\end{tabular}
\caption{\label{fig:goldensilverd3}\sl%
      Sensitivity limit at 90\% CL in the ($\sin^2 2\theta_{13}$, $\theta_{14}$) plane, marginalizing over $\theta_{24},\theta_{34},\delta_2$    
      and $\delta_3$.
      The solid lines refer to the golden channel results, only.  Dashed lines stand for the 
      sum of golden and silver channel results. The colors are: blue for the shortest baseline; red for 
      longest baseline; green for the combination of the two baselines; the horizontal dashed grey line represents
      the present bound on $\theta_{14}$.
      Left panel: 50 GeV Neutrino Factory; Right panel: 20 GeV Neutrino Factory.
            }
\end{center}
\end{figure}

We eventually compare the results obtained using the 50 GeV Neutrino Factory setup, Fig.~\ref{fig:goldensilverd3}(left), 
with those that can be achieved using the 20 GeV Neutrino Factory ISS-inspired setup, Fig.~\ref{fig:goldensilverd3}(right). 
As before, solid lines stand for golden channel data, only, whereas dashed lines stand for the combination of golden
and silver channel data. Blue lines stand for the shortest baseline (3000 km or 4000 km); red lines for the longest
baseline (7500 km in both setups); green lines for the combination of the two; eventually, the grey dashed line represents the present bound on $\theta_{14}$. 
In addition to $\theta_{24}$ and $\theta_{34}$, we have marginalized over the CP-violating phases 
$\delta_2$ and $\delta_3$, with $\delta_2,\delta_3 \in [0,360^\circ]$.
The rest of the parameters have been kept fixed to the values given above. 

First of all notice that for both setups the  $\theta_{13}^{\rm (4fam)}$-sensitivity that can be reached using golden channel data, only, is much poorer at the short baseline than at the long baseline. We have found $\theta_{13}^{(\rm 4fam)} \leq 6 \times 10^{-4} (9 \times 10^{-4})$ for any value of $\theta_{14}$ at the short baseline for the 50 GeV (20 GeV) setup, to be compared with $\theta_{13}^{(\rm 4fam)} \leq 2 \times 10^{-4} (3 \times 10^{-4}$) at the long baseline for the 50 GeV (20 GeV) setup. The second relevant point is that the silver channel significantly improves the $\theta_{13}^{\rm (4fam)}$-sensitivity for the 50 GeV setup, but only marginally for the 20 GeV one. We can see that the $\theta_{13}^{\rm (4fam)}$-sensitivity at the short baseline using the combination of golden and silver channels becomes
$\theta_{13}^{(\rm 4fam)} \leq 2 \times 10^{-4} (6 \times 10^{-4})$ for the 50 GeV (20 GeV) setup. 
When considering the long baseline we can see, however, that the impact of the silver channel becomes marginal
(though it can still be observed for the 50 GeV setup). 

The $\theta_{13}^{\rm (4fam)}$-sensitivity that can be reached using the combination of the two baselines
is: $\theta_{13}^{(\rm 4fam)} \leq 6 \times 10^{-5} (1.5 \times 10^{-4})$ for any value of $\theta_{14}$ at the 
50 GeV (20 GeV) setup. The inclusion of the silver channel data does not modify significantly these results. 

Our conclusion is that the higher energy setup has a much greater ultimate sensitivity to $\theta_{13}^{(\rm 4fam)}$
than the ISS-inspired 20 GeV one. On the other hand, both setups are not able to improve the present bounds on 
$\theta_{14}$ when marginalizing over $\theta_{24}, \theta_{34}$ (see, however, Sec.~\ref{sec:discrimination}).
The silver channel can significantly improve the $\theta_{13}^{\rm (4fam)}$-sensitivity when only one baseline 
is considered at the 50 GeV setup. However, it has a negligible impact when the combination of the golden channel data at the two baselines is considered. 

Eventually, we have studied the impact of the correlated systematic errors $\alpha_s,\beta_s$
in eq.~(\ref{chi2-13-14}) on our results, For both the 50 GeV and 20 GeV setups, we have found that
the results shown in Fig.~\ref{fig:goldensilverd3} do not change for $\alpha_s = \beta_s = 0$. Golden and silver
channels are indeed dominated by statistical errors. 

\subsubsection{Sensitivity to $U_{e4}U_{\mu4}$ and $U_{e4}U_{\tau4}$}
\label{sec:ue4umt4}
\begin{figure}[t]
\begin{center}
\begin{tabular}{cc}
\hspace{-0.5cm}
\includegraphics[width=7.5cm]{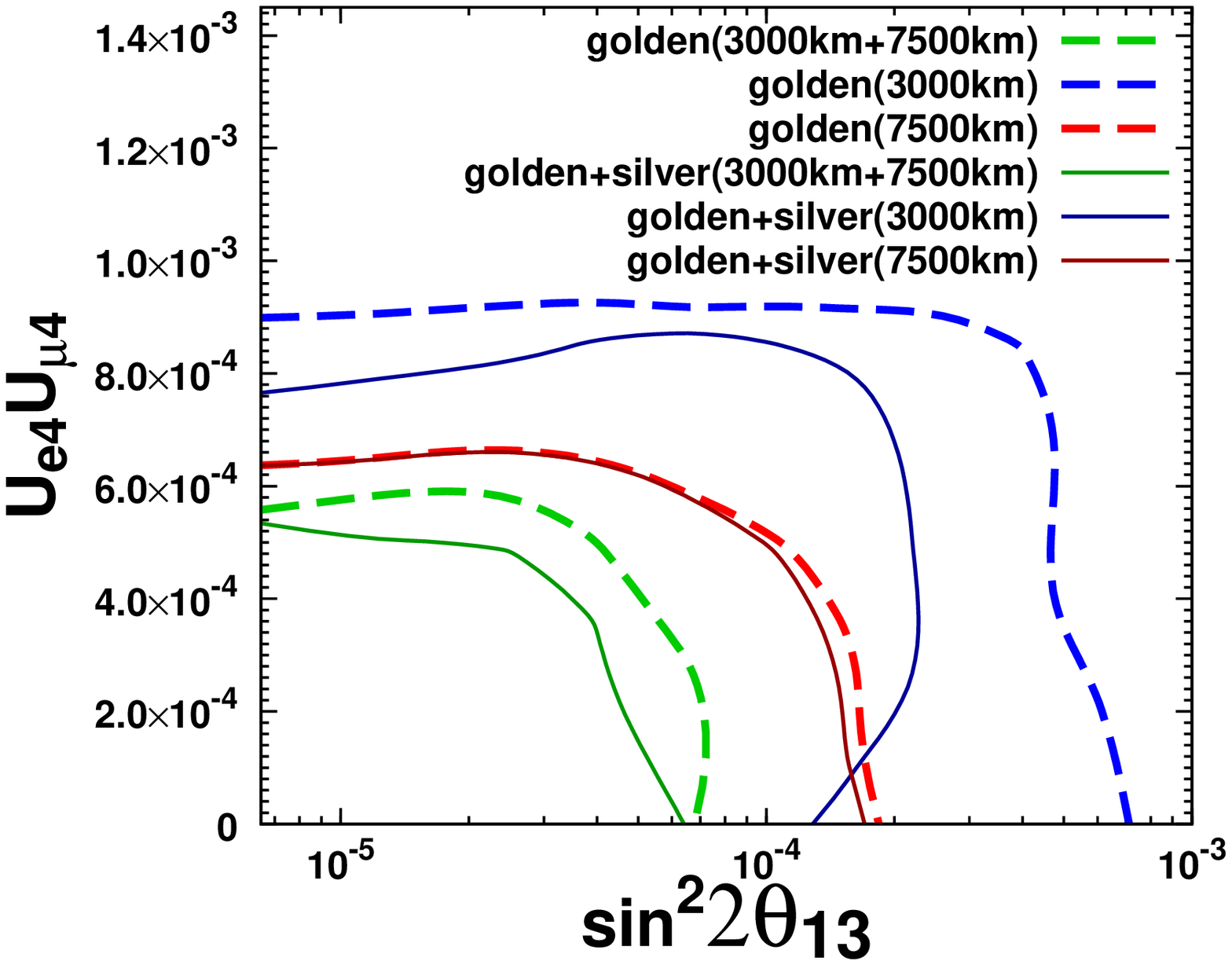} &
\includegraphics[width=7.5cm]{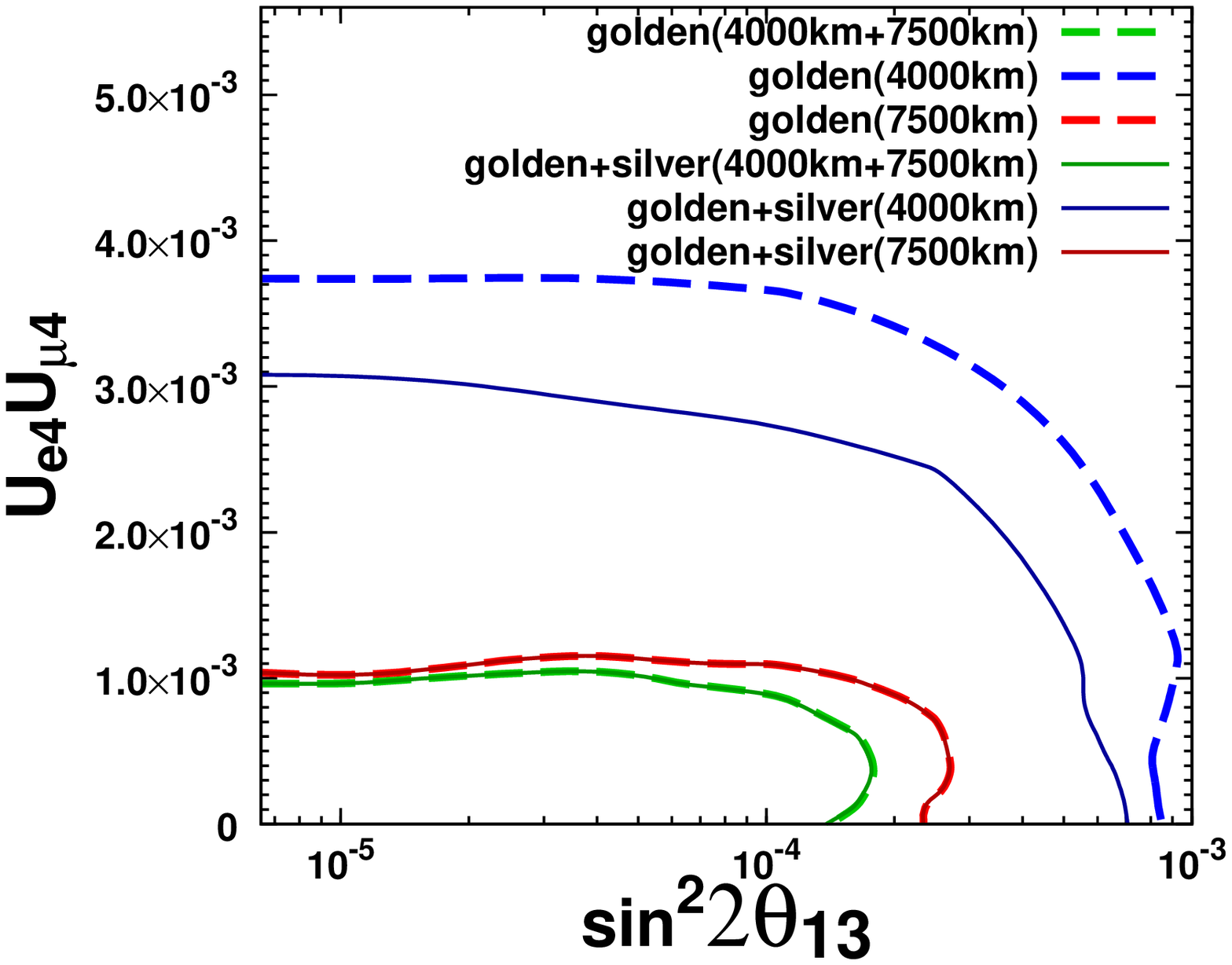}\\
\includegraphics[width=7.5cm]{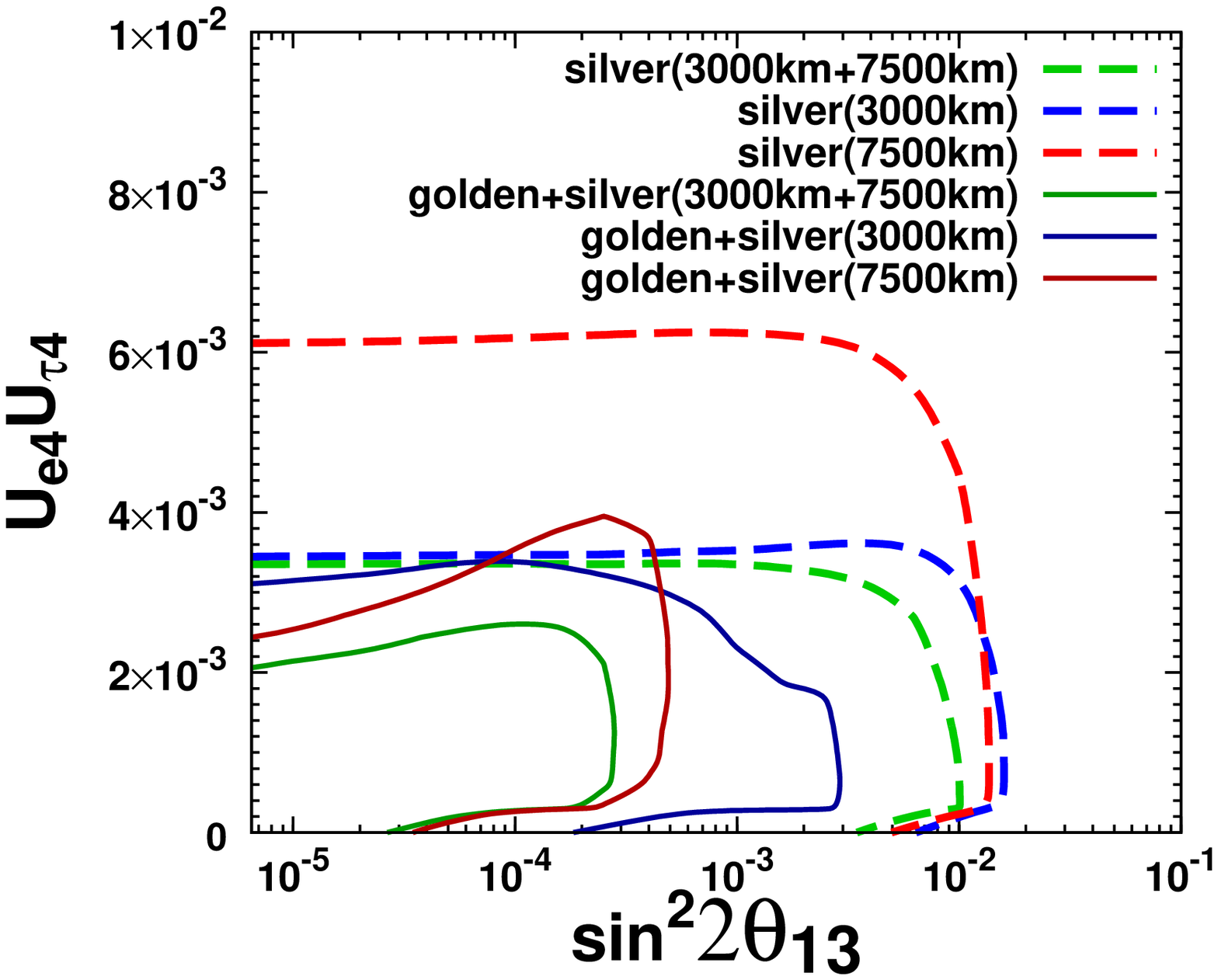} &
\includegraphics[width=7.5cm]{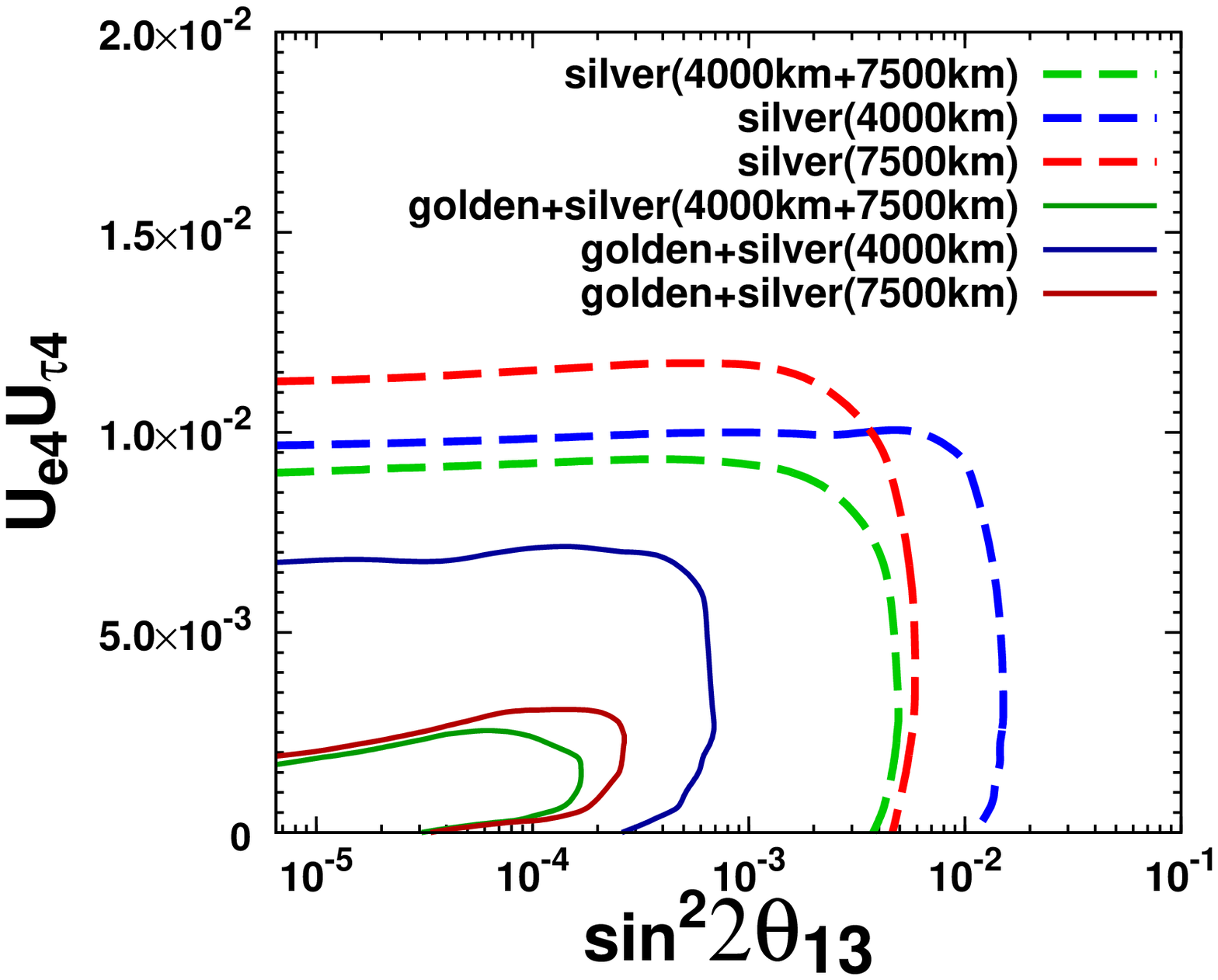}
\end{tabular}  
\caption{\label{fig:ue4umt4}\sl%
Sensitivity limit at 90\% CL to $U_{e4}U_{\mu4}$ and $U_{e4}U_{\tau4}$.
$\Delta\chi^2$ is evaluated for a fixed set of values of ($\sin^22\theta_{13}$, $U_{e4}U_{\mu4}$)
(upper panels) or a fixed set of values of ($\sin^22\theta_{13}$, $U_{e4}U_{\tau4}$) (lower panels),
marginalizing over $s_{14}/s_{24}$, $\theta_{34}$, $\delta_2$ and $\delta_3$
(upper panels), or over $s_{14}/s_{34}$, $\theta_{24}$, $\delta_2$ and $\delta_3$ (lower panels).
Left panels: 50 GeV Neutrino Factory; Right panels: 20 GeV Neutrino Factory.
The current bound on $U_{e4}U_{\mu4}$ ($U_{e4}U_{\tau4}$) is 0.02 (0.06).}
\end{center}
\end{figure}

Since sensitivity to $\theta_{14}$ at the Neutrino Factory is poor,
it is worth investigating whether the Neutrino Factory has sensitivity
to other combinations of the mixing matrix elements.
From the form of the appearance oscillation probability
\begin{eqnarray}
P(\nu_\alpha\to\nu_\beta)=4\mbox{\rm Re} \left[U_{\alpha 3} U_{\beta 3}^\ast
(U_{\alpha 3}^\ast U_{\beta 3} + U_{\alpha 4}^\ast U_{\beta 4})\right]
\sin^2\left(\frac{\Delta m^2_{31}L}{4E}\right)+\cdots,
\nonumber
\end{eqnarray}
we can expect that the golden and silver channels have some
sensitivity to $U_{e4}U_{\mu4}$ and $U_{e4}U_{\tau4}$.
In the present parametrization (\ref{eq:3+1param2})
of the mixing matrix, we have
$U_{e4}U_{\mu4}=s_{14}c_{14}s_{24}= s_{14}s_{24} + O(\epsilon^6)$
and $U_{e4}U_{\tau4}=s_{14}c_{14}c_{24}s_{34}= s_{14}s_{34} + O (\epsilon^5)$,
where we have used the fact that $|\theta_{14}|$ and $|\theta_{24}|$
are small.  In the analysis of sensitivity to
$U_{e4}U_{\mu4}$ ($U_{e4}U_{\tau4}$), instead of using
$|\theta_{14}|$ and $|\theta_{24}|$
($|\theta_{14}|$ and $|\theta_{34}|$) as the independent variables,
it is convenient to take $s_{14}s_{24}$ and $s_{14}/s_{24}$
($s_{14}s_{34}$ and $s_{14}/s_{34}$) as the independent ones,
respectively.
Taking this basis we have performed analysis on sensitivity to
$U_{e4}U_{\mu4}$ and $U_{e4}U_{\tau4}$.
The sensitivity is defined as in Sect.~\ref{sec:subth13th14} with straightforward replacements.
The results are shown in Fig.\ref{fig:ue4umt4} and they indicate that the combination of the
golden and silver channels has good sensitivity to these variables.
To see how much improvement we have, we take the
square root of $U_{e4}U_{\mu4}$ or $U_{e4}U_{\tau4}$,
so that these factors correspond roughly to sine of some
angle.  Then the upper bound for $\sqrt{U_{e4}U_{\mu4}}$
by the 50 GeV (20 GeV) Neutrino Factory is $\sqrt{5\times10^{-4}}$
($\sqrt{1\times10^{-3}}$), which is about 15\% (20\%) of the current bound, $\sqrt{U_{e4}U_{\mu4}} \leq \sqrt{0.02}$.  
Similarly, the upper bound for $\sqrt{U_{e4}U_{\tau4}}$ by both 50 GeV and 20 GeV Neutrino Factory is
$\sqrt{2\times10^{-3}}$, which is about 20\% of the current bound, $\sqrt{U_{e4}U_{\tau4}} \leq 0.06$.

For both $U_{e4}U_{\mu4}$ and $U_{e4}U_{\tau4}$ plots,
we see that both the golden and silver channels play a role in giving
the constraints.  As it is expected from statistics, the
result by the 50 GeV Neutrino Factory is better
for $U_{e4}U_{\mu4}$, but the sensitivity to $U_{e4}U_{\tau4}$ is almost the same for the two setups (notice that the 
scale on the vertical axis for the left and right panels are different).
For the 20 GeV case, the data at 7500 km perform very well
and the combined data of 4000 km and 7500 km give a result
almost comparable to that of 50 GeV.

It is interesting to note that the golden channel also plays
a role in improving sensitivity to $U_{e4}U_{\tau4}\propto s_{34}$.
We have obtained a lengthy analytic formula
for the golden channel in matter with some approximation, i.e.,
to quartic order in $\epsilon$ while keeping all orders in $\theta_{34}$
and we have found that dependence on $\theta_{34}$ appears
through the form of $O(\epsilon^4)\times s_{34}^2A_e/\Delta E_{31}$.
This explains why the golden channel has some sensitivity to
$U_{e4}U_{\tau4}$
through the matter effect at the longer baseline,
once we choose a suitable set of the independent parameters to
vary.

\subsection{Sensitivity to $(\theta_{24},\theta_{34})$}
\label{sec:th24th34}

The sensitivity is defined as in the previous section: we first compute the expected number of events  for 
$\nu_\mu \to \nu_\mu$ and $\nu_\mu \to \nu_\tau$ oscillations  for the input values $\theta^{\rm (4fam)}_{13} = 0$ 
and $\theta_{14} = \theta_{24} = \theta_{34} = 0$, $N^0$;
we then compute the expected number of events in the ($\theta_{24},\theta_{34}$)-plane for the same oscillation channels in the four-family model.
The $\Delta \chi^2$, computed as in eq.~(\ref{chi2-13-14}) with respect to the "true" value $N^0$,  is eventually evaluated.

Contrary to the case of the golden and silver channels, however, 
the number of expected background events is much smaller than the signal for both the disappearance and discovery channels. The effect of $\alpha_b$ and $\beta_b$ is, thus, negligible.  Hence, we will not perform minimization with respect to
$\alpha_b$ and $\beta_b$ for these channels and put these parameters to zero hereafter.
As in the case of the golden and silver channels,
the variance is defined by eq. (\ref{variance}),
where $N^0_j$ in this case are the number of events of
the disappearance or discovery channels,
and the uncorrelated bin-to-bin systematic error $f_j$
is 5\% for the $\nu_\mu$ disappearance channel and 10\% for the 
discovery channel, irrespectively of the energy bin, of the baseline and of the stored muon polarity. 

In the numerical analysis, the central values of the following parameters in common between three- and four-family models 
are: $\theta_{12} = 34^\circ$, $\theta_{13} = 5.7^\circ$, $\theta_{23} = 45^\circ$; $\Dmq_\Sol = 7.9 \times 10^{-5}~\eVq$, 
$\Dmq_{31} = 2.4 \times 10^{-3}~\eVq$; $\delta_2 = \delta = 0$ (where $\delta$ is the three-family CP-violating phase). 
The central values for the following three parameters specific to the four-family model are: 
$\Dmq_\Sbl = 1~\eVq$, $\theta_{14} = 0$ and $\delta_1 = 0$. 
Matter effects have been included considering a constant matter density $\rho = 3.4$ g/cm$^3$
for the shortest baseline and $\rho = 4.3$ g/cm$^3$ for the longest one, 
computed averaging over the density profile in the PREM \cite{Dziewonski:1981xy} along the neutrino path.
We have checked that marginalization over a 10\% matter density uncertainty does not modify our results.

\begin{figure}[t]
\begin{center}
\begin{tabular}{cc}
\hspace{-0.5cm}
\includegraphics[width=7.5cm]{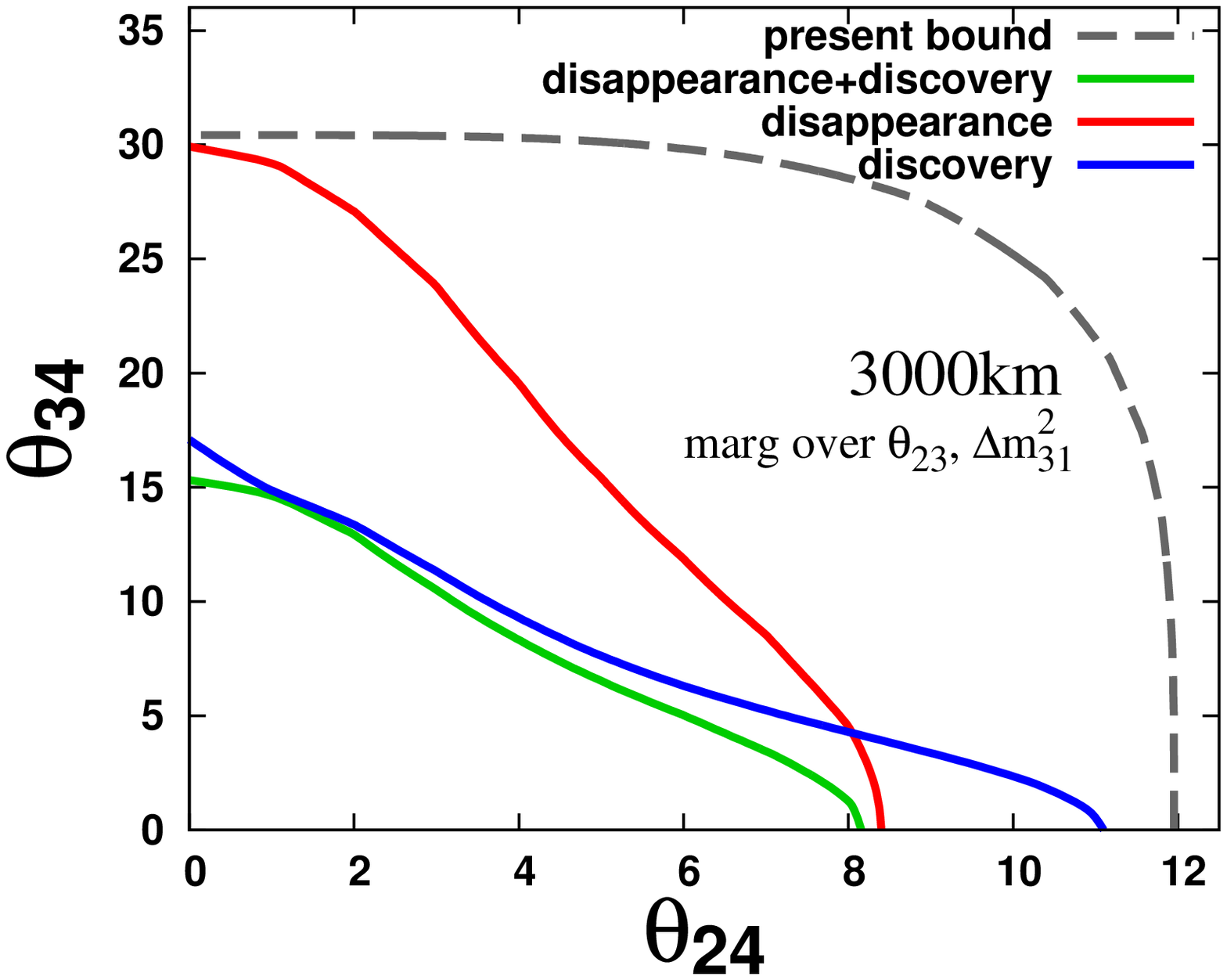} &
\includegraphics[width=7.5cm]{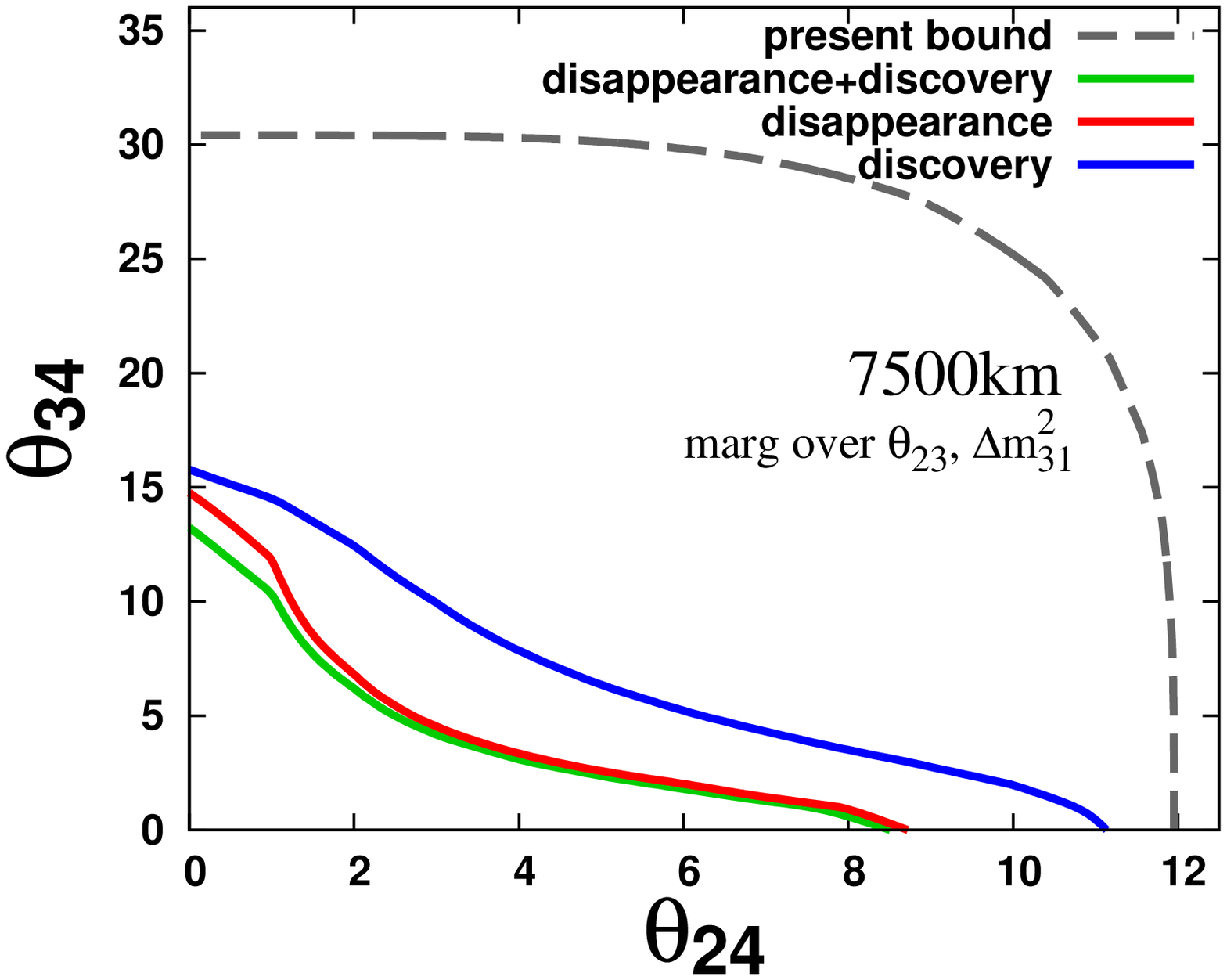}\\
\end{tabular}  
\caption{\label{fig:marginalizingt23dm31}\sl%
      Sensitivity limit at 90\% CL to $\theta_{24}$ and $\theta_{34}$, for fixed $\delta_3 = 0$, 
      whilst marginalizing over $\theta_{23}$ in the range $\theta_{23} \in [40^\circ,50^\circ]$
      and  $\Dmq_{31}$ in the range  $\Dmq_{31} \in [2.0,2.8] \times 10^{-3}$ eV$^2$. 
      Red lines stand for the disappearance channel $\nu_\mu \to \nu_\mu$; blue lines stand for the discovery channel
      $\nu_\mu \to \nu_\tau$;  green lines stand for the combination of both; the dashed grey line represents the present bound
       on $\theta_{24}$ and $\theta_{34}$. Left: $L=3000$ km baseline, Right: $L=7500$ km baseline.}
\end{center}
\end{figure}

\begin{figure}[t]
\begin{center}
\begin{tabular}{cc}
\hspace{-0.5cm}
\includegraphics[width=7.5cm]{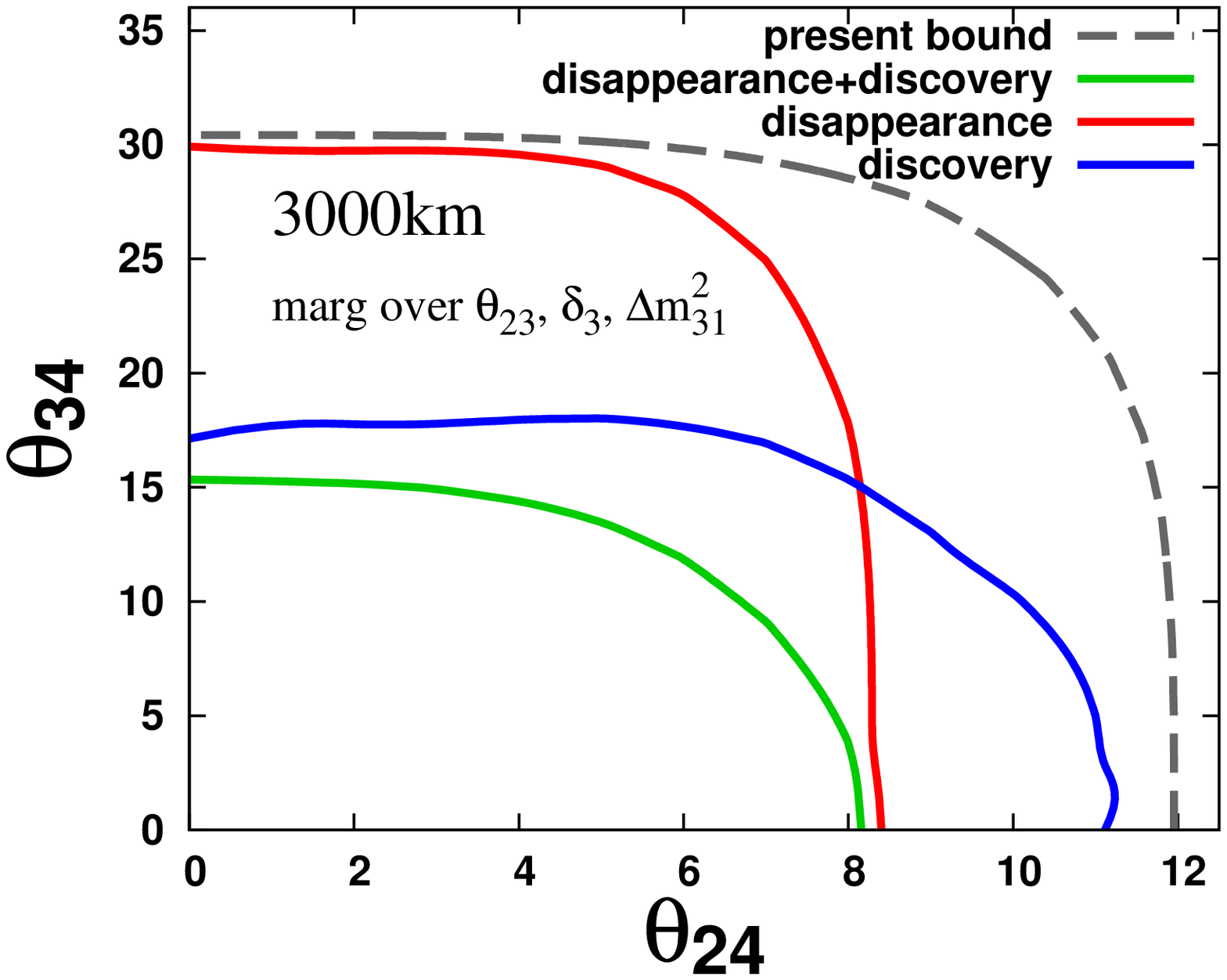} &
\includegraphics[width=7.5cm]{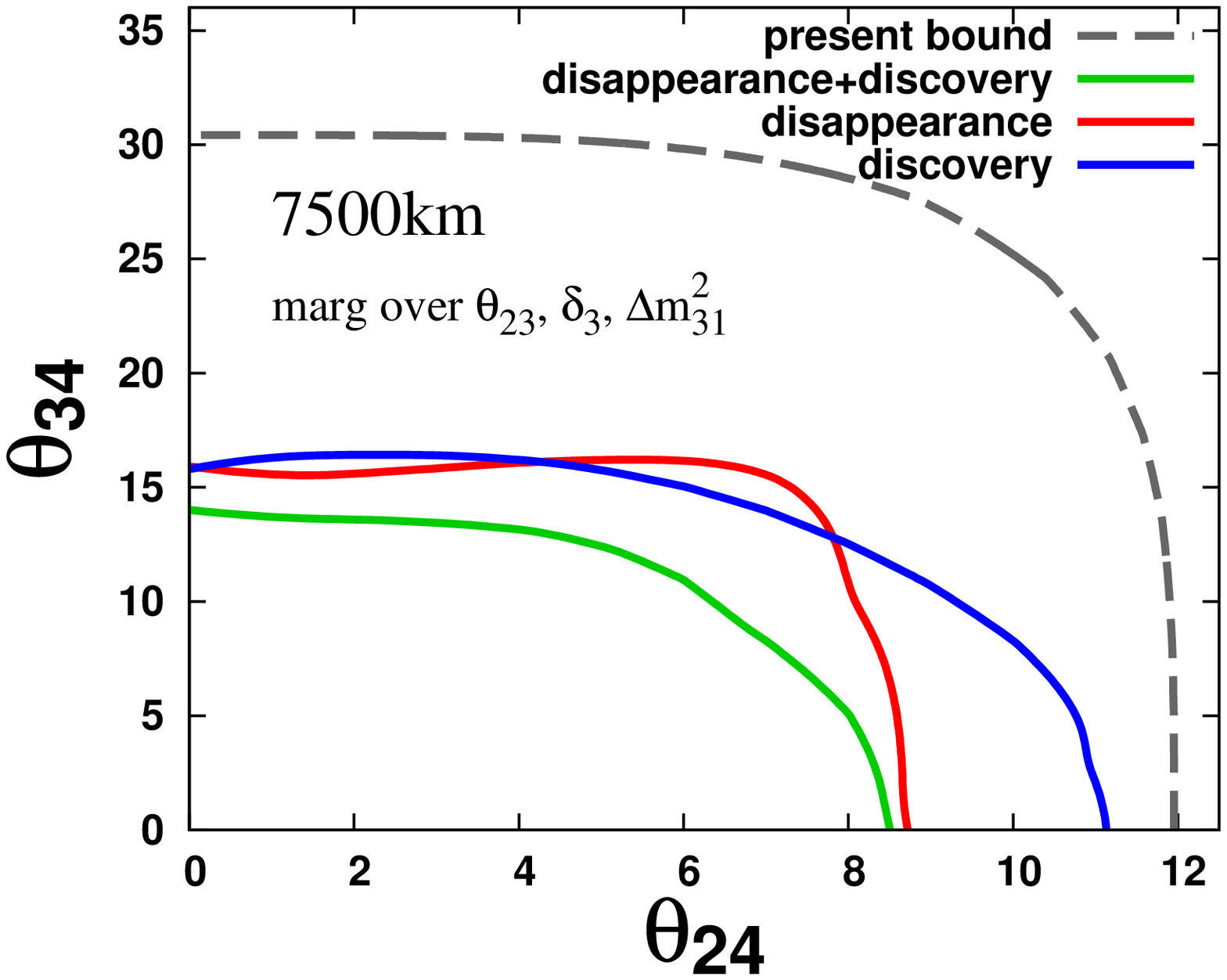}\\
\end{tabular}  
\caption{\label{fig:marginalizingt23dm31d3}\sl%
      Sensitivity limit at 90\% CL to $\theta_{24}$ and $\theta_{34}$,
      marginalizing over $\theta_{23} \in [40^\circ,50^\circ]$, the
      CP-violating phase $\delta_3 \in [0,360^\circ]$ and  $\Dmq_{31} \in [2.0,2.8] \times 10^{-3} \eVq$.
      Red lines stand for the disappearance channel $\nu_\mu \to \nu_\mu$; blue lines stand for the discovery channel
      $\nu_\mu \to \nu_\tau$;  green lines stand for the combination of both; the grey dashed line represents the present
      bound on $\theta_{24}$ and $\theta_{34}$. 
      Left: $L=3000$ km baseline, Right: $L=7500$ km baseline.}
\end{center}
\end{figure}

The sensitivity to (3+1)-sterile neutrinos at the 90\% CL in the ($\theta_{24},\theta_{34}$)-plane for the
50 GeV Neutrino Factory setup is shown in Figs.~\ref{fig:marginalizingt23dm31} and \ref{fig:marginalizingt23dm31d3}.  
In these figures, we have studied how the marginalization over $\theta_{23},\Dmq_{31}$ 
and $\delta_3$ modify our results by varying them in the ranges $\theta_{23} \in [40^\circ,50^\circ]$,
$\Dmq_{31} \in [2.0,2.8] \times 10^{-3} \eVq$ and $\delta_3 \in [0,360^\circ]$. We have also checked the impact
of the marginalization  over all the other parameters (otherwise fixed to their central values, given above), by
studying them one by one in combination with $\theta_{23},\Dmq_{31}$ and $\delta_3$. 
The considered marginalization ranges are: $\theta_{12} \in [30^\circ,36^\circ]; \theta_{13} \in [0,10^\circ]; \theta_{14} \in [0,10^\circ]; \Dmq_{21} \in [7.0,8.3] \times 10^{-5} \eVq; \delta_1 \in [0,360^\circ]$ and $\delta_2 \in [0,360^\circ]$.
We have found that none of these parameters modify significantly our results, contrary to the case of $\theta_{23},\Dmq_{31}$ 
and $\delta_3$. Eventually, we have checked that changing the sign of $\Dmq_{31}$ does not modify our 
results, either.\footnote{Notice that we are not sensitive to the sign of the largest mass difference, $\Dmq_{41}$.}

First of all, in Fig.~\ref{fig:marginalizingt23dm31} we show the sensitivity limit at 90\% CL to $\theta_{24}$ and $\theta_{34}$, for fixed $\delta_3 = 0$, whilst marginalizing over $\theta_{23}$ in the range 
$\theta_{23} \in [40^\circ,50^\circ]$ and $\Dmq_{31}$ in the range $\Dmq_{31} \in [2.0,2.8] \times 10^{-3}$ eV$^2$. 
Notice that the considered allowed range for $\theta_{23}$ is a bit smaller than its 
present allowed range from the three-family global analysis. The Neutrino Factory, however, has an enormous potential for improving the measurement of the three-family atmospheric parameters through the 
$\nu_\mu$ disappearance channel as it was shown, for example, in Ref.~\cite{Donini:2005db}. It is, therefore, absolutely reasonable to vary $\theta_{23}$ over a reduced range. In the two plots, red lines stand for the $\nu_\mu \to \nu_\mu$ disappearance channel data; blue lines stand for the  $\nu_\mu \to \nu_\tau$ discovery channel data;  green lines stand for the combination of both; the dashed grey line represents the present bound on $\theta_{24}$ and $\theta_{34}$. 

We discuss first the left panel of Fig.~\ref{fig:marginalizingt23dm31}, that refers to  the results at the shortest baseline, 
$L=3000$ km. Notice that the disappearance channel has the strongest sensitivity to $\theta_{24}$, that can be excluded
above $\theta_{24} \sim 8^\circ$ for any value of $\theta_{34}$. 
On the other hand, the $\nu_\mu \to \nu_\tau$ channel gives the ultimate sensitivity to $\theta_{34}$ for vanishing 
$\theta_{24}$, $\theta_{34} \leq 15^\circ$. In both channels, we can see a strong correlation between $\theta_{24}$ 
and $\theta_{34}$, with a bound on $\theta_{34}$ that is strongly dependent on the specific value of $\theta_{24}$ 
considered. This behavior can be understood by looking at eqs.~(\ref{eq:pmumu},\ref{eq:pmutau}) in Sec.~\ref{sec:schemes}. 
The strong correlation between the two mixing angles is induced by the subleading $O(\epsilon^3)$ terms, 
proportional to $(A_n L) s_{24} s_{34} \cos \delta_3$ (since we are considering $\delta_3 = 0$ the term proportional to 
$\sin \delta_3$ in $P_{\mu\tau}$ vanishes). The $(\theta_{24},\theta_{34})$-correlation in the $\nu_\mu$ disappearance
channel is softer than in the $\nu_\mu \to \nu_\tau$ one: a consequence of the different statistical significance of this term in 
appearance and disappearance channels. Eventually, the combination of the two channels gives a very good 
sensitivity to both mixing angles. 

A similar combined sensitivity is achieved at the longest baseline, $L = 7500$ km, whose results are shown in the 
right panel of Fig.~\ref{fig:marginalizingt23dm31}. We can see comparing the blue lines on both panels that the 
$\nu_\mu \to \nu_\tau$ channel sensitivity to ($\theta_{24},\theta_{34}$) is  substantially the same when changing baseline
(the expected number of events for this channel at the two baselines is very similar, see Tab.~\ref{tab:rates3}). 
On the other hand, the $\nu_\mu$ disappearance channel data (the red line) has a much stronger sensitivity than at the 
short baseline: a consequence of the increased significance of the subleading term 
with respect to the dominant one in the disappearance channel due to larger matter effects at 7500 km.

The sensitivity of the disappearance channel to $\theta_{34}$ arises at
a higher order in the expansion that we have presented in eqs.~(\ref{eq:pmumu},\ref{eq:pmutau}).
If we introduce terms to fourth order in $\epsilon$ (under the assumption $\theta_{13} = \theta_{14} = 0$,
and taking into account the deviations from maximality of $\theta_{23}$), we get:
\bea
\label{eq:pmumu2}
P_{\mu\mu}&=&1-2\,\theta_{24}^2 - \left[ 1 - 4  (\delta \theta_{23})^2 - 2\theta_{24}^2 + 
\theta_{34}^2  \frac{A_n}{\Delta_{31}}  \left ( 4 \delta \theta_{23}- \theta_{34}^2 \frac{A_n}{\Delta_{31}} \right )
\right]\sin^2\frac{\Delta_{31} L}{2}
\NO\\
&-&\left( A_nL\right)\left\lbrace 2\theta_{24}\,\theta_{34}\cos\delta_3- \frac{\theta_{34}^2}{2}
\left ( 4 \delta\theta_{23} - \theta_{34}^2 \frac{A_n}{2 \Delta_{31}} \right )
\right\rbrace \sin \Delta_{31} L+ O(\epsilon^5) \; ,
\eea
\bea
\label{eq:pmutau2}
P_{\mu\tau}&=&\left \{ 1-4 (\delta\theta_{23})^2-\theta_{24}^2-\theta_{34}^2
\left [ 1- \frac{\theta_{34}^2}{3}   - \frac{A_n}{\Delta_{31}} \left (4\delta\theta_{23} - \theta_{34}^2 \frac{A_n}{\Delta_{31}} \right )
\right ] \right \} \sin^2\frac{\Delta_{31} L}{2} \NO \\
&+&\left\lbrace \theta_{24}\,\theta_{34}\sin\delta_3+\left( A_nL\right)\left[2\theta_{24}\,\theta_{34}\cos\delta_3
-  \frac{\theta_{34}^2}{2} \left ( 4 \delta\theta_{23} - \theta_{34}^2 \frac{A_n}{2 \Delta_{31}} \right )
\right] \right\rbrace \sin\Delta_{31}L \NO \\
&+& O(\epsilon^5) \;, 
\eea
\bea
\label{eq:pmus2}
P_{\mu s}&=&2\,\theta_{24}^2+\left[\theta_{34}^2 \left (1 -\frac{\theta_{34}^2}{3} \right )
-\theta_{24}^2\right]\sin^2\frac{\Delta_{31} L}{2}-\theta_{24}\,\theta_{34}\sin\delta_3 \sin\Delta_{31}L \NO \\
&+& O(\epsilon^5) \; .
\eea

Most of the $\theta_{34}$-dependent terms in eq.~(\ref{eq:pmumu2}) are proportional to the matter parameter $(A_n L)$.
This means that the impact of these terms will be more important at the longest baseline, 
than at the shortest one (as we have seen in Fig.~\ref{fig:marginalizingt23dm31}). 
On the other hand, the $\theta_{24}$-sensitivity arises from the 
$\theta_{24}^2$ term at $O(\epsilon^4)$ in eq.~(\ref{eq:pmumu2}).

These behaviors are strongly modified if we marginalize over the CP-violating phase $\delta_3$, as it can be seen in 
Fig.~\ref{fig:marginalizingt23dm31d3}, where we present the sensitivity limit at 90\% CL to $\theta_{24}$ and $\theta_{34}$, whilst marginalizing over $\theta_{23}$ in the range $\theta_{23} \in [40^\circ,50^\circ]$, over $\delta_3$ in the range $\delta_3 \in [0,360^\circ]$ and over $\Dmq_{31}$ in the range $\Dmq_{31} \in [2.0, 2.8]  \times 10^{-3} \eVq$.
As before, red lines stand for the  $\nu_\mu \to \nu_\mu$ disappearance channel data; blue lines stand for the $\nu_\mu \to \nu_\tau$ discovery channel data;  green lines stand for the combination of both; the grey dashed line represents
the present bound on $\theta_{24}$ and $\theta_{34}$.

Notice that the correlation between $\theta_{24}$ and $\theta_{34}$ in the $\nu_\mu$ disappearance data (red lines)
has vanished. This is a straightforward consequence of the marginalization over $\delta_3$, that removes the
term proportional to  $(A_n L) s_{24} s_{34} \cos \delta_3$ in eq.~(\ref{eq:pmumu}),  responsible for that correlation.
We have found that the sensitivity on the two active-sterile mixing angles $\theta_{24}$ and $\theta_{34}$ is
now represented by vertical and horizontal lines (similar to the results in Fig.~\ref{fig:goldensilverd3}).
A similar effect is found for the $\nu_\mu \to \nu_\tau$ appearance channel data (blue lines), though softer. We can 
see that, at both baselines (but more significantly at the longest one) some remnants of the ($\theta_{24},\theta_{34}$)-correlation can still be observed for this channel. Comparing the results of the two channels, we have found again that
at the short baseline the $\nu_\mu \to \nu_\mu$ data give the best sensitivity to $\theta_{24}$, whereas
the $\nu_\mu \to \nu_\tau$ data  give the best sensitivity to $\theta_{34}$. At the long baseline, $\nu_\mu \to \nu_\mu$
is as good as $\nu_\mu \to \nu_\tau$ in the $\theta_{34}$-sensitivity. 

\begin{figure}[t]
\begin{center}
\hspace{-0.5cm}
\begin{tabular}{cc}
\includegraphics[width=7.5cm]{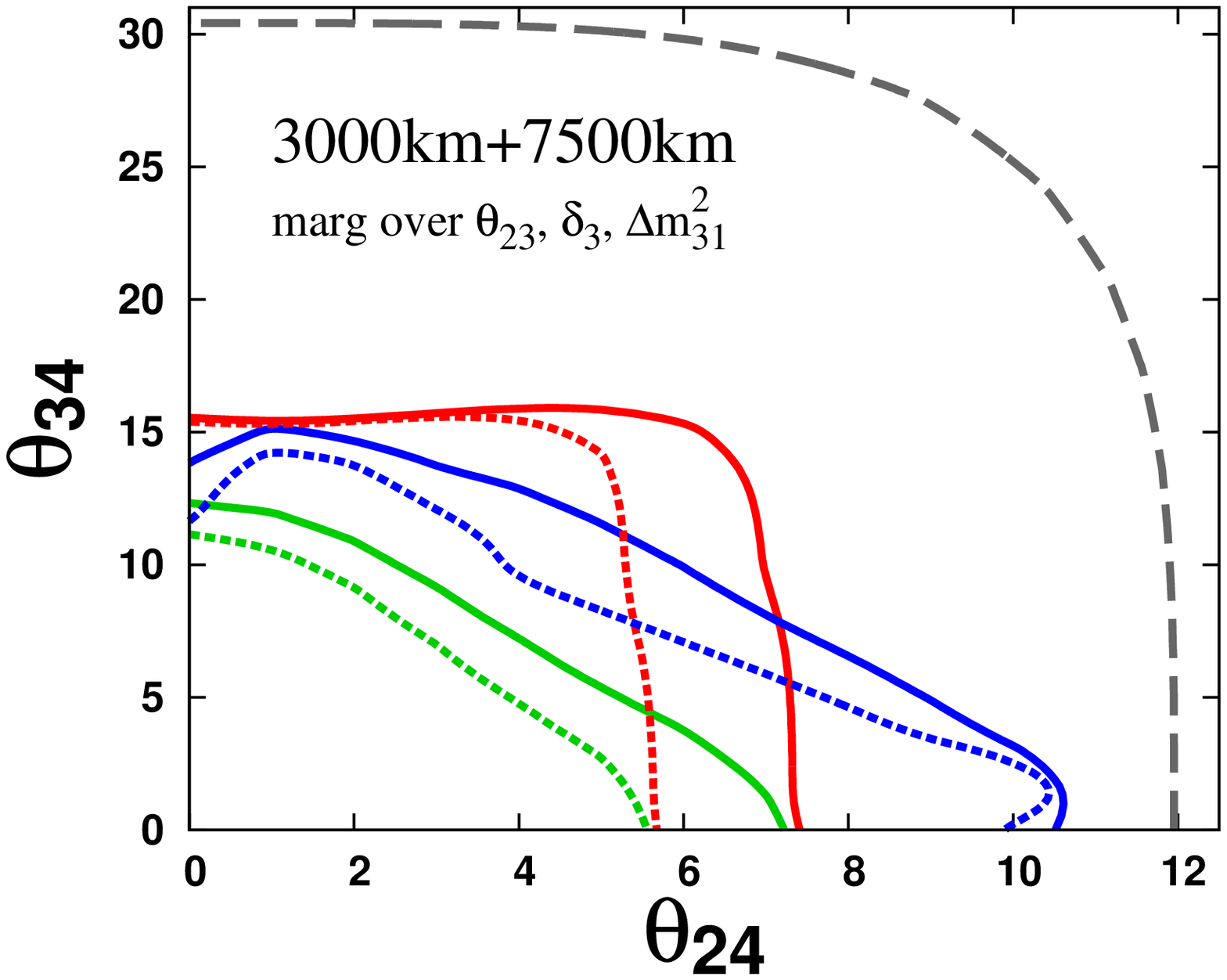} &
\includegraphics[width=7.5cm]{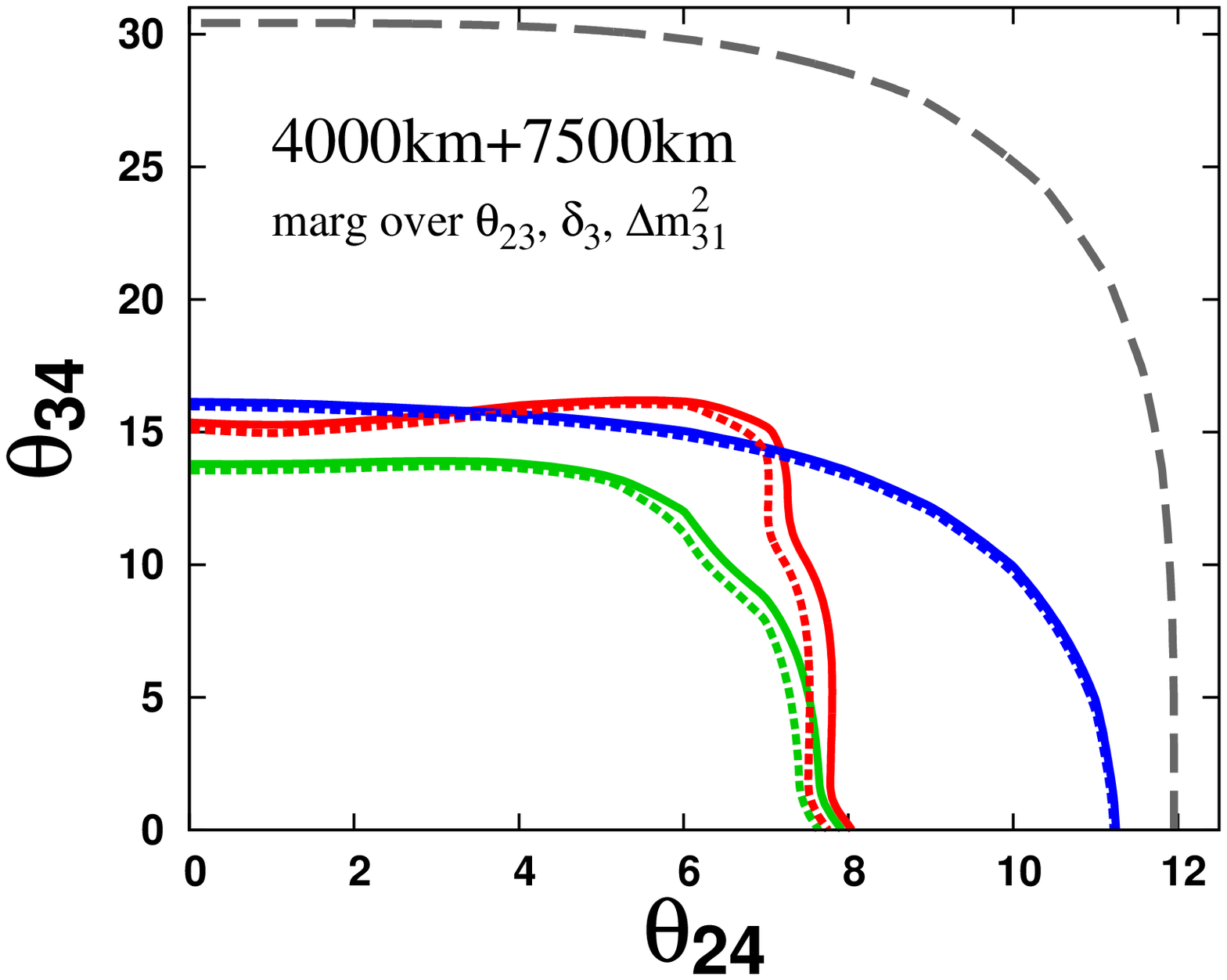} 
\end{tabular}  
\caption{\label{fig:marginalizing3075}\sl%
      Sensitivity limit at 90\% CL to $\theta_{24}$ and $\theta_{34}$,
      marginalizing over $\theta_{23} \in [40^\circ,50^\circ]$, the
      CP-violating phase $\delta_3 \in [0,360^\circ]$ and  $\Dmq_{31} \in [2.0,2.8] \times 10^{-3} \eVq$, 
      for the combination of the two baselines.
      Dashed lines stand for the results without the correlated systematic errors.
      Red lines stand for the disappearance channel $\nu_\mu \to \nu_\mu$; blue lines stand for the discovery channel
      $\nu_\mu \to \nu_\tau$;  green lines stand for the combination of both. 
     Left panel: 50 GeV Neutrino Factory; Right panel: 20 GeV Neutrino Factory.}
\end{center}
\end{figure}

We, eventually, compare the 90\% CL ($\theta_{24},\theta_{34}$)-sensitivity that can be obtained using 
the combination of the two channels and of the two baselines at the 50 GeV Neutrino Factory setup, Fig.~\ref{fig:marginalizing3075}(left), with the one that can be achieved using the 20 GeV Neutrino Factory ISS-inspired setup, Fig.~\ref{fig:marginalizing3075}(right). 
As before, red lines stand for $\nu_\mu$ disappearance channel data; blue lines for the $\nu_\mu \to \nu_\tau$
discovery channel data; green lines for the combination of both channels; the grey dashed line represents
the present bound on $\theta_{24}$ and $\theta_{34}$. Data for the two baselines are always summed.
In these plots, we have marginalized over $\theta_{23} \in [40^\circ,50^\circ]$, $\delta_3 \in [0,360^\circ]$ and 
$\Dmq_{31} \in [2.0,2.8] \times 10^{-3} \eVq$.
The rest of the parameters have been kept fixed to the values given previously.\footnote{
We have checked that the effect of the marginalization on the rest of the parameters 
do not affect the results.}

First of all, we can see by comparing red lines (the disappearance channel data) between Fig.~\ref{fig:marginalizing3075}(left) and Fig.~\ref{fig:marginalizing3075}(right) that the ultimate sensitivities to $\theta_{24}$ and $\theta_{34}$
at the two setups are very similar: the upper bounds $\theta_{24} \leq 7.5^\circ (8^\circ)$ and $\theta_{34} \leq 15^\circ$
can be inferred from the data for the 50 GeV (20 GeV) setup. 
When we eventually combine the results for the disappearance and the discovery channels, however,
we find that the 50 GeV Neutrino Factory outperforms the 20 GeV ISS-inspired one, 
as it can be seen comparing the green lines in Fig.~\ref{fig:marginalizing3075}.
This can be easily explained pointing out that the discovery channel data (blue lines) are able to exclude 
a significantly larger region of the parameter space when going to higher energy, a straightforward consequence 
of the higher statistics due to the higher $\nu_\tau N$ cross-section. 

In Fig.~\ref{fig:marginalizing3075} we have also studied the impact of the correlated systematic errors $\alpha_s,\beta_s$
in eq.~(\ref{chi2-13-14}) on our results. Dashed blue, red and green lines represent the 90\% CL sensitivities to 
($\theta_{24},\theta_{34}$) when the correlated systematic errors $\alpha_s, \beta_s$
are not included and only the uncorrelated bin-to-bin systematic errors ($f_j$= 5\% for the $\nu_\mu$ disappearance channel and $f_j$=10\% for the discovery channel, irrespectively of the energy bin, of the baseline and of the stored muon polarity)
are taken into account.

Comparing dashed and solid lines at the 20 GeV Neutrino Factory 
(right panel), we can see that the inclusion of the correlated systematic errors  has a negligible impact when analyzing the data at this setup. On the other hand, when studying the data for both channels at the 50 GeV setup, 
we can see that the inclusion of these errors modify our results. In particular, 
the sensitivity to $\theta_{24}$ through the $\nu_\mu$ disappearance channel goes from $\theta_{24} \leq 6^\circ$ 
when only $f_j$ is considered to $\theta_{24} \leq 7.5^\circ$ when $\alpha_s, \beta_s$ are also taken into account. 

Our final conclusion is the following: using the 20 GeV ISS-inspired Neutrino Factory, we get two roughly independent
limits on the two angles $\theta_{24}$ and $\theta_{34}$: $\theta_{24} \leq 8^\circ$ for any value of $\theta_{34}$
and $\theta_{34} \leq 14^\circ$ for any value of $\theta_{24}$ at the 90 \% CL. 
Slightly stronger ultimate limits are obtained at the 50 GeV Neutrino Factory: $\theta_{24} \leq 7.5^\circ$ for vanishing $\theta_{34}$ and $\theta_{34} \leq 12^\circ$ for vanishing $\theta_{24}$. The significantly large discovery channel statistics at the 50 GeV Neutrino Factory with respect to the 20 GeV one, however, strongly increase the sensitivity of this
setup to the combination of $\theta_{24}$ and $\theta_{34}$, such that a roughly diagonal line in the ($\theta_{24},\theta_{34}$) plane connecting ($\theta_{24},0$) and ($0,\theta_{34}$) can be drawn. 

\subsection{Discrimination of the four neutrino schemes}
\label{sec:discrimination}

In the subsections \ref{sec:th13th14} and \ref{sec:th24th34} we have
discussed the sensitivity to $\theta_{13}$, $\theta_{14}$,
$\theta_{24}$, $\theta_{34}$ by looking at statistical significance of
deviation of a four-flavor scheme from that with a certain set of reference
values of the oscillation parameters.  Here we will discuss whether
the Neutrino Factory setup can distinguish our four-neutrino scheme
from the three-flavor scenario.

We introduce the "sterile neutrino discovery potential", defined as follows:
\begin{eqnarray}
\Delta \chi^2 ({\rm 4fam}) =  
&
\min_{\rm marg \, par}
&
 \left [
\sum_{{\rm pol., (chan.)}, (L)}
\min_{\alpha's, \,\beta's}
\left \{
\sum_j \frac{1}{\sigma^2_j}
        \left (
                  (1+\alpha_s+ x_j\beta_s)  N_j ({\rm 3fam})
       \right.\right.
  \right.
 \nonumber\\
&{\ }&
+ (1+\alpha_b+ x_j\beta_b) B_j ({ \rm 3fam})
- N_j ({\rm 4fam})
-\left. B_j ({ \rm 4fam})
\right )^2
\nonumber\\
&{\ }& + \left(\frac{\alpha_s}{\sigma_{\alpha s}}\right)^2
 + \left(\frac{\alpha_b}{\sigma_{\alpha b}}\right)^2
+ \left. \left(\frac{\beta_s}{\sigma_{\beta s}}\right)^2
+\left(\frac{\beta_b}{\sigma_{\beta b}}\right)^2 \right \}
+ \Delta \chi^2_{\rm atm+re}({\rm 4fam})
\left. \right ] \, ,
\nonumber\\
\label{chi2-13-14-3-4}
\end{eqnarray}
where $\Delta \chi^2_{\rm atm+re} ({\rm 4fam})$, defined in eq.~(\ref{chi2-atmre4}),
is the prior from the four-flavor oscillation analysis of the atmospheric and reactor data, and
the errors of the oscillation parameters in the prior $\Delta \chi^2_{\rm atm+re} ({\rm 4fam})$
are given by eq.~(\ref{sigmas}).

A remark is in order. The definition  of the $\Delta\chi^2$ in the present case, although looking similar, is
slightly different from that used in the previous sections. In Secs.~\ref{sec:th13th14} 
and \ref{sec:th24th34} we assumed that the minimum of the $\chi^2$ corresponds to the "true" values of the 
four-family model, and therefore $\chi^2_{\rm min, 4fam} = 0$. 
The $\Delta \chi^2$ is then computed in the same model in which data are generated, and CL contours define
the region of parameter space compatible at a given CL with the "true" values.
 In this Section we also assume that data are generated in the four-family model, but we try to fit them in the three-family
 model. 
The minimum 
of the $\chi^2$ in the four-family model is located at the "true" values of the parameters and $\chi^2_{\rm min, 4fam} = 0$. 
On the other hand, when we try to fit the four-family--generated data in the three-family model we will in general find 
$\chi^2_{\rm min,3fam} \neq 0$, since a "wrong" model is used to fit the data, except 
for the special case defined by $\theta_{j4}=0~(j=1,2,3)$,
$\theta_{ij}^{\rm (4fam)}=\theta_{ij}|_{\rm best fit}~((i,j)=(1,2),(1,3),(2,3))$,
$\Delta m_{j1}^{2~\rm (4fam)}=\Delta m^2_{j1}|_{\rm best fit}~(j=2,3)$, where the two models coincide and $\Delta \chi^2 = 0$. 
In the rest of the four-flavor parameter space, the $\Delta \chi^2$ defined in eq.~(\ref{chi2-13-14-3-4}) corresponds to 
$\chi^2_{\rm min,3fam} - \chi^2_{\rm min,4fam}$.
CL contours define, then,  regions in the four-family parameter space for which a three-family fit to the data
is worse than a four-family fit to the data of a quantity $\Delta \chi^2 $.
For example, a point with $\Delta\chi^2 = 4.61$ is a point that is fitted by the four-family model much better
than by the three-family model. We will define points outside this contour as points for which the hypothesis that data 
can be fitted in the three-family model is "excluded at 90\% CL".
Under these premises, we can use eq. (\ref{chi2-13-14-3-4}) to determine regions in which we are able to distinguish 
four- from three-family models in the four-flavor parameter space
in the same manner as in the previous subsections.

Since the excluded region is expected to depend
little on the solar neutrino oscillation parameters
in the three-flavor scheme,
we will not marginalize $\Delta \chi^2$ with respect to
$\theta_{12}^{\rm (3fam)}$ and $\Delta m^{2~\rm (3fam)}_{21}$
in eq. (\ref{chi2-13-14-3-4}).
Moreover, while $\Delta\chi^2$ is naively
expected to depend on all of the parameters
$\theta_{13}^{\rm (3fam)}$, $\theta_{23}^{\rm (3fam)}$, $|\Delta m_{31}^{2~\rm (3fam)}|$,
$\delta^{\rm (3fam)}$, we have found numerically that
it suffices to vary some of the parameters and put other parameters
to the best-fit values in most analyses.
Namely, in the case of the golden and silver
(disappearance and discovery) channels,
a dominant role is played by 
$\theta_{13}^{\rm (3fam)}$ and $\delta^{\rm (3fam)}$
($\theta_{23}^{\rm (3fam)}$ and $|\Delta m_{31}^{2~\rm (3fam)}|$)
which are the only three-family parameters that we vary,
and the other three-family ones are fixed in the analysis.

\begin{figure}[t]
\begin{center}
\hspace{-0.5cm}
\begin{tabular}{cc}
\includegraphics[width=7.5cm]{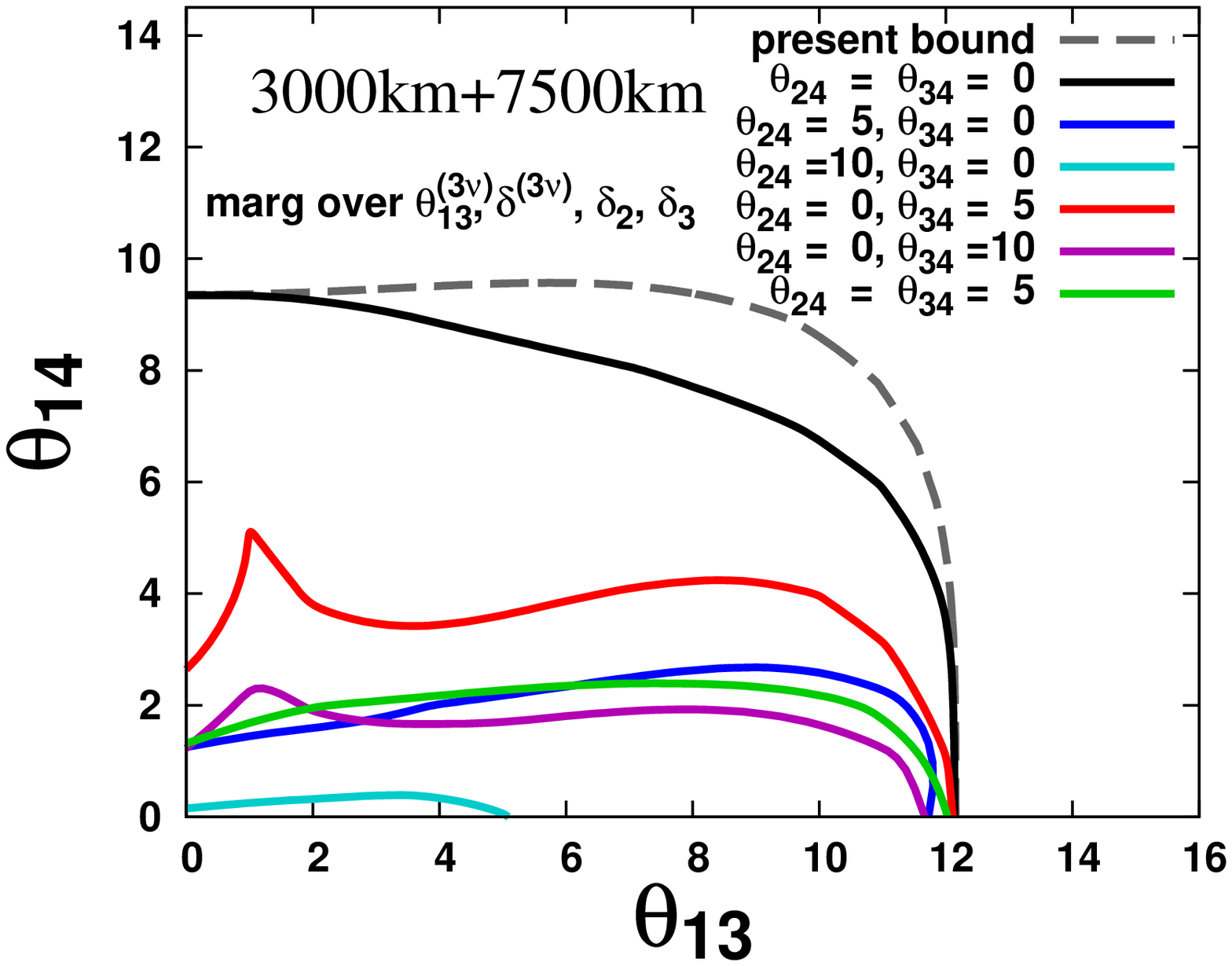} & 
\includegraphics[width=7.5cm]{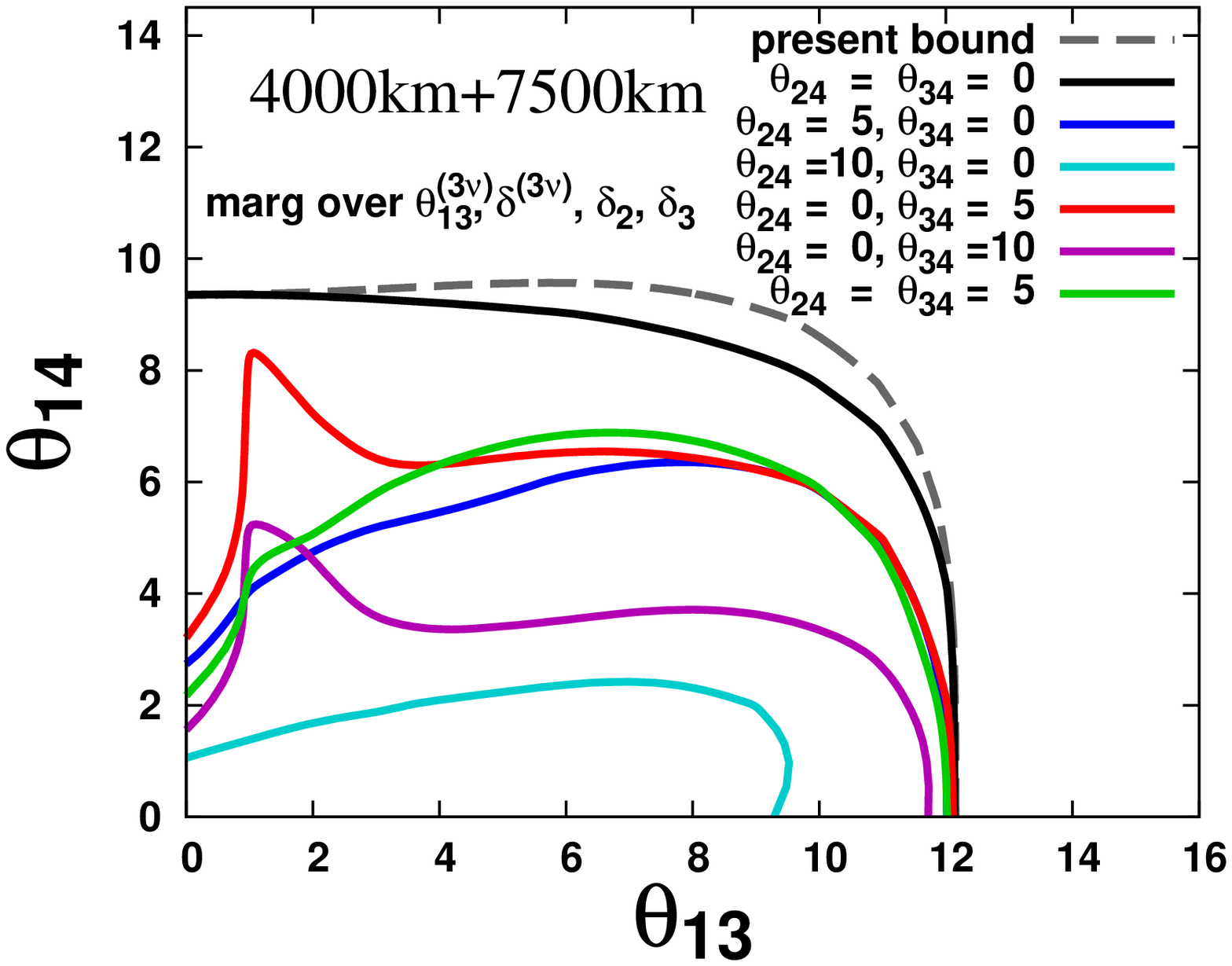}\\
\includegraphics[width=7.5cm]{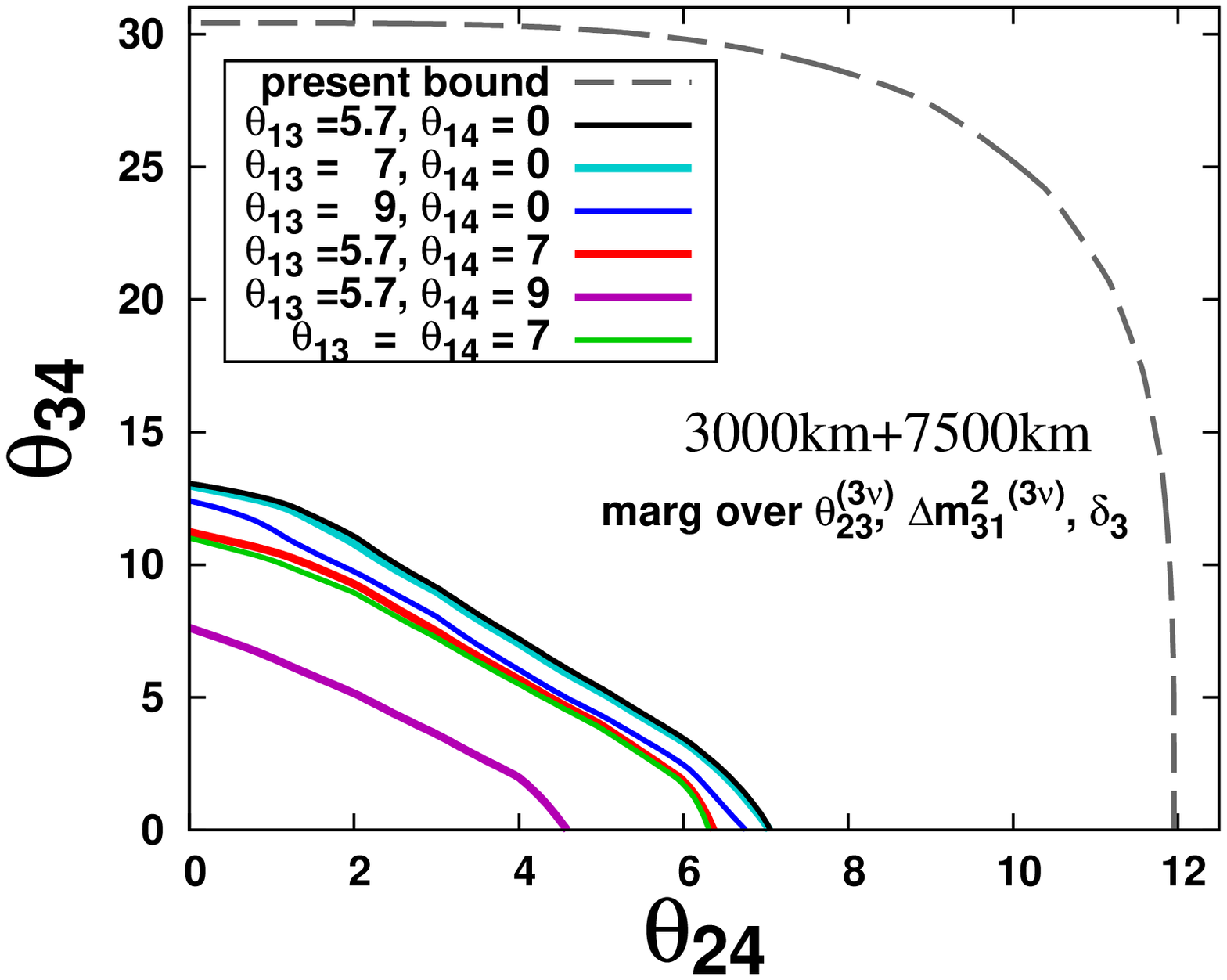} & 
\includegraphics[width=7.5cm]{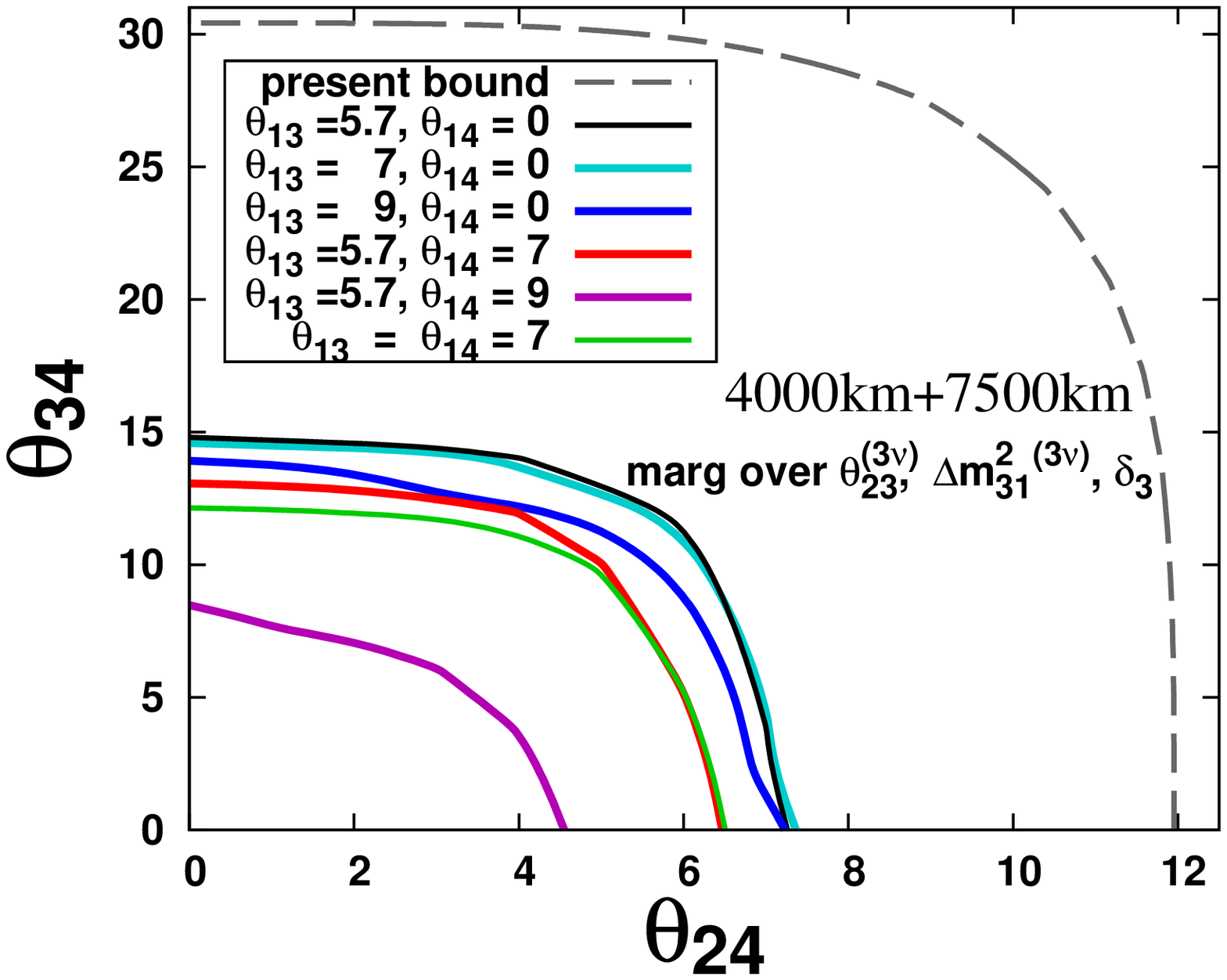}\\
\end{tabular}  
\caption{\label{fig:marginalizing3}\sl%
Left-upper panel: The right upper part of each line
is the region projected onto the ($\theta_{13}$, $\theta_{14}$)
plane, in which the hypothesis of the three flavor
scheme is excluded at 90\% CL at the 50 GeV Neutrino Factory.
It is obtained by
marginalizing over $\theta_{13}^{\rm (3fam)}$ and $\delta^{\rm (3fam)}$
as well as $\delta_2$ and $\delta_3$.
The grey dashed
lines stand for the excluded region obtained only from the
prior $\Delta \chi^2_{\rm atm+re} ({\rm 4fam})$ where
terms other than
$(s^{2~\rm (4fam)}_{13}-0.01)^2/\sigma^2(s^2_{13})+
(s^2_{14})^2/\sigma^2(s^2_{14})$ are assumed to be zero
in eq. (\ref{chi2-atmre4}).
The excluded regions for nonvanishing $\theta_{24}$ or
$\theta_{34}$, which are always larger than
the case for $\theta_{24}=\theta_{34}=0$,
are also depicted for information.
Left-lower panel: Excluded region at 90\% CL projected onto the
($\theta_{24}$, $\theta_{34}$) plane.
It is obtained by
marginalizing over $\theta_{23}^{\rm (3fam)}$, $|\Delta m^{2~\rm (3fam)}_{31}|$ 
as well as $\delta_3$.
The grey dashed
lines stand for the excluded region obtained only from the
prior $\Delta \chi^2_{\rm atm+re} ({\rm 4fam})$ where
terms other than
$(s^2_{24})^2/\sigma^2(s^2_{24})+(s^2_{34})^2/\sigma^2(s^2_{34})$
are assumed to be zero in eq. (\ref{chi2-atmre4}).
Right-upper(lower) panel: The same figure as the left-upper(lower) panel for
the 20 GeV ISS-inspired setup.
}
\end{center}
\end{figure}

To compare the results with those in the subsections
\ref{sec:th13th14} and \ref{sec:th24th34},
we project the excluded region either in
the $(\theta_{13}^{\rm (4fam)}, \theta_{14})$ plane, or in
the $(\theta_{24}, \theta_{34})$ plane.
In these projections,
we would like to obtain the most conservative excluded region,
i.e., the common excluded region in the
$(\theta_{24}, \theta_{34})$ plane irrespective of the values of
$\theta_{13}^{\rm (4fam)}$ and $\theta_{14}$,
or the common excluded region in the
$(\theta_{13}^{\rm (4fam)}, \theta_{14})$ plane
irrespective of the values of $\theta_{24}$ and $\theta_{34}$.
Notice that the four angles can in principle be measured simultaneously if we use informations from the 
four channels at the same time. To obtain them,
in principle we have to marginalize $\Delta \chi^2$ not only
with respect to the three-family parameters described above
but also with respect to the four-family ones,
such as $\theta_{12}^{\rm (4fam)}, \theta_{23}^{\rm (4fam)}$,
$\Delta m^{2~\rm (4fam)}_{21}, |\Delta m^{2~\rm (4fam)}_{32}|$,
$\delta_1, \delta_2, \delta_3$,
as well as ($\theta_{24}, \theta_{34}$) in the former case
and ($\theta_{13}^{\rm (4fam)}, \theta_{14}$) in the latter.
In marginalizing over the four-family parameters, however,
we do not have to vary all the parameters
for a couple of reasons.  First of all,
since the excluded region is expected to depend
little on the solar neutrino oscillation parameters
in the four-flavor scheme,
we can fix the solar parameters
$\theta_{12}^{\rm (4fam)}, \Delta m^{2~\rm (4fam)}_{21}, \delta_1$.
Secondly, because of the prior
$\Delta \chi^2_{\rm atm+re} ({\rm 4fam})$,
in practice we can fix the following parameters
to the best fit values:
$s_{13}^{2~\rm (4fam)}\simeq0.01$,
$s_{23}^{2~\rm (4fam)}\simeq0.5$,
$|\Delta m_{31}^{2~\rm (4fam)}|\simeq 2.4\times10^{-3}$eV$^2$,
$(\theta_{24}, \theta_{34})\simeq (0,0)$
in the case of the $(\theta^{(\rm 4fam)}_{13}, \theta_{14})$ plane,
and $(\theta^{(\rm 4fam)}_{13}, \theta_{14})\simeq (5.7^\circ,0)$
in the case of the $(\theta_{24}, \theta_{34})$ plane.
Thus, the only non-trivial four-family parameters to be marginalized over
are
$\delta_2$ and $\delta_3$
in the case of the $(\theta^{(\rm 4fam)}_{13}, \theta_{14})$ plane,
and $\delta_3$
in the case of the $(\theta_{24}, \theta_{34})$ plane.

The results obtained are presented in Fig.~\ref{fig:marginalizing3}, where the dashed black line stands for the region which 
is excluded by the prior, i.e., by the present data of the atmospheric and reactor experiments.
Upper panels show the "sterile neutrinos discovery potential" of golden and silver channels; 
lower panels the "discovery potential" of $\nu_\mu$ disappearance and $\nu_\mu \to \nu_\tau$ appearance channel. 
On the left, we show results obtained for the 50 GeV Neutrino Factory; on the right, using the 20 GeV ISS-inspired setup. 
We can see in Fig. ~\ref{fig:marginalizing3}(upper panels) that at both setups,
 if we have no information on $\theta_{24}$ and $\theta_{34}$,
then the golden and silver channels cannot discriminate between the three- and four-family models for both setups
within the presently allowed region.
If we can previously measure $\theta_{24}$ and $\theta_{34}$ using the $\nu_\mu \to \nu_\mu$ and $\nu_\mu \to \nu_\tau$
channels, finding that one or both angles are non-vanishing, then we can see in both upper panels of Fig.~\ref{fig:marginalizing3} that the "sterile neutrino discovery potential" of the combination of golden and silver channels
is significantly improved. For example, for $\theta_{34} \sim 10^\circ$, the 50 GeV (20 GeV) setup can distinguish the (3+1)-neutrino model from standard three-family oscillations for $\theta_{14}$ as small as $\theta_{14} \sim 2^\circ (4^\circ)$.
This is because of the same reason
that the Neutrino Factory has good sensitivity to
$U_{e4}U_{\mu4}$ and $U_{e4}U_{\tau4}$,
as was discussed in sect. \ref{sec:ue4umt4}.
We point out that the silver channel gives a non-trivial contribution to the "discovery potential" $\Delta\chi^2$ in some
regions of the parameter space, although the contributions of the golden and silver channels are not shown separately.

In Fig. ~\ref{fig:marginalizing3}(lower panels) we can see that, irrespectively of prior
knowledge of $\theta_{13}^{(\rm 4fam)}$ and $\theta_{14}$, the combination of the $\nu_\mu$ disappearance and the 
$\nu_\mu \to \nu_\tau$ discovery channels permits discrimination between the four- and three-neutrino oscillation models in a significant region of the presently allowed parameter space. The additional information from 
golden and silver channels increases the region in which discrimination is possible. 

We conclude that combination of the four channels is extremely effective 
to tell the difference between the four- and three-flavor schemes in a significant region of the presently allowed
parameter space. Notice that the synergy between the four channels is not symmetric: whereas a previous knowledge of $(\theta_{24},\theta_{34})$ strongly increase the "discovery potential" in the ($\theta_{13}^{(\rm 4fam)},\theta_{14}$)-plane, 
the measurement of ($\theta_{13}^{(\rm 4fam)},\theta_{14}$) has a much smaller impact in the 
"discovery potential" in the $(\theta_{24},\theta_{34})$-plane. Eventually, the 50 GeV setup has a greater
"sterile neutrino discovery potential"  than the 20 GeV one.

\subsection{Dependence of sensitivity on the systematic errors}
\label{sec:systematics}

\begin{figure}[t]
\begin{center}
\hspace{-0.5cm}
\begin{tabular}{cc}
\includegraphics[width=7.5cm]{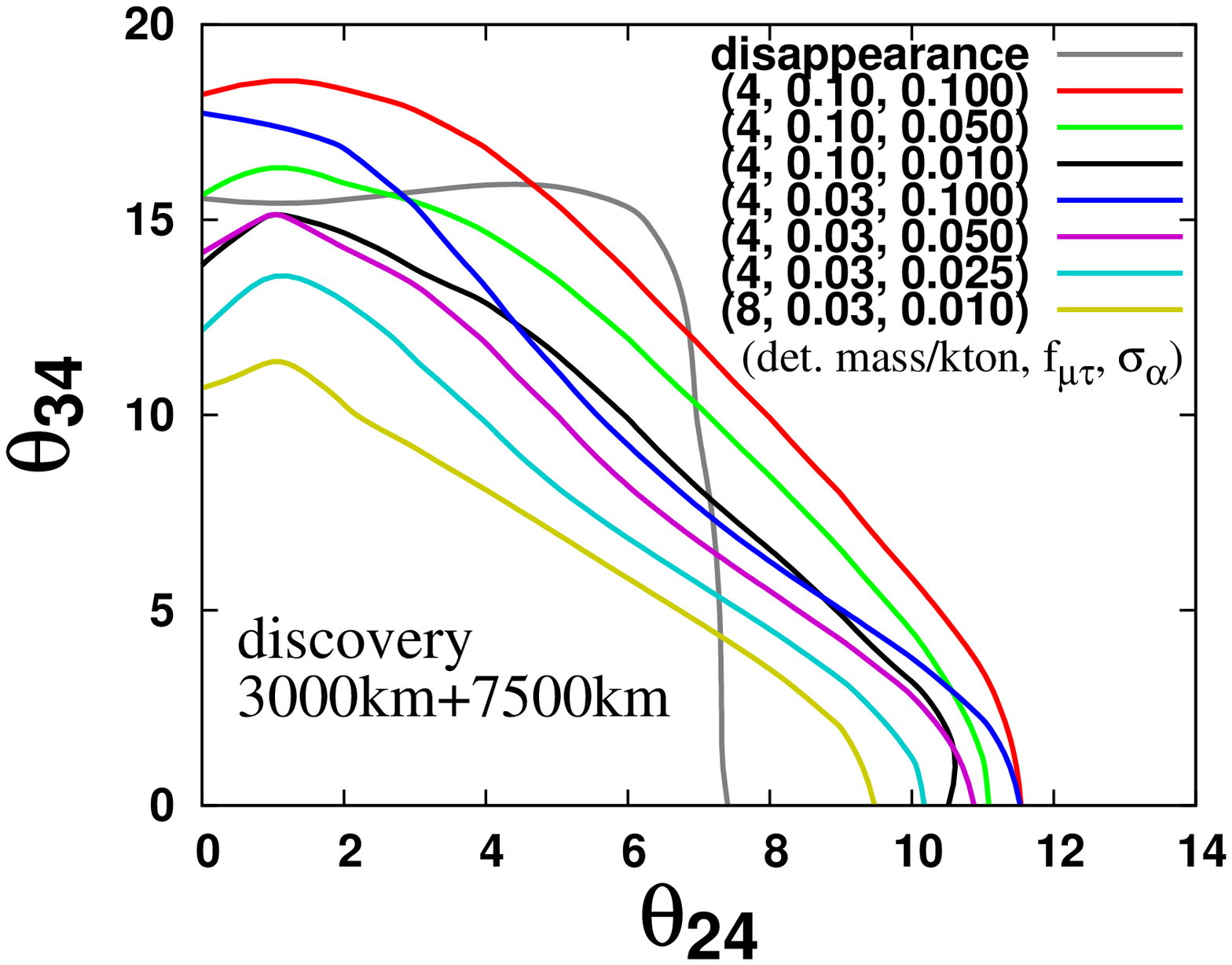} & 
\includegraphics[width=7.5cm]{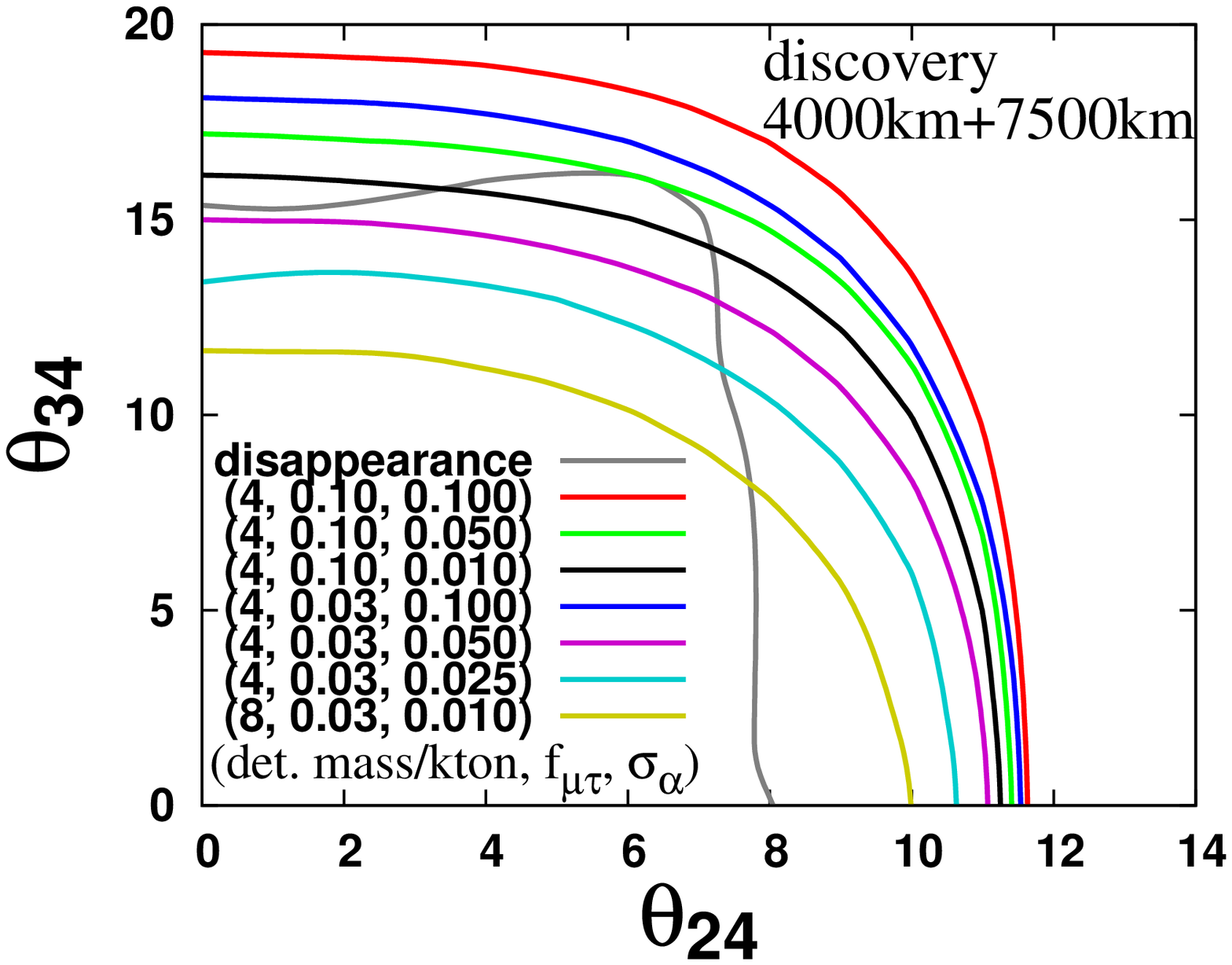}\\
\includegraphics[width=7.5cm]{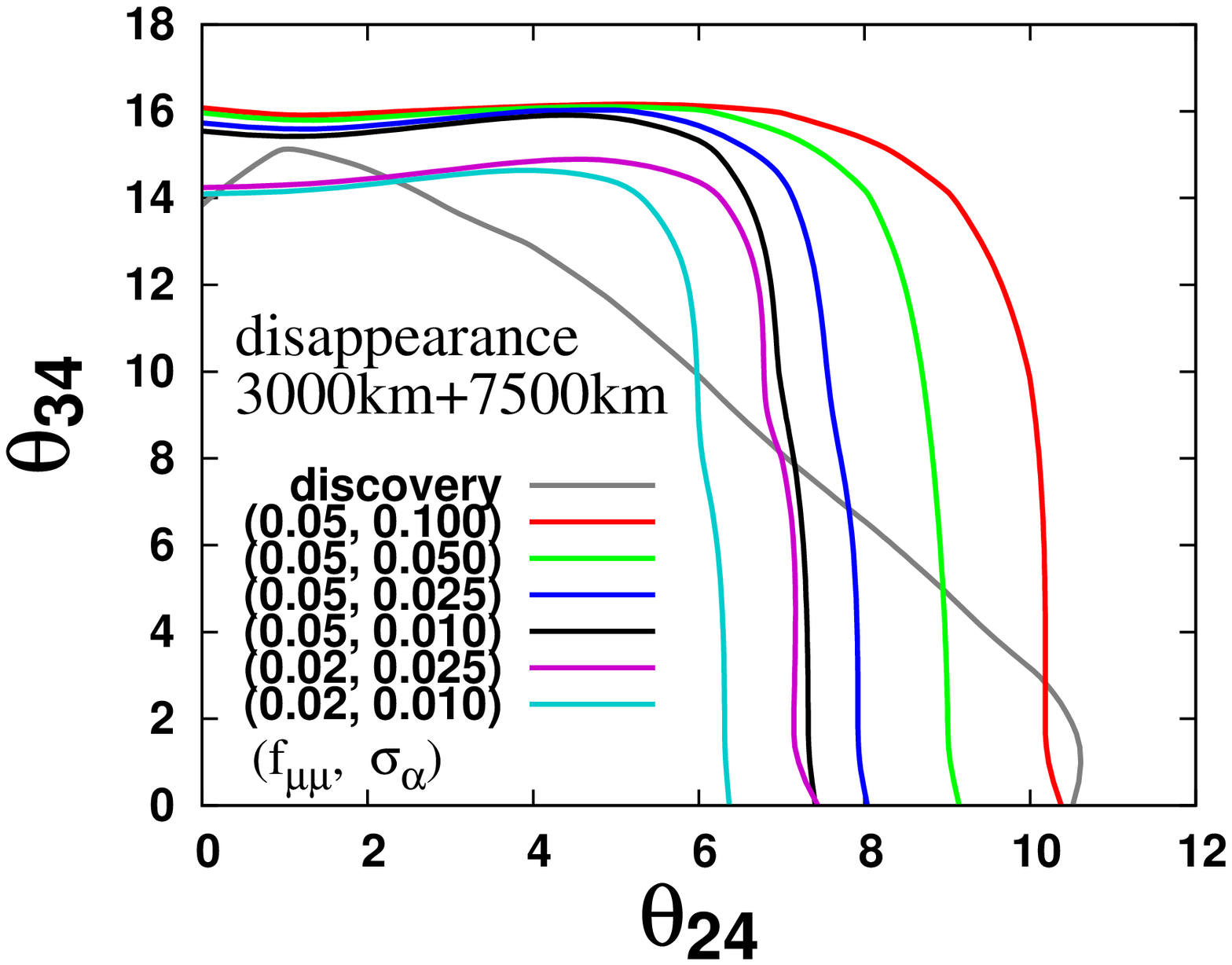} & 
\includegraphics[width=7.5cm]{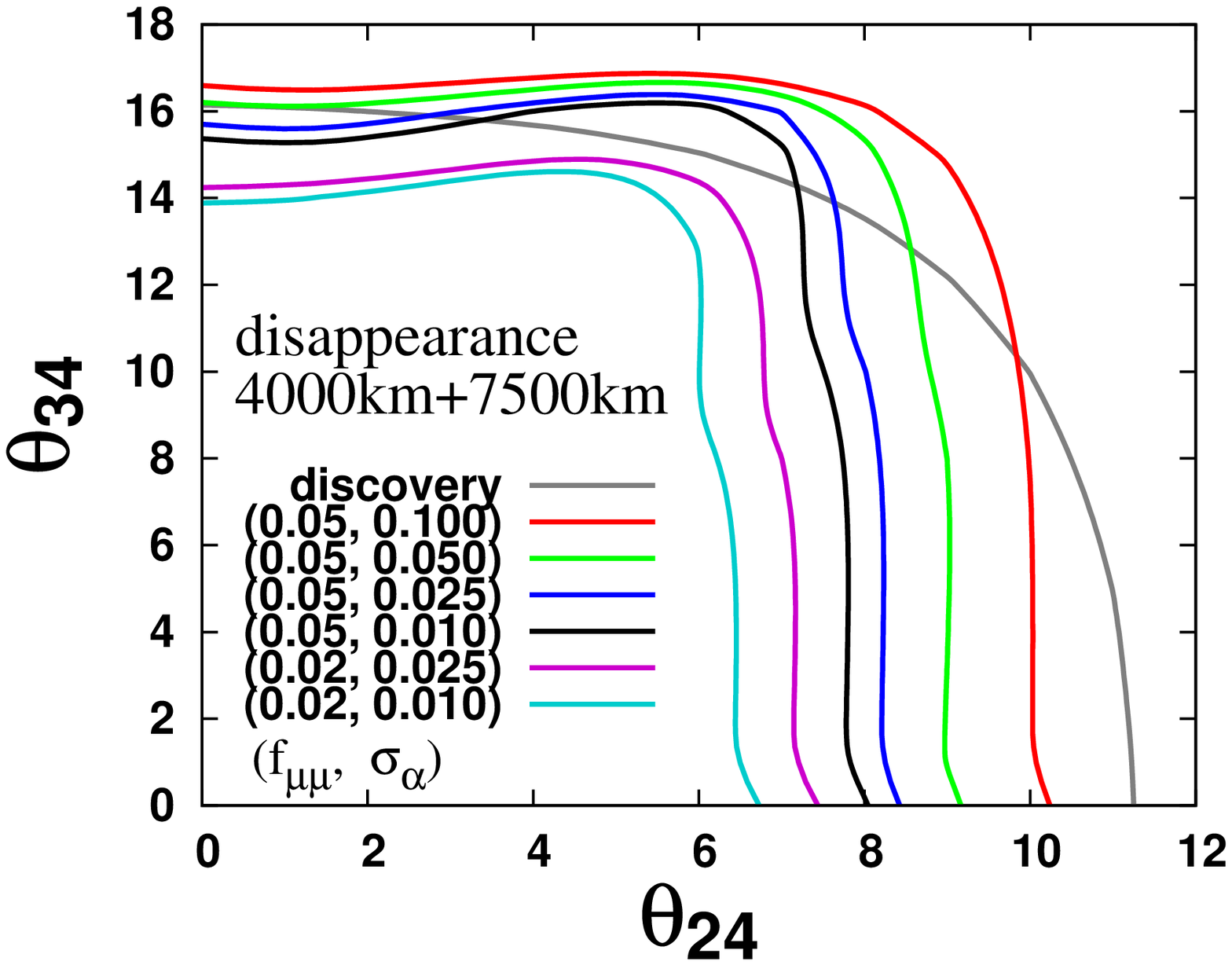}
\end{tabular}  
\caption{\label{fig:systematics}\sl%
The 90\%CL sensitivity to ($\theta_{24}$, $\theta_{34}$) using
the $\nu_\mu \to \nu_\tau$ (the upper panels)
and $\nu_\mu \to \nu_\mu$ (the lower panels) channels
for different values of the uncorrelated bin-to-bin systematic error
$f_j \equiv f_{\mu\tau}=0.1, 0.03$
and  $f_j \equiv f_{\mu\mu}=0.05, 0.02$,
of the correlated systematic error on the overall normalization
$\sigma_{\alpha_s}=0.1, 0.05, 0.025, 0.01$ and
of the MECC mass (= 4, 8 kton in the case of  $\nu_\mu \to \nu_\tau$).
Left panels:  at the 50 GeV setup; Right panels: at the 20 GeV ISS-inspired setup.
In all the figures black lines stand for the excluded region
for the reference values used in the calculations in other sections.
The grey lines stand for the excluded region using the disappearance
(discovery) channel with $f_{\mu\mu}=0.05$, $\sigma_{\alpha_s} = 0.01$
(with 4 kton, $f_{\mu\tau}=0.1$, $\sigma_{\alpha_s} = 0.01$).
}
\end{center}
\end{figure}

We have also investigated the dependence of the performance of the
four channels on the systematic errors.

As for the golden and silver channels, for which statistical
errors are dominant, we have found from numerical calculations that
the sensitivities of these two channels to $\theta_{13}^{(\rm 4fam)}$ and $\theta_{14}$
depend  only to some extent  on $\sigma_\alpha$ (the correlated systematic error on the overall normalization),
and they depend very little on $f_j$ (the bin-to-bin uncorrelated error)
and $\sigma_\beta$ (the correlated systematic error in the
linear distortion of the spectral shape).

The dependence of the discovery and disappearance channels on $f_j$, $\sigma_\alpha$ and $\sigma_\beta$
(and on the MECC volume, in the case of the former) is shown in Fig. \ref{fig:systematics}. 
The upper panels present the discovery channel ($\theta_{24},\theta_{34}$)-sensitivity,
where we consider $f_j \equiv f_{\mu\tau}= 0.10, 0.03$, $\sigma_\alpha=0.100, 0.050, 0.025, 0.010$,
and the MECC mass either 4 or 8 kton. The lower panels present the disappearance channel
($\theta_{24},\theta_{34}$)-sensitivity, assuming $f_j \equiv f_{\mu\mu}=0.05, 0.02$ and $\sigma_\alpha=0.100, 0.050, 0.025, 0.010$.
We have checked numerically that the impact of the systematic error $\sigma_\beta$ in the linear distortion of the spectral shape is small and it will not be discussed here. For both channels we have considered the 50 GeV setup performance
(left panels) and the 20 GeV setup one (right panels).
From Fig.~ \ref{fig:systematics}(upper panels), we see that $f_{\mu\tau}$ is the most important factor to improve
the performance of the discovery channel for both the 50 GeV and 20 GeV neutrino factories. 
On the other hand, an increase of the MECC mass from 4 to 8 kton improves only marginally the discovery channel 
sensitivity. In Fig.~\ref{fig:systematics}(lower panels), we see that a reduction of both $f_{\mu\mu}$ and $\sigma_\alpha$ 
should be pursued to increase the disappearance channel sensitivity.
One important conclusion from Fig. ~\ref{fig:systematics} is that an improvement of $f_{\mu\tau}$ below 10\% is mandatory
in order to take full advantage of the discovery channel at the Neutrino Factory, particularly for the 20 GeV setup.
This error represents indeed our incomplete knowledge of the MECC detector.
To improve our analysis on the discovery channel in the future,
we need detailed information on the correlated and uncorrelated systematic errors,
such as the uncertainties of the detection efficiency which depend
only on the nature of MECC or depend on the characteristics of the individual detectors located at each of the two baselines.
Although there has been no study on these systematic errors so far,
they are expected to be better understood after the first years of data
taking of the OPERA experiment (that started operation in 2008).

\subsection{A  CP-violating sterile neutrino signal}
\label{sec:measurement}

In Secs.~\ref{sec:th13th14} and \ref{sec:th24th34} we have considered the case of a null result for sterile neutrino searches 
at the Neutrino Factory after 4 years running for both muon polarities, showing exclusion plots both in the 
($\theta_{13}^{\rm (4fam)},\theta_{14}$)- and in the ($\theta_{24},\theta_{34}$)-planes. 
However, due to the impressive statistics achievable at the Neutrino Factory, it could well be possible that a positive signal 
is found (if sterile neutrinos with $O(1\eVq)$ mass difference with respect to active ones do exist). 
For this reason, in Sec.~\ref{sec:discrimination} we have shown the region of the parameter space
for which it is possible to distinguish the (3+1)-model from three-family oscillations.

Eventually, we will present in this section a first analysis of the precision achievable in our setup in the simultaneous measurement of mixing angles and CP-violating phases. We first focus on $\theta_{24},\theta_{34}$ and on the CP-violating phase $\delta_3$.
Notice that each of the three possible CP-violating signals in a four-family model is related to a different Jarlskog invariant, 
proportional to a different combination of the mixing angles.  The Jarlskog invariant that depends on $\sin \delta_3$ is, 
in our parametrization, proportional to the combination $\sin 2 \theta_{23} s_{24} s_{34} \sin \delta_3$, as it can 
be seen in eqs.~(\ref{eq:pmumu2},\ref{eq:pmutau2}). A measurement of $\delta_3$ is thus possible only if both 
$\theta_{24}$ and $\theta_{34}$ are simultaneously non-vanishing. We will thus show
99\% CL contours in the $(\theta_{34},\delta_3$)-plane for particular input pairs $(\bar \theta_{34},\bar \delta_3$)
for fixed non-vanishing values of $\theta_{24}$. 

The measurement of $(\theta_{34},\delta_3)$ is achieved combining data from the $\nu_\mu$ disappearance channel 
and the $\nu_\mu \to \nu_\tau$ discovery channel.
This analysis, of course, does not pretend to be as exhaustive as those that have been presented in the framework 
of the three-family model. In particular, we will not address within a comprehensive approach
the problem of degeneracies in four-family models. Notice that this problem, extremely severe in the three-family oscillation studies at the Neutrino Factory (see, for example, Refs.~\cite{BurguetCastell:2001ez,Barger:2001yr} and 
\cite{Donini:2003vz}), is expected to be even more complicated in a four-neutrino model. 
In the particular case of the $\delta_3$-dependent CP-violating signal, that can be extracted using 
the $\nu_\mu \to \nu_\mu$ and $\nu_\mu \to \nu_\tau$ channels, we do expect to observe at least
degeneracies due to the ($\theta_{34},\delta_3$)-correlation (the so-called "intrinsic degeneracies", 
\cite{BurguetCastell:2001ez}); 
those dependent on the wrong reconstruction of the sign of the atmospheric mass difference\footnote{
At long baselines we are not sensitive to the sign of the SBL mass difference $\Delta m^2_{41}$.}  
$\Delta m^2_{31}$ (known as "sign degeneracies",  \cite{Minakata:2001qm}); 
and, eventually, those dependent on a wrong reconstruction of the "atmospheric" mixing angle
$\theta_{23}$ octant (known as "octant degeneracies",\cite{Fogli:1996pv}). 

The contours in the  ($\theta_{34},\delta_3$)-plane have been obtained as follows: 
we have first computed the expected number of events  for $\nu_\mu \to \nu_\mu$ and $\nu_\mu \to \nu_\tau$ oscillations 
in the four-family model for particular choices of the relevant parameters,
$\theta_{24} = \bar \theta_{24},\theta_{34} = \bar \theta_{34}$ and $\delta_3 = \bar \delta_3$.
We have then computed the expected number of events in the ($\theta_{34},\delta_{3}$)-plane for the same oscillation channels in the four-family model, varying $\theta_{34} \in [0,35^\circ]$ and $\delta_3 \in [0,360^\circ]$. 
The $\Delta \chi^2$ is then computed as follows:
\begin{equation}
\Delta \chi^2 =  \left (\sum_j \left [ N_j (\bar \theta_{24},\theta_{34}, \delta_3) - 
N_j (\bar \theta_{24},\bar \theta_{34}, \bar \delta_3) \right ]^2/\sigma^2_j \right )
\end{equation}
where the minimum of the $\chi^2$ is, trivially, obtained for $\theta_{34} = \bar \theta_{34}$; $\delta_3 = \bar \delta_3$.
As before, $j$ runs over the different signals: the $\nu_\mu$ disappearance and the $\nu_\mu \to \nu_\tau$ discovery channels data, divided into 10 energy bins, for the two baselines and the two possible stored muons polarities. 
The variance $\sigma_j$ is defined by eq. (\ref{variance}), with $f_j = 5\%$ for the $\nu_\mu$ disappearance channel and 
$10\%$ for the $\nu_\mu \to \nu_\tau$ discovery channel. No correlated systematic errors have been considered in the plots
of this section.
The region in the ($\theta_{34},\delta_3$)-plane compatible with the input values ($\bar \theta_{34},\bar \delta_3$)
at the 2 d.o.f.'s  99\% CL is eventually defined by drawing the contour line corresponding to $\Delta \chi^2 = 9.21$.
Notice that we have also studied the simultaneous measurement of $\theta_{24},\theta_{34}$ and $\delta_3$ using the combination of the $\nu_\mu \to \nu_\mu$ and $\nu_\mu \to \nu_\tau$ channels, finding that sensitivity to $\delta_3$ is lost
for values of the product $s_{24} s_{34}$ smaller than $(s_{24} s_{34})_{\rm min} \sim 0.01$. 
The results that we show have been obtained for choices of the input parameters $(\bar \theta_{24},\bar \theta_{34})$
such that $s_{24} s_{34} \geq (s_{24} s_{34})_{\rm min}$.

\begin{figure}[t]
\begin{center}
\vspace{-0.2cm}
\hspace{-0.5cm}
\begin{tabular}{cc}
\includegraphics[width=7.5cm]{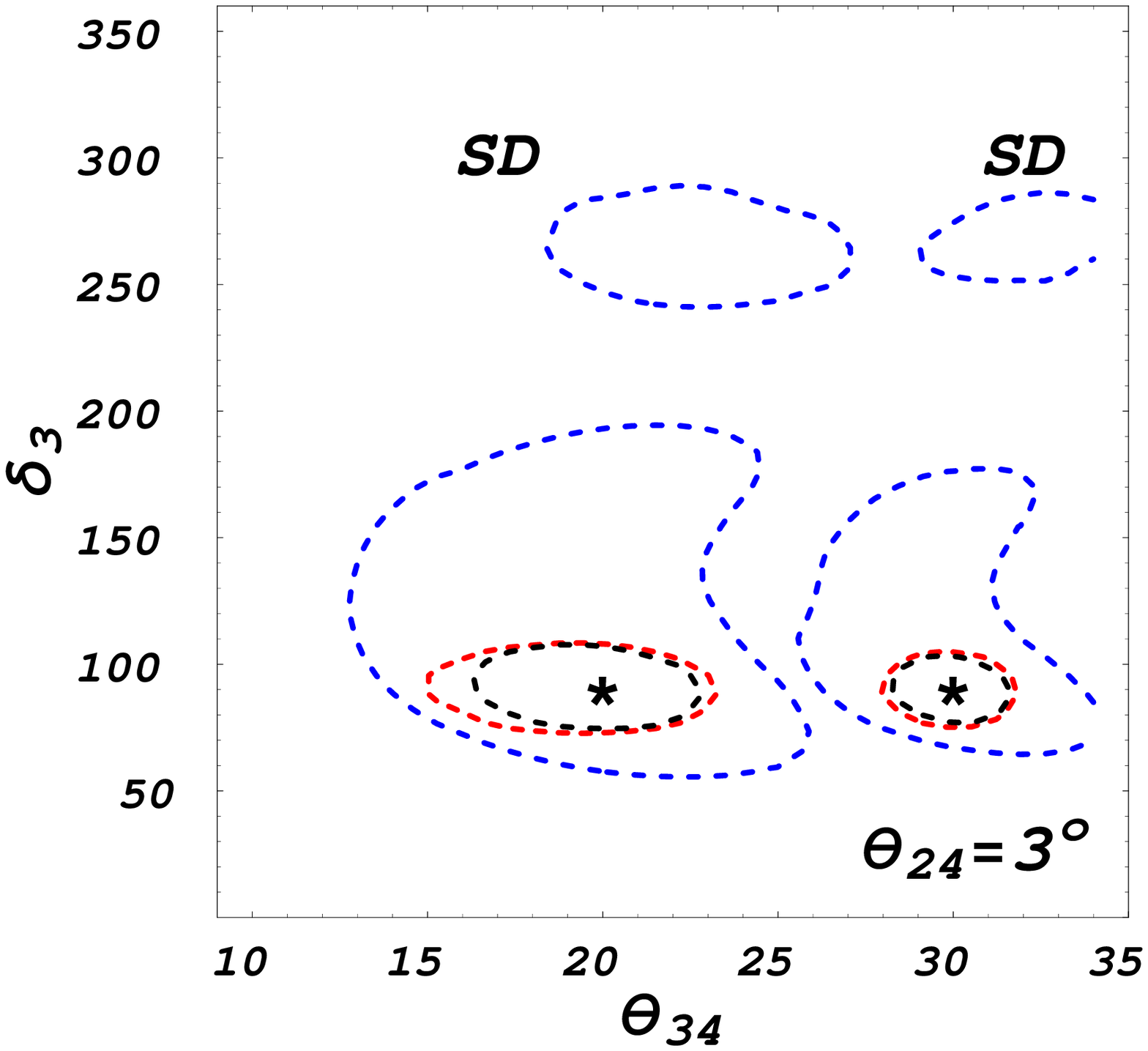}  &
\includegraphics[width=7.5cm]{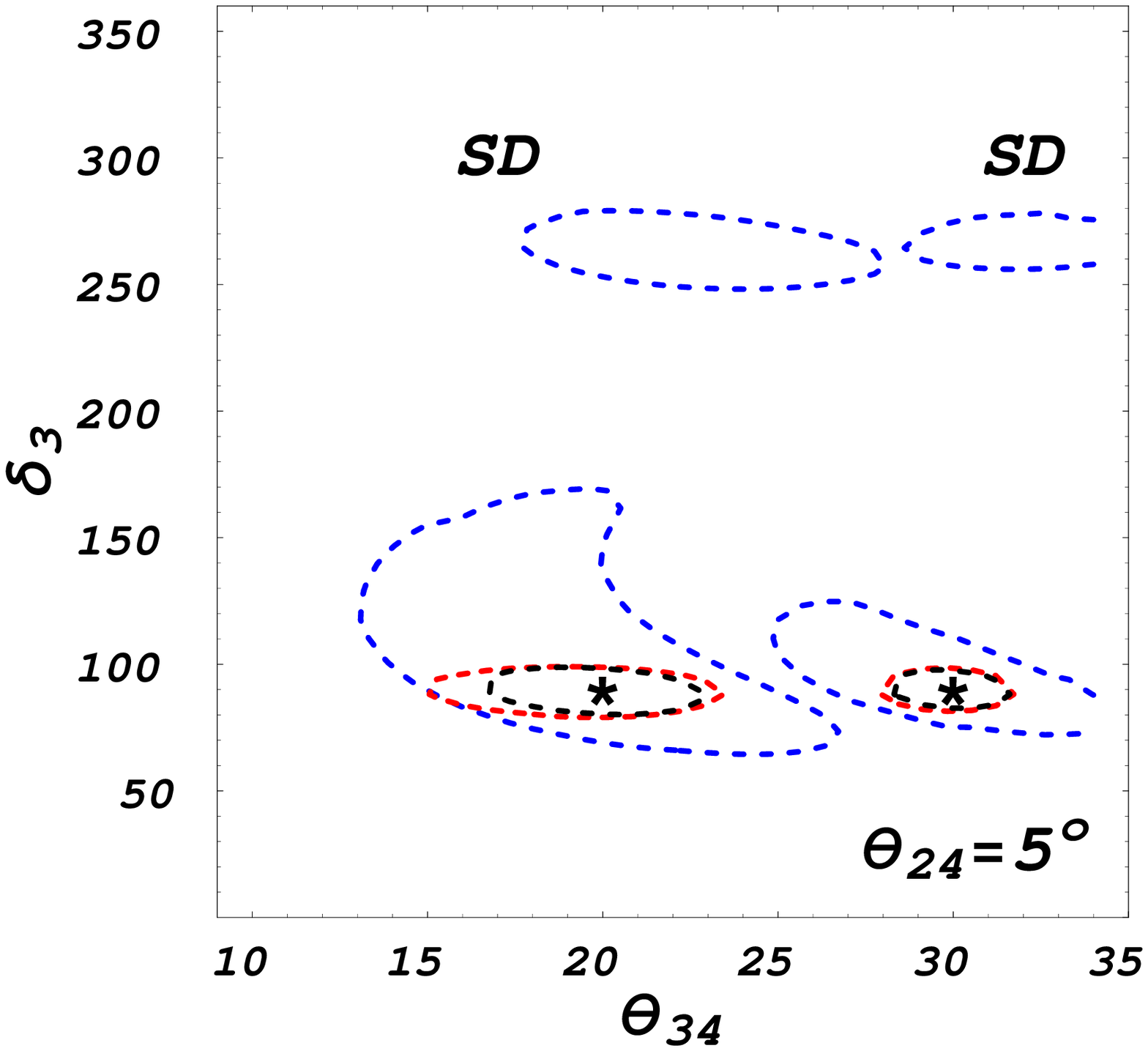} \\
\includegraphics[width=7.5cm]{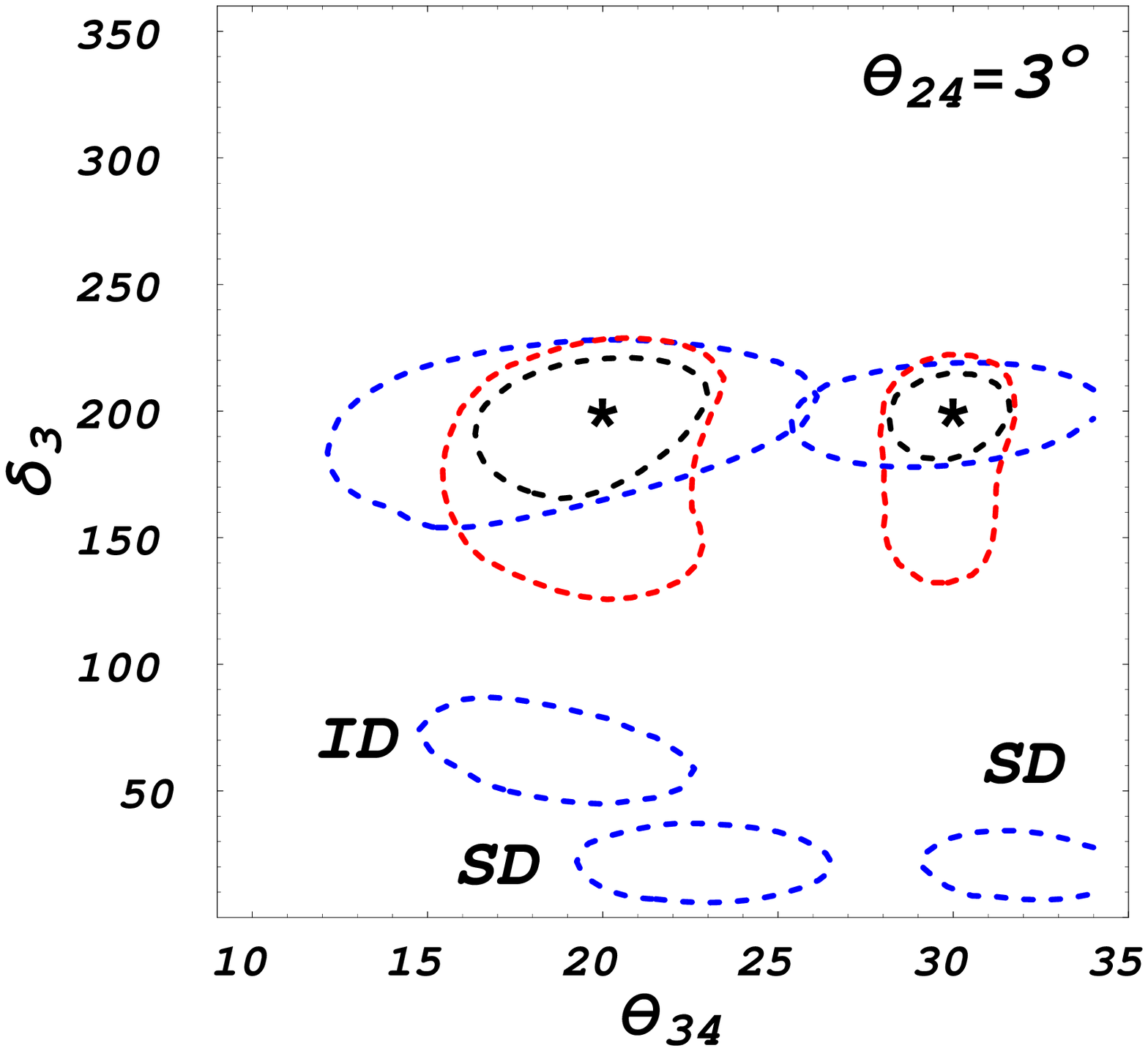}  &
\includegraphics[width=7.5cm]{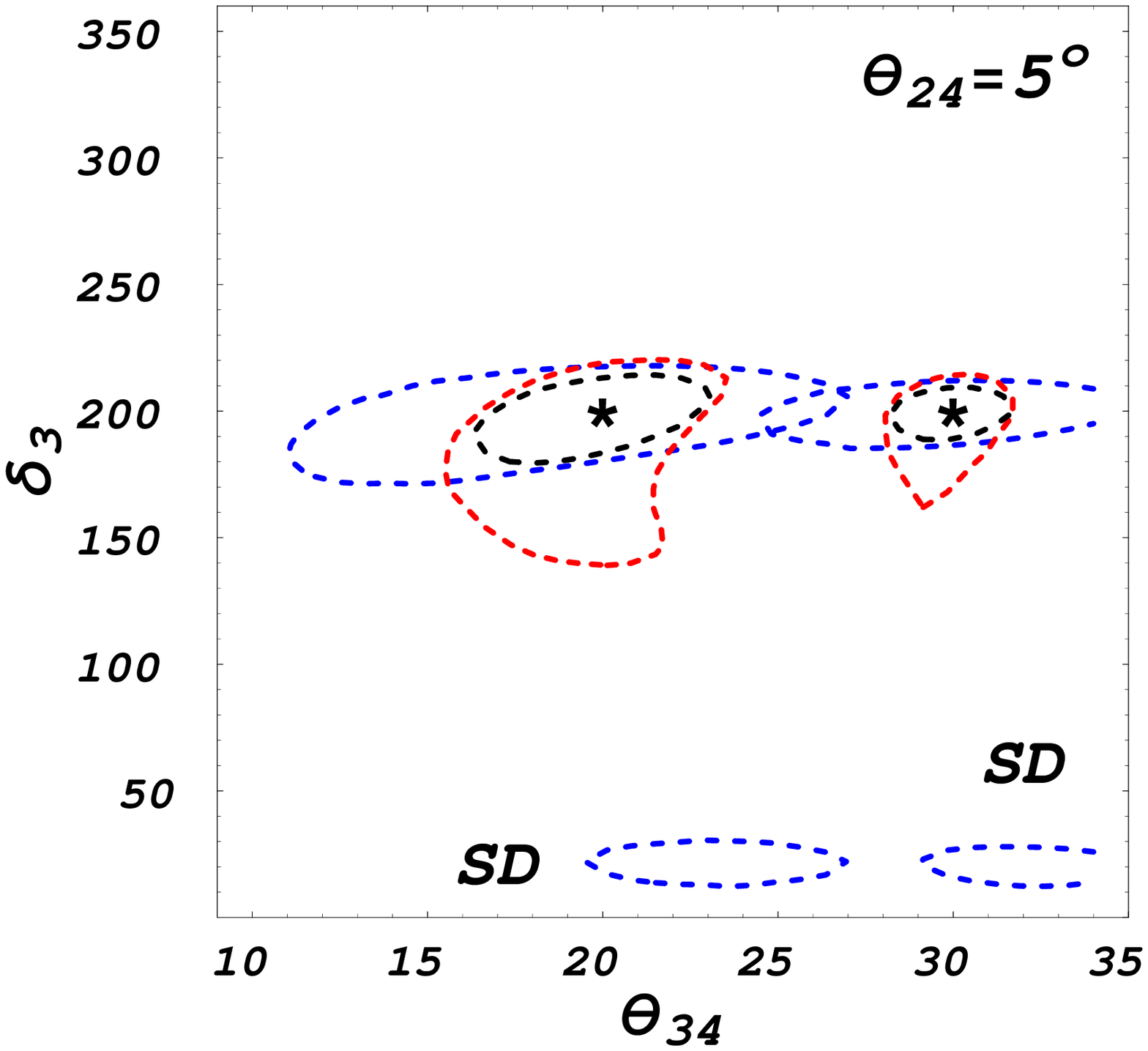} \\
\end{tabular}  
\caption{\label{fig:potatoes}\sl%
      99\% CL contours for the simultaneous measurement of $\theta_{34}$ and $\delta_3$ using the combined data   
      from the $\nu_\mu$ disappearance and the $\nu_\mu \to \nu_\tau$ discovery channels. Two
      different values of $\theta_{24}$ have been considered: $\theta_{24} = 3^\circ$ (left panels);
      $\theta_{24} = 5^\circ$ (right panels). The input pairs ($\bar \theta_{34},\bar \delta_3$), marked by a star
      in the plots, are: $\bar \theta_{34} = 20^\circ,30^\circ; \bar \delta_3 = 90^\circ$ (upper panels) and $200^\circ$ (lower
      panels).
      In the plots, "ID" stands for "Intrinsic Degeneracy"; "SD" stands for "Sign Degeneracy".
      Blue dashed lines represent the $L = 3000$ km baseline data; red dashed lines the $L = 7500$ km baseline data;
     black dashed lines stand for the combination of both baselines. 
           }
\end{center}
\end{figure}

We show in Fig.~\ref{fig:potatoes} the 2 d.o.f.'s  99 \%CL contours for the simultaneous measurement of $\theta_{34}$ and $\delta_3$ using the combined data from the disappearance and the discovery  channels for two representative values
 of $\theta_{24}$: $\bar \theta_{24} = 3^\circ$ (left panels) and $\bar \theta_{24} = 5^\circ$ (right panels). 
Blue dashed lines stand for the $L = 3000$ km baseline; red dashed lines stand for the $L = 7500$ km baseline;
black dashed lines stand for the combination of both baselines. 

In the numerical analysis, the following parameters in the four-family model have been kept fixed to their central values: 
$\theta_{12} = 34^\circ$, $\theta_{13} = 0$; $\Dmq_\Sol = 7.9 \times 10^{-5}~\eVq$; $\Dmq_{31} = \Dmq_\Atm = 
2.4 \times 10^{-3} \eVq$; $\delta_1 = \delta_2 = 0$. 
Eventually, $\Dmq_\Sbl = 1~\eVq$ and $\theta_{14} = 0^\circ$. 
For simplicity, we have fixed $\theta_{23} = 45^\circ$. We do not expect any "octant degeneracies", thus.
The input values that we have studied to illustrate the discovery potential of our setup are: 
$\bar \theta_{34} = 20^\circ, 30^\circ$; $\bar \delta_3 = 90^\circ$ (upper panels) and 
$\bar \delta_3 = 200^\circ$ (lower panels).
Matter effects have been included considering, as always, a constant matter density $\rho = 3.4$ g/cm$^3$
for the shortest baseline and $\rho = 4.3$ g/cm$^3$ for the longest one, 
computed averaging over the density profile in the PREM \cite{Dziewonski:1981xy} along the neutrino path.

First of all, we can see that the combination of the two channels at the shortest baseline (blue lines) is not enough
to solve the sign degeneracies (labeled with "SD" in the plot), that can be observed for all of the choices of the three input
parameters ($\bar \theta_{24},\bar \theta_{34},\bar \delta_3$). The sign clones are located at the point $(\theta_{34}^{\rm SD},\delta_3^{\rm SD})$, where $\theta_{34}^{\rm SD} \sim \bar \theta_{34}$ and
$\delta_3^{\rm SD}$ satisfies the relation $\sin \bar \delta_3 \sin \Delta_{31} L = - \sin \delta_3^{\rm SD} \sin \Delta_{31} L$, with $\delta_3^{\rm SD} \sim - 90^\circ$ for $\bar \delta_3 = 90^\circ$ and $\delta_3^{\rm SD} \sim 20^\circ$ for $\bar \delta_3 = 200^\circ$. The intrinsic degeneracy is also found for one 
specific choice of the input parameter ($\bar \theta_{24} = 3^\circ,\bar \theta_{34} = 20^\circ, \bar \delta_3 = 200^\circ$).
On the other hand, no intrinsic or sign degeneracy are found at the longest baseline (red lines). 
When combining the two baselines we see that the degeneracies are solved and that a very good precision 
on the simultaneous measurement of $\theta_{34}$ and $\delta_3$ is achieved for all the choices of the input parameters
 that we have considered. In particular, the error in $\delta_3$ at the 99\% CL is of the order of a few tens of degrees.
At the same time, the mixing angle $\theta_{34}$ can be measured for these particular inputs with a precision of
a few degrees. 

\begin{figure}[t]
\begin{center}
\vspace{-0.2cm}
\hspace{-1cm}
\begin{tabular}{cc}
\includegraphics[width=7.5cm]{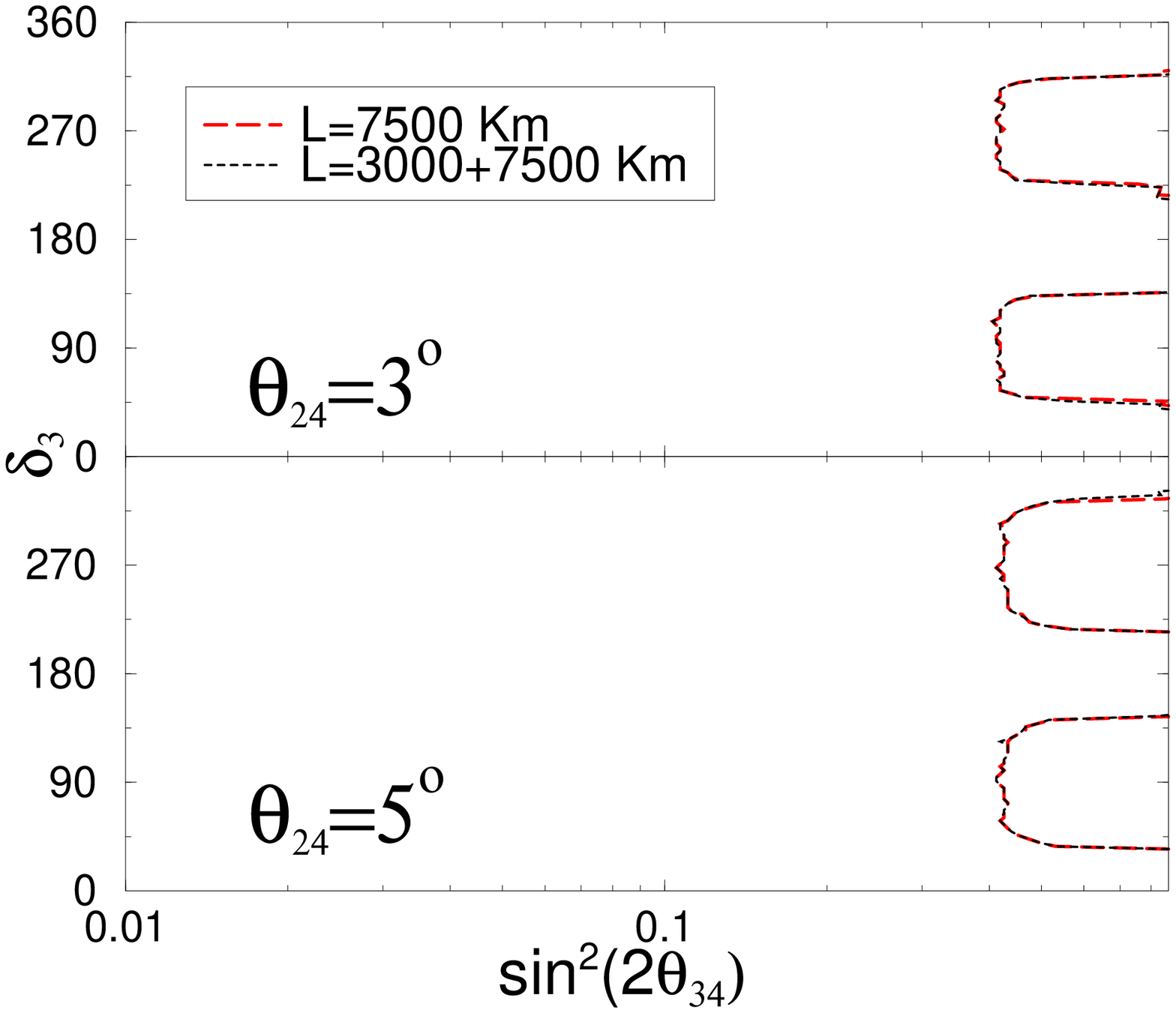}  &
\includegraphics[width=7.5cm]{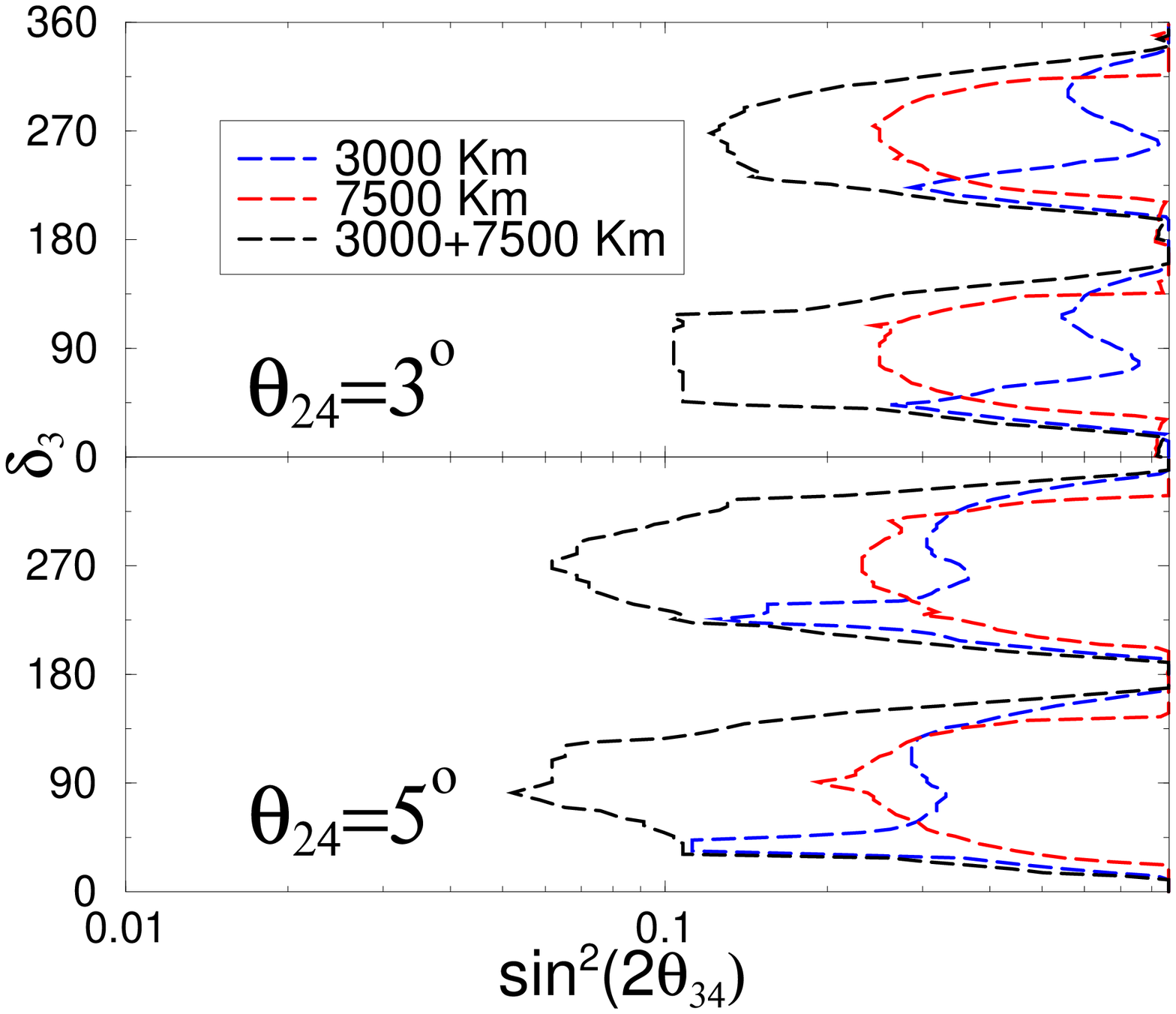} 
\end{tabular}  
\caption{\label{fig:discoverypotential}\sl%
 The 99 \% CL "$\delta_3$-discovery potential" in the ($\theta_{34},\delta_3$)-plane.
 Left: only $\nu_\mu \to \nu_\mu$ disappearance channel data; Right: combination of $\nu_\mu \to \nu_\mu$ disappearance 
 and $\nu_\mu \to \nu_\tau$ appearance channels data. Upper panels have been obtained for $\bar \theta_{24} = 3^\circ$; 
 lower panels for $\bar \theta_{24} = 5^\circ$.
 Blue dashed lines stand for $L = 3000$ km baseline data; red dashed lines stand for $L = 7500$ km baseline data; 
 black dashed lines stand for the combination of the two baselines. 
                   }
\end{center}
\end{figure}

We summarize our results for the simultaneous measurement of $\theta_{34}$ and $\delta_3$ in Fig.~\ref{fig:discoverypotential}, where the 99\% CL "$\delta_3$-discovery potential" in the ($\theta_{34},\delta_3$)-plane
for different values of $\bar \theta_{24}$ is shown.\footnote{Notice that, as we stressed at the beginning of this section, sensitivity to $\delta_3$ is lost when the product $s_{24} s_{34}$ is smaller than $(s_{24} s_{34})_{\rm min} \sim 0.01$.} 
We define the "$\delta_3$-discovery potential" as the region in the ($\sin^2  2\theta_{34},\delta_3$)-plane for which a given 
(non-zero) value of the CP-violating phase $\delta_3$ can be distinguished at the 99\% CL (for 2 d.o.f.'s)
from the CP-conserving case, i.e.,  $\delta_3= 0,\pi$.
Note that we have also taken into account the effects of the sign degeneracy in this analysis.

In the left panel, only data from the $\nu_\mu \to \nu_\mu$ disappearance channel are shown. In the right panel, 
we have combined data from the $\nu_\mu$ disappearance channel with those from the $\nu_\mu \to \nu_\tau$
appearance channel. Upper panels refer to $\bar \theta_{24} = 3^\circ$; lower panels to $\bar \theta_{24} = 5^\circ$.
Blue dashed lines stand for the $L=3000$ km baseline; red dashed lines stand for the $L=7500$ km baseline; 
eventually, black dashed lines stand for the combination of the two baselines. 

We can see from Fig.~\ref{fig:discoverypotential}(left) that, using $\nu_\mu$ disappearance channel only, 
we are able to measure a non-vanishing $\delta_3$ for values of $\theta_{34}$ above $\sin^2 2 \theta_{34} \geq 0.4 (\theta_{34} \geq 18^\circ)$. The CP-coverage\footnote{The CP-coverage is the fraction of the $\delta_3$-parameter space for which we are able to exclude $\delta_3 = 0,\pi$ at the 99\% CL for a given value of $\theta_{34}$.} is $\sim 50\%$,
with a very smooth dependence on $\theta_{34}$, being a bit larger for larger $\bar \theta_{24}$. 
We can also see that the detector at $L = 3000$ km have no $\delta_3$-sensitivity whatsoever. 

The situation is completely different when the $\nu_\mu \to \nu_\tau$ discovery channel data are added to the 
$\nu_\mu$ disappearance ones, Fig.~\ref{fig:discoverypotential}(right). First of all, we see that the $L=3000$ km 
detector is no longer useless to measure $\delta_3$: spikes of $\delta_3$-sensitivity for particular values of $\delta_3$
can be observed, in some cases outperforming the far detector results. However, it is in the combination of the two
baselines where we can see that a dramatic improvement in the $\delta_3$-discovery potential is achievable. 
When the $\nu_\mu \to \nu_\tau$ data are included, a non-vanishing $\delta_3$ can be measured for 
values of $\theta_{34}$ as small as $\sin^2 2 \theta_{34} = 0.06 (\theta_{34} = 7^\circ)$ for $\bar \theta_{24} = 5^\circ$
and $\sin^2 2 \theta_{34} = 0.10 (\theta_{34} = 9^\circ)$ for $\bar \theta_{24} = 3^\circ$. For $\sin^2 2 \theta_{34} \geq 0.4
(\theta_{34} \geq 20^\circ)$, roughly 80\% (60\%) of CP-coverage is achieved for $\bar \theta_{24} = 5^\circ (3^\circ)$. 
The striking improvement in the $\delta_3$-discovery potential is a consequence of the synergy of the two channels
and of the two baselines, whose combination is able to solve most of the correlations that otherwise strongly limits
the potential of the $\nu_\mu$ disappearance channel. 

\begin{figure}[t]
\begin{center}
\vspace{-0.2cm}
\hspace{-0.5cm}
\begin{tabular}{cc}
\includegraphics[width=7.5cm]{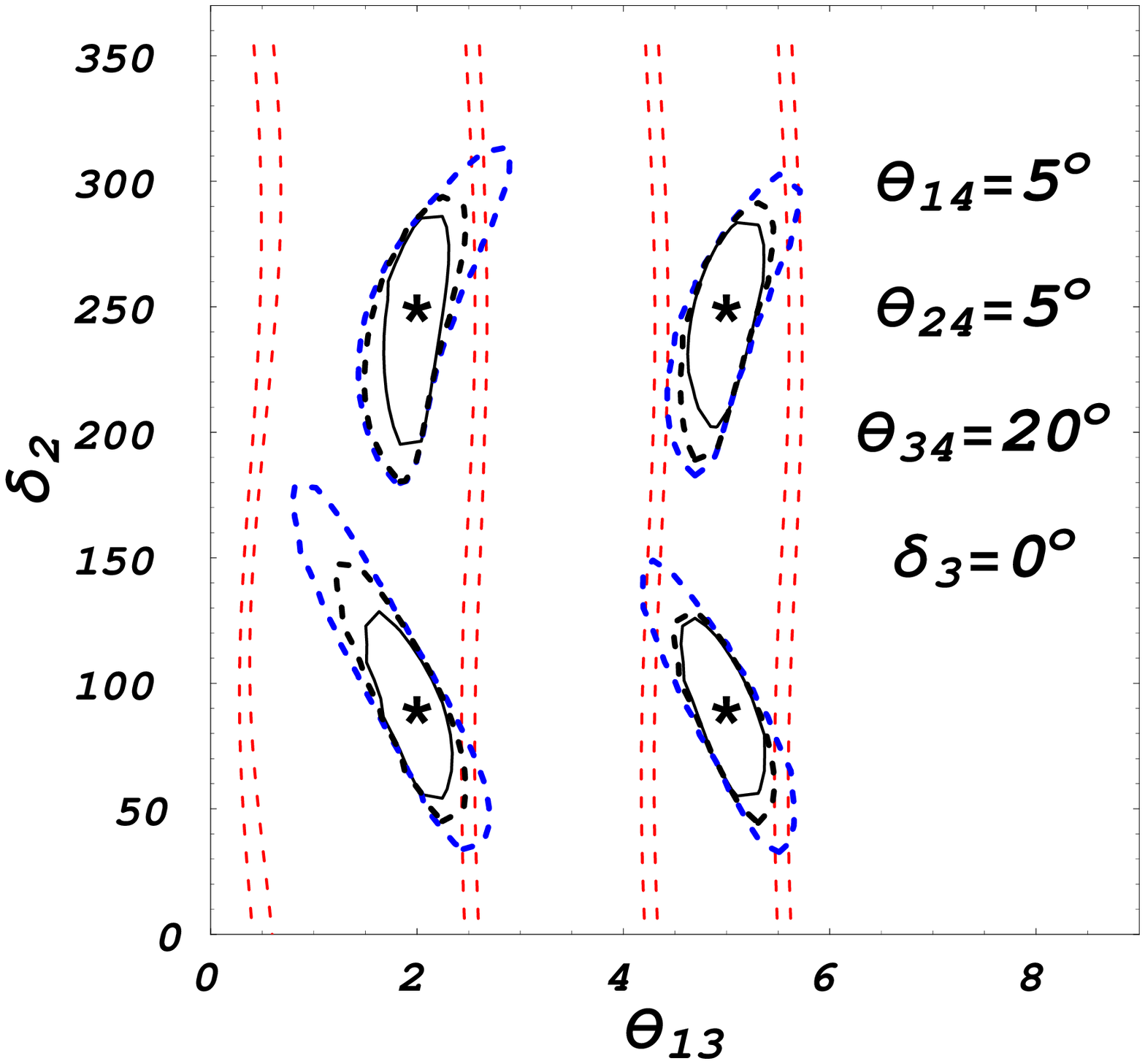}  &
\includegraphics[width=7.5cm]{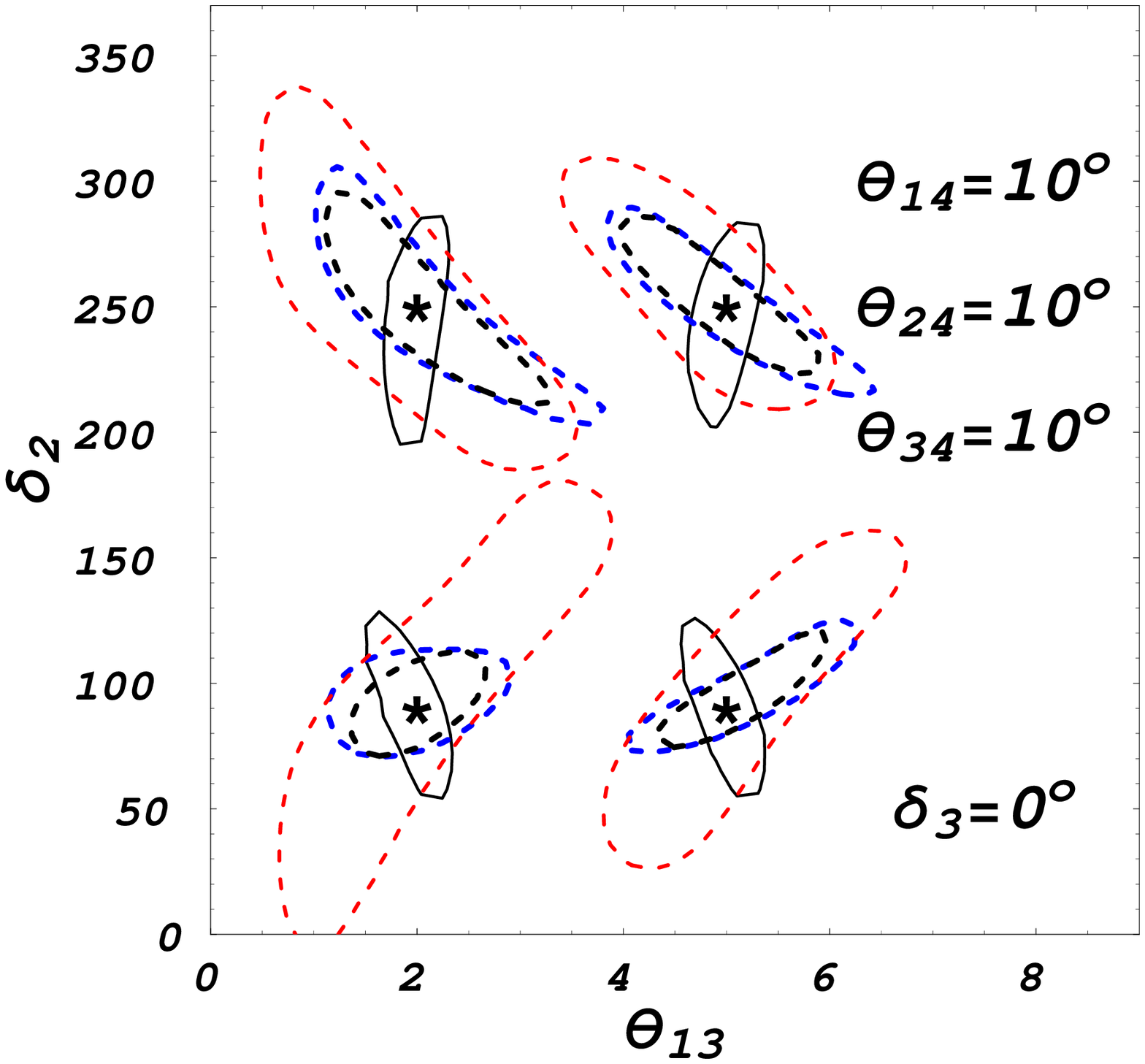} 
\end{tabular}  
\caption{\label{fig:delta2potatoes}\sl%
      99\% CL contours for the simultaneous measurement of $\theta_{13}$ and $\delta_2$ using the combined data   
      from the $\nu_e \to \nu_\mu$ and the $\nu_e \to \nu_\tau$ golden and silver channels. 
      Two different choices of the active-sterile mixing angles have been considered: 
      $\theta_{14} = \theta_{24} = 5^\circ; \theta_{34}= 20^\circ$ (left panel);
      $\theta_{14} = \theta_{24} = \theta_{34}=10^\circ$ (right panel). 
           The input pairs ($\theta_{13},\delta_2$), marked by a star
      in the plots, are: $\bar \theta_{13} = 2^\circ,5^\circ; \bar \delta_2 = 90^\circ,250^\circ$.
      Golden and silver channels data are always summed. 
      Four-family results are shown for the two baseline separately and summed: 
      blue dashed lines stand for the $L=3000$ km baseline data; red dashed lines stand for the $L=7500$ km baseline
      data; black dashed lines stand for the combination of all data. Eventually, black solid lines stand for the 
      three-family results for the combination of the baselines. 
           }
\end{center}
\end{figure}

For completeness, we also present in Fig.~\ref{fig:delta2potatoes} results for the sensitivity of the golden and silver channels to the phase $\delta_2 $, that reduces to the three-family CP-violating phase $\delta$ in the limit $\theta_{i4} \to 0$. A comment is in order: as it can be seen from eq.~(\ref{eq:emuprobeps6}), using the parametrization in eq.~(\ref{eq:3+1param2}), the golden channel oscillation probability in vacuum depends on the combination $(\delta_2 - \delta_3)$ up to the eighth-order in $\epsilon$. 
This means that, in the (3+1)-model, a CP-conserving result in the golden channel may be found for non-vanishing values
of $\delta_2$ and $\delta_3$ if $\delta_1 =0; (\delta_2 - \delta_3) = 0, \pi$.
This degeneracy can be broken only by adding new information, such as that obtained using the silver channel, see
eq.~(\ref{eq:etauprobeps6}), $\nu_\mu$ disappearance or $\nu_\mu \to \nu_\tau$ appearance data.
Golden channel data at the $L=3000$ km baseline may not be able, thus, to detect a non-vanishing CP-violating
signal even when both $\delta_2, \delta_3$ are different from $0,\pi$.

The contours in the  ($\theta_{13}^{(\rm 4fam)},\delta_2$)-plane have been obtained as follows: 
we have first computed the expected number of events  for $\nu_e \to \nu_\mu$ and $\nu_e \to \nu_\tau$ oscillations 
in the four-family model for particular choices of the relevant parameters,
$ \theta_{13}^{(\rm 4fam)} = \bar \theta_{13}^{(\rm 4fam)}$ and $\delta_2 = \bar \delta_2$.
We have then computed the expected number of events in the ($ \theta_{13}^{(\rm 4fam)},\delta_{2}$)-plane for the same oscillation channels in the four-family model, varying $ \theta_{13}^{(\rm 4fam)} \in [0,10^\circ]$ and 
$\delta_2 \in [0,360^\circ]$. The $\Delta \chi^2$ is then computed as follows:
\begin{equation}
\Delta \chi^2 =  \left (\sum_j \left [ N_j ( \theta_{13}^{(\rm 4fam)}, \delta_2) - 
N_j (\bar  \theta_{13}^{(\rm 4fam)},\bar \delta_2) \right ]^2/\sigma^2_j \right )
\end{equation}
where the minimum of the $\chi^2$ is, trivially, obtained for $ \theta_{13}^{(\rm 4fam)} = \bar  \theta_{13}^{(\rm 4fam)}$; 
$\delta_2 = \bar \delta_2$. As before, $j$ runs over the different signals: the $\nu_e \to \nu_\mu$ and the $\nu_e \to \nu_\tau$ data, divided into 10 energy bins, for the two baselines and the two possible stored muons polarities. 
The variance $\sigma_j$ is defined by eq. (\ref{variance}), with $f_j = 2\%$ for the golden channel and 
$10\%$ for the silver channel. No correlated systematic errors have been considered in the plots of this section.
The region in the ($ \theta_{13}^{(\rm 4fam)},\delta_2$)-plane compatible with the input values 
($\bar  \theta_{13}^{(\rm 4fam)},\bar \delta_2$) at the 2 d.o.f.'s  99\% CL is eventually defined by drawing the contour line corresponding to $\Delta \chi^2 = 9.21$.

Results have been obtained for $\bar \theta_{13}^{\rm (4fam)} = 2^\circ,5^\circ$ and $\bar \delta_2 = 90^\circ,250^\circ$. The other parameters are: $\theta_{12} = 34^\circ; \theta_{23} = 45^\circ$; $\Delta m^2_{21}  = 7.9 \times 10^{-5} $ eV$^2$, $\Delta m^2_{31} = 2.4 \times 10^{-3}$ eV$^2$ and $\Delta m^2_{41} = 1$ eV$^2$. For simplicity, we show results for $\delta_1 = \delta_3 = 0$ in these plots (remember that the measured phase
should be interpreted as $\delta = (\delta_2 - \delta_3)$ for $\delta_1 = 0$). 
Eventually, the three active-sterile mixing angles are: $\theta_{14} = \theta_{24} = 5^\circ;  \theta_{34} = 20^\circ$
in Fig.~\ref{fig:delta2potatoes}(left); and $\theta_{14} = \theta_{24} = \theta_{34} = 10^\circ$
in Fig.~\ref{fig:delta2potatoes}(right).
In the plots,  golden and silver channels data are always summed. Four-family results are shown for the two baselines separately and summed: blue dashed lines stand for the $L=3000$ km baseline data; red dashed lines stand for the $L=7500$ km baseline data; black dashed lines stand for the combination of all data. 
For comparison, black solid lines stand for the three-family results for the combination of the baselines. 

We can see from the plots in Fig.~\ref{fig:delta2potatoes} (left) that, when the active-sterile mixing angles 
$\theta_{14}$ and $\theta_{24}$ are "small", the four-family results are extremely similar to those obtained by
a fit in the three-family model.\footnote{ Results do not depend significantly on $\theta_{34}$: as it can be seen in 
eqs.~(\ref{eq:emuprobeps6}) and (\ref{eq:etauprobeps6}), this angle only enters in the silver channel oscillation probability, 
statistically less relevant than the golden channel data.} 
The 99\%  CL contours in the four-family model are slightly larger than the three-family
ones. The shape of contours in the $(\theta_{13}^{\rm (4fam)},\delta_2)$-plane is identical for the four- and three-family. 
This means that the correlation between the two parameters is not modified by contributions proportional to 
$\theta_{14},\theta_{24}$ with respect to three-family expressions. 

In Fig.~\ref{fig:delta2potatoes} (right) we see that, when $\theta_{14}$ and $\theta_{24}$ assume values near 
their upper bound, the results for the four-family contours can significantly differ from the three-family ones. 
In particular, the four-family contours in the $(\theta_{13}^{\rm (4fam)},\delta_2)$-plane are orthogonal to the three-family
ones. This is easily understood by looking at the approximated expressions for the oscillation probabilities
of the golden and silver channels expanded to order $\epsilon^8$ in Sec.~\ref{sec:th13th14}. As we can see, the two terms
with a $\delta_2$-dependence in the second and third lines of eq.~(\ref{eq:emuprobeps6}) are proportional
to $s_{13} s_{14} s_{24}  \sin (\delta_2 +  \Delta_{31} L/2) $ and to $s_{13} (\Delta_{21} L)  \cos(\delta_2 +  \Delta_{31} L/2) $, 
respectively. When the first term becomes as large as the solar-suppressed one, the two terms give a destructive interference. 
Eventually, if the first term becomes larger than the second, the $(\theta_{13}^{\rm (4fam)},\delta_2)$-correlation 
changes, becoming orthogonal to the three-family--like one, as it can be seen in the figure. 
Notice that, when $\theta_{14}$ and $\theta_{24}$ assume values near their upper 
bound, there is a little sensitivity to the CP-phases at the longer 
baseline. Since this baseline corresponds to the magic baseline, we can 
conclude that this is an effect characteristic to the four neutrino scheme.

\section{Conclusions}
\label{sec:concl}

We have studied the potential of a Neutrino Factory to search for
signature of the (3+1)-scheme, where the largest mass squared
difference $\Dmq_\Sbl$ can be any value larger than $0.1~\eVq$, as
long as oscillations driven by $\Dmq_\Sbl$ are averaged for the range
of the energy and the baseline at a Neutrino Factory.  From the
analytic expressions of the oscillation probabilities, we have seen
that the disappearance channel $\nu_\mu \to\nu_\mu$ and the
``{\it discovery channel}" $\nu_\mu \to\nu_\tau$ are extremely powerful in 
constraining the parameter space of the (3+1)-scheme.

We have performed a numerical analysis of the
sensitivity to sterile neutrinos in two setups: (a) A Neutrino
Factory with the muon energy $E_\mu=50$ GeV, $2\times 10^{20}$ useful
muon decays per year per baseline, with two detectors located at $L=3000$
km and $L=7500$ km from the source, respectively; (b) The setup suggested in the
final ISS Physics Report, i.e., a Neutrino Factory with $E_\mu=20$
GeV, $5\times 10^{20}$ useful muon decays per year per baseline, with two
detectors located at $L=4000$ km and $L=7500$ km from the source, respectively.  
The two detectors are of the Hybrid-MIND type (50 kton MIND + 4 kton MECC).
In both cases we have assumed data taking for 4 years for both muon polarities.
We have carefully taken into account the relevant backgrounds,
efficiencies and systematic errors and our analysis is, thus, 
much more detailed than in previous studies in the literature.

We have looked at the sensitivity to
$\theta_{13}$, $\theta_{14}$, analyzing the golden and silver channels.
While Neutrino Factories do
not give any useful constraint on $\theta_{14}$ when marginalizing over
$\theta_{24}$ and $\theta_{34}$,
as in the case of the three-flavor framework, they give quite a
strong constraint on $\sin^22\theta_{13}$, down to $7\times 10^{-5}$
($2\times 10^{-4}$) for the 50 GeV (20 GeV) case
with a slight dependence on $\theta_{14}$.
In the present work we did not assume any near detector,
since the issue has not yet been studied in detail (see
\cite{Laing:2008zz}).  If we put a near detector, then we should be
able to give a constraint on $\theta_{14}$ because the near
detector will give us information on how much neutrino oscillations
occur due to the largest mass squared difference $\Dmq_\Sbl$.  Further
studies on the near detector will be necessary to obtain quantitative
results on $\theta_{14}$.

Combining the disappearance and
discovery channels, we have found that both the 50 GeV (20 GeV) Neutrino
Factories can constrain $\theta_{34}$ and $\theta_{24}$ down to
12$^\circ$ (14$^\circ$) and 7.5$^\circ$ (8$^\circ$) , respectively,
marginalizing over all other four-family parameters.
If we closely look at the details, then
we found some difference between the performances of
the 50 GeV and 20 GeV Neutrino Factories, such as
the sensitivity plot in the ($\theta_{24},\theta_{34}$)-plane
or the regions in the ($\theta_{24},\theta_{34}$)-
and ($\theta_{13}^{(\rm 4fam)}$, $\theta_{14}$)-planes where the three-flavor
mixing hypothesis is excluded, where the sensitivity of
the 50 GeV Neutrino Factory is always a bit better
because the $\nu_\mu\to\nu_\tau$ channel, which
becomes more important at higher energies, makes this difference.
The present results should be compared to the previous
study~\cite{Donini:2007yf} on the CNGS experiment, which cannot improve the present bound on $\theta_{24}$ and $\theta_{34}$, if running at the nominal luminosity.
The sensitivity  to $\theta_{24}$ and $\theta_{34}$ of the Neutrino Factory setups proposed in this paper
is better than the potential sensitivity of the CNGS experiment even if the CNGS beam intensity would be
increased by a factor 10.

We have also discussed dependence of the sensitivity of
the four channels on the systematic errors.
The sensitivities of the golden and silver channels depend mainly
on the correlated systematic error $\sigma_\alpha$
on the overall normalization.
As for the disappearance and discovery channels,
the sensitivity of the former depends
on the bin-to-bin uncorrelated error $f_j$ and $\sigma_\alpha$,
whereas that of the latter depends mainly on $f_j$.
To take advantage of the discovery channel, it is necessary
to reduce $f_j$ below 10\%. To this purpose, the expertise on ECC technology that will be gained
after the first years of OPERA data taking will be of great importance.

On the other hand, in the case where we have a positive signal of the
sterile neutrino mixing, we have found that we can measure the new CP
phase $\delta_3$ by combining the disappearance and discovery channels
at the two baselines, where information from the $L=7500$ km baseline
plays a dominant role in resolving parameter degeneracy. 
After combination of the two channels and the two baselines, the
CP-violating phase $\delta_3$
can be measured at 99\% CL for values of $\sin^2 2 \theta_{34} \geq 0.06$
(notice that, using only the disappearance channel at the two baseline,
$\delta_3$ can be measured at 99\% CL for $\sin^2 2 \theta_{34} \geq 0.4$, only).
It should be emphasized that the measurement of CP-violation in the
$\nu_\mu\to\nu_\tau$ channel is a clear new signal of CP-violation
associated to the sterile neutrino scheme.
We have also found that the measurement of the three-family--like CP-violating
phase $\delta_2$ is modified by the presence of non-vanishing active-sterile mixing angles.

While our works contain some results on parameter degeneracy in the
(3+1)-scheme, the problem on the general structure of parameter
degeneracy in the four neutrino schemes is beyond the scope of the
present paper and should be pursued in future.

Finally, we would like to stress that, while the discovery channel at a
Neutrino Factory is not very useful for the measurements of the three-flavor 
oscillation parameters, it is a very important channel to
search for new physics beyond the standard scenario.  The discovery
channel at a Neutrino Factory deserves, thus, further studies.

\vglue 0.5in
\begin{flushleft}
{\bf\large Appendix}
\end{flushleft}
\appendix
\vspace{-0.2in}

\section{The  mixing matrix elements $U_{\alpha j}$}
\label{app:uaj}
The mixing matrix elements in the parametrization (\ref{eq:3+1param2}) are
given by the following:
 
\begin{eqnarray}
&
\left \{
\begin{array}{lll}
U_{e1} & = & c_{12} c_{13} c_{14} \\
U_{e2} & = & c_{13} c_{14} s_{12} e^{-i \delta_1} \\
U_{e3} & = & c_{14} s_{13} e^{-i \delta_2} \\
U_{e4} & = & s_{14}
\end{array}
\right. 
&
\\
&
\left \{
\begin{array}{lll}
U_{\mu 1} & = & - c_{23} c_{24} s_{12} e^{i \delta_1} 
- c_{12} \left [
c_{24} s_{13} s_{23} e^{i (\delta_2 - \delta_3)}  
+ c_{13} s_{14} s_{24} \right ] \\
U_{\mu 2} & = & c_{12} c_{23} c_{24} - s_{12} e^{- i \delta_1}
\left [ c_{24} s_{13} s_{23} e^{i (\delta_2 - \delta_3)}
+ c_{13} s_{14} s_{24}
\right ] \\
U_{\mu 3} & = & c_{13} c_{24} s_{23} e^{-i \delta_3}
 - s_{13} s_{14} s_{24} e^{-i \delta_2}\\
U_{\mu 4} & = & c_{14} s_{24}
\end{array}
\right .
&
\end{eqnarray}

\begin{eqnarray}
\left \{
\begin{array}{lll}
U_{\tau 1} & = & s_{12} e^{i \delta_1}
\left [ c_{34} s_{23} e^{i \delta_3} + c_{23} s_{24} s_{34} \right ]\\
&{\ }&-c_{12} \left \{ 
c_{13} c_{24} s_{14} s_{34} + s_{13} e^{i \delta_2} \left [
c_{23} c_{34} - s_{23} s_{24} s_{34} e^{-i \delta_3}
\right ]
\right \} \\
U_{\tau 2} & = & - c_{12} \left [ c_{34} s_{23} e^{i \delta_3}
 + c_{23} s_{24} s_{34} \right ]\\
&{\ }&- s_{12} e^{- i \delta_1} \left \{ 
c_{13} c_{24} s_{14} s_{34} + s_{13} e^{i \delta_2} \left [
c_{23} c_{34} - s_{23} s_{24} s_{34} e^{-i \delta_3}
\right ]
\right \} \\
U_{\tau 3} & = & -c_{24} s_{13} s_{14} s_{34} e^{-i \delta_2}
 + c_{13} \left [ 
c_{23} c_{34} - s_{23} s_{24} s_{34} e^{-i \delta_3}
\right ] \\
U_{\tau 4} & = & c_{14} c_{24} s_{34}
\end{array}
\right .
&&
\\
\left \{
\begin{array}{lll}
U_{s1} & = & s_{12} e^{i \delta_1}
\left [ c_{23} c_{34} s_{24} - s_{23} s_{34} e^{i \delta_3} \right ]\\
&{\ }&- c_{12} \left \{ 
c_{13} c_{24} c_{34} s_{14} - s_{13} e^{i \delta_2} \left [
c_{34} s_{23} s_{24} e^{- i \delta_3} + c_{23} s_{34}
\right ]
\right \} \\
U_{s2} & = & - c_{12} \left [ c_{23} c_{34} s_{24}
- s_{23} s_{34} e^{i \delta_3 }\right ]\\
&{\ }&- s_{12} e^{-i \delta_1} \left \{
c_{13} c_{24} c_{34} s_{14} - s_{13} e^{i \delta_2} \left [
c_{34} s_{23} s_{24} e^{- i \delta_3} + c_{23} s_{34}
\right ]
\right \} \\
U_{s3} & = & - c_{24} c_{34} s_{13} s_{14} e^{- i \delta_2} - c_{13}
\left [ 
c_{34} s_{23} s_{24} e^{- i \delta_3} + c_{23} s_{34}
\right ] \\
U_{s4} & = & c_{14} c_{24} c_{34}
\end{array}
\right .
&&
\end{eqnarray}
where $c_{ij} = \cos \theta_{ij}$ and $s_{ij} = \sin \theta_{ij}$.

\section{Oscillation probabilities by the KTY formalism}
\label{app:formulae}

To derive the expressions for the oscillation probabilities in matter,
we use the KTY formalism which has been introduced in Ref.~\cite{Kimura:2002hb,Kimura:2002wd}.\footnote{
Another proof of the KTY formalism was given in Ref.~\cite{Xing:2005gk,Yasuda:2007jp}
and it was extended to four neutrino schemes in Ref.~\cite{Zhang:2006yq}.}
The evolution equation of flavor eigenstates\footnote{Greek
(Latin) indices label the flavor (mass) basis: $\alpha={e,\mu,\tau,s}$
($i={1,2,3,4}$).} is:

\be
i\frac{d}{dt} |\nu_{\alpha}\rangle = {\cal H}_{\alpha \beta} \,|\nu_{\beta}\rangle \equiv
\left[ U \mathcal{E} U^{\dagger} + \mathcal{A}\right] _{\alpha\beta} \,|\nu_{\beta}\rangle,
\NO
\ee
where

\bea
\mathcal{E} &=& \mrm{diag}(0,\ \frac{\Delta m^2_{21}}{2E},
\ \frac{\Delta m^2_{31}}{2E},\ \frac{\Delta m^2_{41}}{2E})
\ \equiv\ \mrm{diag}(0,\ \Delta_{21},\ \Delta_{31},\ \Delta_{41}), 
\label{kty-e}\\
\mathcal{A} &=& \sqrt{2}\,G_{F}\, \mrm{diag}(n_{\mrm{e}},
\ 0,\ 0,\ n_{\mrm{n}}/2)\ \equiv\ \mrm{diag}(A_{\mrm{e}},
\ 0,\ 0,\ A_{\mrm{n}}),\NO
\eea
$\Delta_{ij}=\Delta m_{ij}^2/2E$, and $n_e$ and $n_n$ are respectively
the electron and neutron densities.  In eq. (\ref{kty-e}),
we have subtracted from ${\cal H}$ the term $E_1\,{\bf 1}=\sqrt{E^2+m^2_1}\,{\bf 1}$,
which contributes only to the phase of the oscillation amplitude
and therefore does not affect the probability.
In the KTY
formalism, the oscillation probabilities in matter assume the
following form:
\bea
P_{\alpha\beta}
&=& \delta_{\alpha\beta}
-4\sum_{i<j}\mbox{\rm Re}(\tilde{X}^{\alpha\beta}_i
\tilde{X}^{\alpha\beta *}_j)
\sin^2\left( \dfrac{\Delta \tilde{E}_{ij}L}{2}\right)
+2\sum_{i<j}\mbox{\rm Im}(\tilde{X}^{\alpha\beta}_i
\tilde{X}^{\alpha\beta *}_j)
\sin(\Delta \tilde{E}_{ij}L),
\label{P} \NO \\
&&
\eea
where $\Delta \tilde{E}_{ji}\equiv \tilde{E}_{j}-\tilde{E}_{i}$
and $\tilde{X}^{\alpha\beta}_j\equiv
\tilde{U}_{\alpha j}\tilde{U}^{*}_{\beta j}$~($j=1,2,3,4)$. 
$\tilde E_i$ and $\tilde U_{\alpha i}$ are the ${\cal H}$
eigenvalues and the effective mixing matrix 
in matter, respectively, defined through
\be
{\cal H}=\tilde{U}\,\mrm{diag}(\tilde{E}_j)\,\tilde{U}^\dagger.\NO
\ee
The  $\tilde{X}^{\alpha\beta}_j$ matrices can be expressed as follows:
\be
\tilde{X}^{\alpha\beta}_j\equiv\sum_{l}\left(V^{-1}\right)_{jl} [{\rm {\cal H}}^{l-1}]_{\alpha\beta}=\sum_{l}\left(V^{-1}\right)_{jl}\,\left[ \left( U\mathcal{E} U^\dagger + \mathcal{A}\right)^{l-1} \right]_{\alpha\beta}, 
\label{xtilde}
\ee
where $V$ is the Vandermonde matrix:
\begin{eqnarray}
V=\left(\begin{array}{llll}
1&1&1&1\cr
\tilde{E}_1&\tilde{E}_2&\tilde{E}_3&\tilde{E}_4\cr
\tilde{E}_1^2&\tilde{E}_2^2&\tilde{E}_3^2&\tilde{E}_4^2\cr
\tilde{E}^{3}_1&\tilde{E}^{3}_2&\tilde{E}_3^3&\tilde{E}^{3}_4
\ea\right) \, ,\NO
\end{eqnarray}
whose determinant is $\prod_{i<j}\,\Delta\tilde{E}_{ji}$. 
The inverse of $V$ can then be easily obtained as long as we know the eigenvalues $\tilde{E}_j$ of the effective Hamiltonian in matter ${\cal H}$, expressed in terms of $A_e$, $A_n$ and the vacuum parameters:
\begin{eqnarray}
\hspace{-3.3mm}
V^{-1}=\left(\begin{array}{cccc}
\displaystyle
\frac{{\ }1}{\Delta \tilde{E}_{21} \Delta \tilde{E}_{31} \Delta \tilde{E}_{41}}
(\tilde{E}_2\tilde{E}_3\tilde{E}_4, &
-(\tilde{E}_2\tilde{E}_3+\tilde{E}_3\tilde{E}_4+\tilde{E}_4\tilde{E}_2),&
\tilde{E}_2+\tilde{E}_3+\tilde{E}_4,&
-1)\cr
\displaystyle
\frac{-1}{\Delta \tilde{E}_{21} \Delta \tilde{E}_{32} \Delta \tilde{E}_{42}}
(\tilde{E}_3\tilde{E}_4\tilde{E}_1, &
-(\tilde{E}_3\tilde{E}_4+\tilde{E}_4\tilde{E}_1+\tilde{E}_1\tilde{E}_3),&
\tilde{E}_3+\tilde{E}_4+\tilde{E}_1,&
-1)\cr
\displaystyle
\frac{{\ }1}{\Delta \tilde{E}_{31} \Delta \tilde{E}_{32} \Delta \tilde{E}_{43}}
(\tilde{E}_4\tilde{E}_1\tilde{E}_2, &
 -(\tilde{E}_4\tilde{E}_1+\tilde{E}_1\tilde{E}_2+\tilde{E}_2\tilde{E}_4),&
\tilde{E}_4+\tilde{E}_1+\tilde{E}_2,&
-1)\cr
\displaystyle
\frac{-1}{\Delta \tilde{E}_{41} \Delta \tilde{E}_{42} \Delta \tilde{E}_{43}}
(\tilde{E}_1\tilde{E}_2\tilde{E}_3, &
 -(\tilde{E}_1\tilde{E}_2+\tilde{E}_2\tilde{E}_3+\tilde{E}_3\tilde{E}_1),&
\tilde{E}_1+\tilde{E}_2+\tilde{E}_3,&
-1)\cr
\end{array}\right)\, .\nonumber\\
\label{V}
\end{eqnarray}

Within the KTY formalism, thus, we only need to compute the
eigenvalues of ${\cal H}$ to derive the oscillation probabilities in
matter.

A possible drawback of this approach is that the physical
understanding of the oscillation probabilities (i.e. the dependence on
the vacuum mixing matrix parameters) is encoded in the explicit
expressions for the $\tilde X$ coefficients. To make contact with the
parameters to be measured in a manageable way, we thus need to
introduce some approximations in the computation of the eigenvalues
$\tilde E_i$ and of the corresponding matrices $\tilde X_i^{\alpha
\beta}$.  Now,
considering the present constraints from \cite{Donini:2007yf} in the
standard and sterile small parameters, we see that
$\theta_{13},\theta_{14}$ and $\theta_{24}$ cannot be much larger than
$10^\circ$ while the third active-sterile mixing angle, $\theta_{34}$,
can be as large as $\theta_{34} \sim 35^\circ$.
Notice also that the present constraint on the $\theta_{23}$ deviation
from the maximal mixing, $\delta \theta_{23} \equiv \theta_{23}-\pi/4$,
is of the same order as those on $\theta_{13},\theta_{14}$ and
$\theta_{24}$.
On the other hand, the solar and atmospheric mass differences,
$\Dmq_\Sol$, $\Dmq_\Atm$, are much smaller than $\Dmq_\Sbl$.  In what follows,
therefore, we expand all the quantities
in power of a small parameter $\epsilon$,
and keep terms of cubic order in $\epsilon$, where the small
parameter is defined by
\bea
\epsilon &\equiv&\theta_{34}\sim \sqrt{\theta_{13}} \sim \sqrt{\theta_{14}}\sim\;
\sqrt{\theta_{24}}\sim\sqrt{\delta \theta_{23}}\;\lesssim 4\times 10^{-1},\NO\\
\eta_2&\equiv& \Delta m_{21}^2/\Delta m^2_{41} 
\lesssim 10^{-4},
\NO\\
\eta_3 &\equiv& \Delta m_{31}^2/\Delta m_{41}^2 
\lesssim 10^{-3},
\NO\\
\eta_{e(n)} &\equiv& A_{e(n)}/\Delta E_{41}
\lesssim 10^{-3} \, .
\nonumber
\eea
Notice that, to third order in $\epsilon$, in the expansion in the probabilities
we have neglected all terms proportional to $\eta_{e,n,2,3}$.
Although this can be a rather rough approximation, as we have seen
before, it is very useful in order to understand the different
physics potential of the various oscillation channels.  Thus
we have the following probabilities to third order in $\epsilon$:
\bea
\NO
P_{ee}&\sim& 1+O\left( \epsilon^4 \right), \\
\NO
P_{e\mu}&\sim& P_{e\tau}\;\sim\; P_{e s}\; \sim\;  O\left(\epsilon^4 \right), \\
\NO
P_{\mu\mu}&=&1- \sin^2\frac{\Delta_{31} L}{2}-2\left( A_nL\right)s_{24}\,s_{34}\cos\delta_3\sin \Delta_{31} L
+ O\left( \epsilon^4 \right) \; ,\\
P_{\mu\tau}&=&\left( 1-s_{34}^2 \right) \sin^2\frac{\Delta_{31} L}{2}+\left\lbrace s_{24}\,s_{34}\sin\delta_3+2\left( A_nL\right)s_{24}\,s_{34}\cos\delta_3\right\rbrace \sin\Delta_{31}L \NO \\
&& \NO
+ O\left( \epsilon^4 \right) , \\
\NO
P_{\mu s}&=&s_{34}^2\sin^2\frac{\Delta_{31} L}{2}-s_{24}\,s_{34}\sin\delta_3\sin\Delta_{31}L +O\left( \epsilon^4 \right)\;.
\eea

In Sec.~\ref{sec:th13th14} we had to go beyond $O(\epsilon^3)$ in order to explain our numerical results using the golden 
and silver channels.To this purpose, in the text we have shown the approximated expressions for $P_{e\mu}$ and 
$P_{e\tau}$ to order $\epsilon^8$ in vacuum. To check unitarity of the four-family PMNS matrix at this order in $\epsilon$, 
it is useful to show here the $P_{es}$ oscillation probability, also: 
\begin{eqnarray}
\label{eq:esprobeps6}
P_{e s} &=& 2\, \theta_{14}^2 (1 - \theta_{14}^2 - \theta_{24}^2 - \theta_{34}^2 )
\nonumber \\
 &+& 2 \left\lbrace \theta_{13}^2 (   - 2\,  \theta_{14}^2 +  \theta_{24}^2 + \theta_{34}^2  - 2 \delta \theta_{23} \theta_{34}^2  ) 
+ \theta_{13}^2 \theta_{24} \theta_{34}\cos\delta_3\right\rbrace  \sin^2 \frac{\Delta_{31} L}{2}
\nonumber\\
 &-&2  \sqrt{2}\, \theta_{13} \theta_{14} \theta_{24} (1 + \delta \theta_{23} - \theta_{34}^2)\sin \left (\delta_2 - \delta_3 
 + \frac{\Delta_{31} L}{2} \right )  \sin \frac{\Delta_{31} L}{2}
 \nonumber\\
&-& 2   \sqrt{2}\, \theta_{13} \theta_{14} \theta_{34} \left (1- \delta \theta_{23} - \frac{\theta_{34}^2}{2} \right)\sin \left ( \delta_2 
+ \frac{\Delta_{31} L}{2} \right )  \sin \frac{\Delta_{31} L}{2}
\nonumber \\
&-&  \sin2\theta_{12}\,  \theta_{13} \theta_{34}^2  (\Delta_{21} L) 
\cos \left ( \delta_1-\delta_2+\delta_3 - \frac{\Delta_{31} L}{2} \right ) \sin \frac{\Delta_{31} L}{2}
\nonumber\\
&+&\frac{1}{\sqrt{2}} \sin 2\theta_{12}\, \theta_{14} \theta_{34}  (\Delta_{21}L ) \sin(\delta_1+\delta_3)
- \frac{1}{\sqrt{2}} \sin 2 \theta_{12}\, \theta_{14} \theta_{24}  (\Delta_{21} L) \sin \delta_1 \, .
\nonumber
\end{eqnarray}

As a final analytical contribution, we have calculated approximate 
probabilities associated to the channels under study, $P_{\mu\mu}$ and 
$P_{\mu\tau}$ (together with $P_{\mu s}$), to fourth order in $\epsilon$ but neglecting $\theta_{13}$ and 
$\theta_{14}$:
\bea
P_{\mu\mu}&=&1-2\,\theta_{24}^2 - \left[ 1 - 4  (\delta \theta_{23})^2 - 2\theta_{24}^2 + 
\theta_{34}^2  \frac{A_n}{\Delta_{31}}  \left ( 4 \delta \theta_{23}- \theta_{34}^2 \frac{A_n}{\Delta_{31}} \right )
\right]\sin^2\frac{\Delta_{31} L}{2}
\NO\\
&-&\left( A_nL\right)\left\lbrace 2\theta_{24}\,\theta_{34}\cos\delta_3- \frac{\theta_{34}^2}{2}
\left ( 4 \delta\theta_{23} - \theta_{34}^2 \frac{A_n}{2 \Delta_{31}} \right )
\right\rbrace \sin \Delta_{31} L+ O(\epsilon^5) \; , \NO
\eea
\bea
P_{\mu\tau}&=&\left \{ 1-4 (\delta\theta_{23})^2-\theta_{24}^2-\theta_{34}^2
\left [ 1- \frac{\theta_{34}^2}{3}   - \frac{A_n}{\Delta_{31}} \left (4\delta\theta_{23} - \theta_{34}^2 \frac{A_n}{\Delta_{31}} \right )
\right ] \right \} \sin^2\frac{\Delta_{31} L}{2} \NO \\
&+&\left\lbrace \theta_{24}\,\theta_{34}\sin\delta_3+\left( A_nL\right)\left[2\theta_{24}\,\theta_{34}\cos\delta_3
-  \frac{\theta_{34}^2}{2} \left ( 4 \delta\theta_{23} - \theta_{34}^2 \frac{A_n}{2 \Delta_{31}} \right )
\right] \right\rbrace \sin\Delta_{31}L \NO \\
&+& O(\epsilon^5) \;,  \NO
\eea
\bea
P_{\mu s}&=&2\,\theta_{24}^2+\left[\theta_{34}^2 \left (1 -\frac{\theta_{34}^2}{3} \right )
-\theta_{24}^2\right]\sin^2\frac{\Delta_{31} L}{2}-\theta_{24}\,\theta_{34}\sin\delta_3 \sin\Delta_{31}L \NO \\
&+& O(\epsilon^5) \; . \NO
\eea

\section*{Acknowledgements}

We acknowledge H. Minakata for useful comments, M.~Maltoni and
P.~Migliozzi for useful discussions, and P. Lipari and M.~Lusignoli for discussions.
The work has been partially supported by the JSPS-CSIC Bilateral Joint Projects (Japan- Spain), 
a Grant-in-Aid for Scientific Research of the Ministry of Education, Science and Culture of Japan, \#19340062.
K.F. thanks the support by the MEXT program "Support Program for Improving Graduate School Education";
J. L.-P. acknowledges partial financial support by the Ministry of Science and Innovation of Spain (MICIINN) 
through an FPU grant, ref.~AP2005-1185;
A.D. acknowledges partial financial support by the Istituto Nazionale di Fisica Nucleare (INFN) through the 
Iniziativa Specifica RM21 for foreign guests;  and by the MICINN
 through the research project FPA2006-05423 and through the INFN-MICINN-08 Bilateral Agreement (Italy-Spain)
 "Flavour as a window for new physics";
D.M. acknowledges partial financial support by Ministry of University and of Scientific Research of Italy, through 
the 2007-08 COFIN program.
A.D. and J.L.-P. acknowledge also financial support from the Comunidad
Aut\'onoma de Madrid through the project P-ESP-00346.
Eventually, A.D., J.L.-P. and D.M. acknowledge the financial support of the European Community under the European Commission Framework Programme 7 Design Study: EUROnu, Project Number 212372; and under the Framework Programme 6 BENE-CARE networking activity MRTN-CT-2004-506395. 
The EC is not liable for any use that may be made of the information contained herein.

A.D. and J.L.-P. thank the Physics Department of the Universit\`a di Roma "La Sapienza", where this work has been completed.


\begin{thebibliography}{99}

\bibitem{Cleveland:1998nv}
  B.~T.~Cleveland {\it et al.},
  Astrophys.\ J.\  {\bf 496} (1998) 505.

\bibitem{Abdurashitov:1999zd}
  J.~N.~Abdurashitov {\it et al.}  [SAGE Collaboration],
  Phys.\ Rev.\  C {\bf 60} (1999) 055801
  [arXiv:astro-ph/9907113].

\bibitem{Hampel:1998xg}
  W.~Hampel {\it et al.}  [GALLEX Collaboration],
  Phys.\ Lett.\  B {\bf 447} (1999) 127.

\bibitem{Fukuda:2001nj}
  S.~Fukuda {\it et al.}  [Super-Kamiokande Collaboration],
  Phys.\ Rev.\ Lett.\  {\bf 86} (2001) 5651
  [arXiv:hep-ex/0103032].

\bibitem{:2008zn}
  J.~P.~Cravens {\it et al.}  [Super-Kamiokande Collaboration],
  Phys.\ Rev.\  D {\bf 78} (2008) 032002
  [arXiv:0803.4312 [hep-ex]].

\bibitem{Ahmad:2001an}
  Q.~R.~Ahmad {\it et al.}  [SNO Collaboration],
  Phys.\ Rev.\ Lett.\  {\bf 87} (2001) 071301
  [arXiv:nucl-ex/0106015].

\bibitem{Ahmed:2003kj}
  S.~N.~Ahmed {\it et al.}  [SNO Collaboration],
  Phys.\ Rev.\ Lett.\  {\bf 92} (2004) 181301
  [arXiv:nucl-ex/0309004].

\bibitem{Aharmim:2008kc}
  B.~Aharmim {\it et al.}  [SNO Collaboration],
  Phys.\ Rev.\ Lett.\  {\bf 101} (2008) 111301
  [arXiv:0806.0989 [nucl-ex]].

\bibitem{Fukuda:1998mi}
  Y.~Fukuda {\it et al.}  [Super-Kamiokande Collaboration],
  Phys.\ Rev.\ Lett.\  {\bf 81} (1998) 1562
  [arXiv:hep-ex/9807003].

\bibitem{Ambrosio:2001je}
  M.~Ambrosio {\it et al.}  [MACRO Collaboration],
  Phys.\ Lett.\  B {\bf 517} (2001) 59
  [arXiv:hep-ex/0106049].
  
\bibitem{Apollonio:1999ae}
  M.~Apollonio {\it et al.}  [CHOOZ Collaboration],
  Phys.\ Lett.\  B {\bf 466} (1999) 415
  [arXiv:hep-ex/9907037].

\bibitem{Apollonio:2002gd}
  M.~Apollonio {\it et al.}  [CHOOZ Collaboration],
  Eur.\ Phys.\ J.\  C {\bf 27} (2003) 331
  [arXiv:hep-ex/0301017].
  
\bibitem{Boehm:2001ik}
  F.~Boehm {\it et al.},
  Phys.\ Rev.\  D {\bf 64} (2001) 112001
  [arXiv:hep-ex/0107009].

\bibitem{Eguchi:2002dm}
  K.~Eguchi {\it et al.}  [KamLAND Collaboration],
  Phys.\ Rev.\ Lett.\  {\bf 90} (2003) 021802
  [arXiv:hep-ex/0212021].

\bibitem{Ahn:2002up}
  M.~H.~Ahn {\it et al.}  [K2K Collaboration],
  Phys.\ Rev.\ Lett.\  {\bf 90} (2003) 041801
  [arXiv:hep-ex/0212007].

\bibitem{Aliu:2004sq}
  E.~Aliu {\it et al.}  [K2K Collaboration],
  Phys.\ Rev.\ Lett.\  {\bf 94} (2005) 081802
  [arXiv:hep-ex/0411038].

\bibitem{Michael:2006rx}
  D.~G.~Michael {\it et al.}  [MINOS Collaboration],
  Phys.\ Rev.\ Lett.\  {\bf 97} (2006) 191801
  [arXiv:hep-ex/0607088].

\bibitem{Adamson:2008zt}
  P.~Adamson {\it et al.}  [MINOS Collaboration],
  Phys.\ Rev.\ Lett.\  {\bf 101} (2008) 131802
  [arXiv:0806.2237 [hep-ex]].

\bibitem{Amsler:2008zz}
  C.~Amsler {\it et al.}  [Particle Data Group],
  Phys.\ Lett.\  B {\bf 667} (2008) 1.

\bibitem{Pontecorvo:1957cp}
  B.~Pontecorvo,
  Sov.\ Phys.\ JETP {\bf 6} (1957) 429
  [Zh.\ Eksp.\ Teor.\ Fiz.\  {\bf 33} (1957) 549].

\bibitem{Maki:1962mu}
  Z.~Maki, M.~Nakagawa and S.~Sakata,
  Prog.\ Theor.\ Phys.\  {\bf 28} (1962) 870.

\bibitem{Pontecorvo:1967fh}
  B.~Pontecorvo,
  Sov.\ Phys.\ JETP {\bf 26} (1968) 984
  [Zh.\ Eksp.\ Teor.\ Fiz.\  {\bf 53} (1967) 1717].

\bibitem{Gribov:1968kq}
  V.~N.~Gribov and B.~Pontecorvo,
  Phys.\ Lett.\  B {\bf 28} (1969) 493.

\bibitem{Schwetz:2008er}
  T.~Schwetz, M.~Tortola and J.~W.~F.~Valle,
  New J.\ Phys.\  {\bf 10} (2008) 113011
  [arXiv:0808.2016 [hep-ph]].

\bibitem{Fogli:2008cx}
  G.~L.~Fogli, E.~Lisi, A.~Marrone, A.~Palazzo and A.~M.~Rotunno,
  arXiv:0809.2936 [hep-ph].

\bibitem{Fogli:2008jx}
  G.~L.~Fogli, E.~Lisi, A.~Marrone, A.~Palazzo and A.~M.~Rotunno,
  Phys.\ Rev.\ Lett.\  {\bf 101} (2008) 141801
  [arXiv:0806.2649 [hep-ph]].

\bibitem{Ge:2008sj}
  H.~L.~Ge, C.~Giunti and Q.~Y.~Liu,
  arXiv:0810.5443 [hep-ph].

\bibitem{GonzalezGarcia:2007ib}
  M.~C.~Gonzalez-Garcia and M.~Maltoni,
  Phys.\ Rept.\  {\bf 460} (2008) 1
  [arXiv:0704.1800 [hep-ph]].

\bibitem{Itow:2001ee}
  Y.~Itow {\it et al.}  [The T2K Collaboration],
  arXiv:hep-ex/0106019.

\bibitem{Ayres:2004js}
  D.~S.~Ayres {\it et al.}  [NOvA Collaboration],
  arXiv:hep-ex/0503053.

\bibitem{Ishitsuka:2005qi}
  M.~Ishitsuka, T.~Kajita, H.~Minakata and H.~Nunokawa,
  Phys.\ Rev.\  D {\bf 72} (2005) 033003
  [arXiv:hep-ph/0504026].

\bibitem{Hagiwara:2005pe}
  K.~Hagiwara, N.~Okamura and K.~i.~Senda,
  Phys.\ Lett.\  B {\bf 637} (2006) 266
  [Erratum-ibid.\  B {\bf 641} (2006) 486]
  [arXiv:hep-ph/0504061].

\bibitem{Diwan:2003bp}
  M.~V.~Diwan {\it et al.},
  Phys.\ Rev.\  D {\bf 68} (2003) 012002
  [arXiv:hep-ph/0303081].

\bibitem{Geer:1997iz}
  S.~Geer,
  Phys.\ Rev.\  D {\bf 57} (1998) 6989
  [Erratum-ibid.\  D {\bf 59} (1999) 039903]
  [arXiv:hep-ph/9712290].

\bibitem{Zucchelli:2002sa}
  P.~Zucchelli,
  Phys.\ Lett.\  B {\bf 532} (2002) 166.

\bibitem{belle}
Belle experiment,
{\tt http://belle.kek.jp/}.

\bibitem{babar}
Babar experiment,
{\tt http://www-public.slac.stanford.edu/babar/}.

\bibitem{Grossman:1995wx}
  Y.~Grossman,
  Phys.\ Lett.\  B {\bf 359} (1995) 141
  [arXiv:hep-ph/9507344].

\bibitem{Guzzo:1991hi}
  M.~M.~Guzzo, A.~Masiero and S.~T.~Petcov,
  Phys.\ Lett.\  B {\bf 260} (1991) 154.

\bibitem{Roulet:1991sm}
  E.~Roulet,
  Phys.\ Rev.\  D {\bf 44} (1991) 935.

\bibitem{Antusch:2006vwa}
  S.~Antusch, C.~Biggio, E.~Fernandez-Martinez, M.~B.~Gavela and J.~Lopez-Pavon,
  JHEP {\bf 0610} (2006) 084
  [arXiv:hep-ph/0607020].

\bibitem{Abada:2007ux}
  A.~Abada, C.~Biggio, F.~Bonnet, M.~B.~Gavela and T.~Hambye,
  JHEP {\bf 0712} (2007) 061
  [arXiv:0707.4058 [hep-ph]].

\bibitem{Group:2007kx}
  The ISS Physics Working Group,
  arXiv:0710.4947 [hep-ph].
  
\bibitem{Ota:2001pw}
  T.~Ota, J.~Sato and N.~a.~Yamashita,
  Phys.\ Rev.\  D {\bf 65} (2002) 093015
  [arXiv:hep-ph/0112329].

\bibitem{Ota:2002na}
  T.~Ota and J.~Sato,
  Phys.\ Lett.\  B {\bf 545}, 367 (2002)
  [arXiv:hep-ph/0202145].
  
\bibitem{FernandezMartinez:2007ms}
  E.~Fernandez-Martinez, M.~B.~Gavela, J.~Lopez-Pavon and O.~Yasuda,
  Phys.\ Lett.\  B {\bf 649} (2007) 427
  [arXiv:hep-ph/0703098];
  
\bibitem{Altarelli:2008yr}
  G.~Altarelli and D.~Meloni,
  Nucl.\ Phys.\  B {\bf 809} (2009) 158
  [arXiv:0809.1041 [hep-ph]].

\bibitem{Athanassopoulos:1996jb}
  C.~Athanassopoulos {\it et al.}  [LSND Collaboration],
  Phys.\ Rev.\ Lett.\  {\bf 77} (1996) 3082
  [arXiv:nucl-ex/9605003].

\bibitem{Athanassopoulos:1997pv}
  C.~Athanassopoulos {\it et al.}  [LSND Collaboration],
  Phys.\ Rev.\ Lett.\  {\bf 81} (1998) 1774
  [arXiv:nucl-ex/9709006].

\bibitem{Aguilar:2001ty}
  A.~Aguilar {\it et al.}  [LSND Collaboration],
  Phys.\ Rev.\  D {\bf 64} (2001) 112007
  [arXiv:hep-ex/0104049].

\bibitem{LEPfinal}
  LEP Collaborations (ALEPH, DELPHI, OPAL, L3) {\it et al.},
  Phys.\ Rept.\  {\bf 427} (2006) 257
  [arXiv:hep-ex/0509008].

\bibitem{AguilarArevalo:2007it}
  A.~A.~Aguilar-Arevalo {\it et al.}  [The MiniBooNE Collaboration],
  Phys.\ Rev.\ Lett.\  {\bf 98} (2007) 231801
  [arXiv:0704.1500 [hep-ex]].

\bibitem{Sorel:2003hf}
  M.~Sorel, J.~M.~Conrad and M.~Shaevitz,
  Phys.\ Rev.\  D {\bf 70} (2004) 073004
  [arXiv:hep-ph/0305255].

\bibitem{Maltoni:2007zf}
  M.~Maltoni and T.~Schwetz,
  arXiv:0705.0107 [hep-ph], to appear in PRD.

\bibitem{Barger:2005mh}
  V.~Barger, D.~Marfatia and K.~Whisnant,
  Phys.\ Rev.\  D {\bf 73}, 013005 (2006)
  [arXiv:hep-ph/0509163].

\bibitem{PalomaresRuiz:2005vf}
  S.~Palomares-Ruiz, S.~Pascoli and T.~Schwetz,
  JHEP {\bf 0509} (2005) 048
  [arXiv:hep-ph/0505216].

\bibitem{deGouvea:2006qd}
  A.~de Gouvea and Y.~Grossman,
  Phys.\ Rev.\  D {\bf 74}, 093008 (2006)
  [arXiv:hep-ph/0602237].

\bibitem{Schwetz:2007cd}
  T.~Schwetz,
  JHEP {\bf 0802}, 011 (2008)
  [arXiv:0710.2985 [hep-ph]].

\bibitem{Nelson:2007yq}
  A.~E.~Nelson and J.~Walsh,
  Phys.\ Rev.\  D {\bf 77}, 033001 (2008)
  [arXiv:0711.1363 [hep-ph]].

\bibitem{Donini:2007yf}
  A.~Donini, M.~Maltoni, D.~Meloni, P.~Migliozzi and F.~Terranova,
  JHEP {\bf 0712} (2007) 013
  [arXiv:0704.0388 [hep-ph]].

\bibitem{cngs}
CNGS experiment,
{\tt http://proj-cngs.web.cern.ch/proj-cngs/}.

\bibitem{Kimura:2002hb}
  K.~Kimura, A.~Takamura and H.~Yokomakura,
  Phys.\ Lett.\  B {\bf 537}, 86 (2002)
  [arXiv:hep-ph/0203099].
  
\bibitem{Kimura:2002wd}
  K.~Kimura, A.~Takamura and H.~Yokomakura,
  Phys.\ Rev.\  D {\bf 66}, 073005 (2002)
  [arXiv:hep-ph/0205295].

\bibitem{Donini:2001xy}
  A.~Donini and D.~Meloni,
  Eur.\ Phys.\ J.\  C {\bf 22} (2001) 179
  [arXiv:hep-ph/0105089].

\bibitem{Donini:2001xp}
  A.~Donini, M.~Lusignoli and D.~Meloni,
  Nucl.\ Phys.\  B {\bf 624} (2002) 405
  [arXiv:hep-ph/0107231].

\bibitem{Abe:2007bi}
  T.~Abe {\it et al.}  [ISS Detector Working Group],
  arXiv:0712.4129 [physics.ins-det].

\bibitem{Cervera:2000kp}
  A.~Cervera, A.~Donini, M.~B.~Gavela, J.~J.~Gomez Cadenas, P.~Hernandez, O.~Mena and S.~Rigolin,
  Nucl.\ Phys.\  B {\bf 579}, 17 (2000)
  [Erratum-ibid.\  B {\bf 593}, 731 (2001)]
  [arXiv:hep-ph/0002108].

\bibitem{Donini:2002rm}
  A.~Donini, D.~Meloni and P.~Migliozzi,
  Nucl.\ Phys.\  B {\bf 646} (2002) 321
  [arXiv:hep-ph/0206034].
  
\bibitem{BurguetCastell:2001ez}
  J.~Burguet-Castell {\em et al.},
  Nucl.\ Phys.\  B {\bf 608} (2001) 301
  [arXiv:hep-ph/0103258];
  
\bibitem{Minakata:2001qm}
  H.~Minakata and H.~Nunokawa,
  JHEP {\bf 0110}, 001 (2001)
  [arXiv:hep-ph/0108085].

\bibitem{Donini:1999jc}
  A.~Donini, M.~B.~Gavela, P.~Hernandez and S.~Rigolin,
  Nucl.\ Phys.\  B {\bf 574} (2000) 23
  [arXiv:hep-ph/9909254].

\bibitem{Donini:1999he}
  A.~Donini, M.~B.~Gavela, P.~Hernandez and S.~Rigolin,
  Nucl.\ Instrum.\ Meth.\  A {\bf 451} (2000) 58
  [arXiv:hep-ph/9910516].

\bibitem{Kalliomaki:1999ii}
  A.~Kalliomaki, J.~Maalampi and M.~Tanimoto,
  Phys.\ Lett.\  B {\bf 469} (1999) 179
  [arXiv:hep-ph/9909301].

\bibitem{Dighe:2007uf}
  A.~Dighe and S.~Ray,
  Phys.\ Rev.\  D {\bf 76}, 113001 (2007)
  [arXiv:0709.0383 [hep-ph]].

\bibitem{Boyarsky:2005us}
  A.~Boyarsky, A.~Neronov, O.~Ruchayskiy and M.~Shaposhnikov,
  Mon.\ Not.\ Roy.\ Astron.\ Soc.\  {\bf 370} (2006) 213
  [arXiv:astro-ph/0512509].

\bibitem{Maltoni:2004ei}
  M.~Maltoni, T.~Schwetz, M.~A.~Tortola and J.~W.~F.~Valle,
  New J.\ Phys.\  {\bf 6} (2004) 122
  [arXiv:hep-ph/0405172].

\bibitem{Okada:1996kw}
  N.~Okada and O.~Yasuda,
  Int.\ J.\ Mod.\ Phys.\  A {\bf 12} (1997) 3669
  [arXiv:hep-ph/9606411].

\bibitem{Bilenky:1996rw}
  S.~M.~Bilenky, C.~Giunti and W.~Grimus,
  Eur.\ Phys.\ J.\  C {\bf 1} (1998) 247
  [arXiv:hep-ph/9607372].

\bibitem{Dydak:1983zq}
  F.~Dydak {\it et al.},
  Phys.\ Lett.\  B {\bf 134} (1984) 281.

\bibitem{Declais:1994su}
  Y.~Declais {\it et al.},
  Nucl.\ Phys.\  B {\bf 434} (1995) 503.

\bibitem{Karagiorgi:2006jf}
  G.~Karagiorgi, A.~Aguilar-Arevalo, J.~M.~Conrad, M.~H.~Shaevitz, K.~Whisnant, M.~Sorel and V.~Barger,
  Phys.\ Rev.\  D {\bf 75}, 013011 (2007)
  [arXiv:hep-ph/0609177].

\bibitem{Bilenky:1998ne}
  S.~M.~Bilenky, C.~Giunti, W.~Grimus and T.~Schwetz,
  Astropart.\ Phys.\  {\bf 11}, 413 (1999)
  [arXiv:hep-ph/9804421].

\bibitem{Foot:1996qc}
  R.~Foot and R.~R.~Volkas,
  Phys.\ Rev.\  D {\bf 55}, 5147 (1997)
  [arXiv:hep-ph/9610229].

\bibitem{Cirelli:2004cz}
  M.~Cirelli, G.~Marandella, A.~Strumia and F.~Vissani,
  Nucl.\ Phys.\  B {\bf 708}, 215 (2005)
  [arXiv:hep-ph/0403158].

\bibitem{DeRujula:1979yy}
  A.~De Rujula, M.~Lusignoli, L.~Maiani, S.~T.~Petcov and R.~Petronzio,
  Nucl.\ Phys.\  B {\bf 168} (1980) 54.
  
\bibitem{Dziewonski:1981xy}
  A.~M.~Dziewonski and D.~L.~Anderson,
  Phys.\ Earth Planet.\ Interiors {\bf 25} (1981) 297.
 
\bibitem{Agarwalla:2006vf}
  S.~K.~Agarwalla, S.~Choubey and A.~Raychaudhuri,
  Nucl.\ Phys.\  B {\bf 771} (2007) 1
  [arXiv:hep-ph/0610333].

\bibitem{Bueno:2000fg}
  A.~Bueno, M.~Campanelli and A.~Rubbia,
  Nucl.\ Phys.\  B {\bf 589} (2000) 577
  [arXiv:hep-ph/0005007].

\bibitem{Campanelli:2000we}
  M.~Campanelli, A.~Bueno and A.~Rubbia,
  Nucl.\ Instrum.\ Meth.\  A {\bf 451} (2000) 176.

\bibitem{DeRujula:1998hd}
  A.~De Rujula, M.~B.~Gavela and P.~Hernandez,
  Nucl.\ Phys.\  B {\bf 547} (1999) 21
  [arXiv:hep-ph/9811390].

\bibitem{Broncano:2002hs}
  A.~Broncano and O.~Mena,
  Eur.\ Phys.\ J.\  C {\bf 29} (2003) 197
  [arXiv:hep-ph/0203052].

\bibitem{Autin:1999ci}
  B.~Autin, A.~Blondel and J.~R.~Ellis,
  ``Prospective study of muon storage rings at CERN,''

\bibitem{Gruber:2002tn}
  P.~Gruber {\it et al.},
  ``The study of a European Neutrino Factory complex.''

\bibitem{Finley:2000cn}
  D.~Finley and N.~Holtkamp,
  Nucl.\ Instrum.\ Meth.\  A {\bf 472}, 388 (2000).

\bibitem{Ozaki:2001bb}
  S.~Ozaki {\it et al.}, BNL-52623,
  ``Feasibility study 2 of a muon based neutrino source.''

\bibitem{Alsharoa:2002wu}
  M.~M.~Alsharoa {\it et al.}  [Muon Collider/Neutrino Factory
                  Collaboration],
  Phys.\ Rev.\ ST Accel.\ Beams {\bf 6}, 081001 (2003)
  [arXiv:hep-ex/0207031].

\bibitem{Zisman:2003bh}
  M.~S.~Zisman,
  Nucl.\ Instrum.\ Meth.\  A {\bf 503}, 384 (2003).

\bibitem{Kuno:2001tb}
\fussy
  Y.~Kuno, Y.~Mori, S.~Machida, T.~Yokoi, Y.~Iwashita, J.~Sato and O.~Yasuda,\\
  ``A feasibility study of a neutrino factory in Japan,''
{\tt http://www-prism.kek.jp/nufactj/}.
  
\bibitem{Zisman:2008zz}
  M.~S.~Zisman,
  J.\ Phys.\ Conf.\ Ser.\  {\bf 110} (2008) 112006.

\bibitem{Lipari}
P. Lipari, private communication.

\bibitem{Lipari:1994pz}
  P.~Lipari, M.~Lusignoli and F.~Sartogo,
  Phys.\ Rev.\ Lett.\  {\bf 74} (1995) 4384
  [arXiv:hep-ph/9411341].
  
\bibitem{CerveraVillanueva:2008zz}
  A.~Cervera-Villanueva,
  AIP Conf.\ Proc.\  {\bf 981} (2008) 178.

\bibitem{Cervera:2000vy}
  A.~Cervera, F.~Dydak and J.~Gomez Cadenas,
  Nucl.\ Instrum.\ Meth.\  A {\bf 451}, 123 (2000).
  
\bibitem{Autiero:2003fu}
  D.~Autiero {\it et al.},
  Eur.\ Phys.\ J.\  C {\bf 33} (2004) 243
  [arXiv:hep-ph/0305185].

\bibitem{Huber:2003ak}
  P.~Huber and W.~Winter,
  Phys.\ Rev.\  D {\bf 68} (2003) 037301
  [arXiv:hep-ph/0301257].

\bibitem{Kopp:2008ds}
  J.~Kopp, T.~Ota and W.~Winter,
  Phys.\ Rev.\  D {\bf 78} (2008) 053007
  [arXiv:0804.2261 [hep-ph]].

\bibitem{Fogli:1996pv}
  G.~L.~Fogli and E.~Lisi,
  Phys.\ Rev.\  D {\bf 54}, 3667 (1996)
  [arXiv:hep-ph/9604415].

\bibitem{Barger:2001yr}
  V.~Barger, D.~Marfatia and K.~Whisnant,
  Phys.\ Rev.\  D {\bf 65}, 073023 (2002)
  [arXiv:hep-ph/0112119].

\bibitem{Donini:2005db}
  A.~Donini, E.~Fernandez-Martinez, D.~Meloni and S.~Rigolin,
  Nucl.\ Phys.\  B {\bf 743} (2006) 41
  [arXiv:hep-ph/0512038].

\bibitem{Huber:2008yx}
  P.~Huber and T.~Schwetz,
  Phys.\ Lett.\  B {\bf 669} (2008) 294
  [arXiv:0805.2019 [hep-ph]].

\bibitem{Ables:1995wq}
  E.~Ables {\it et al.}  [MINOS Collaboration],

\bibitem{Huber:2006wb}
  P.~Huber, M.~Lindner, M.~Rolinec and W.~Winter,
  Phys.\ Rev.\  D {\bf 74} (2006) 073003
  [arXiv:hep-ph/0606119].

\bibitem{Indumathi:2004pn}
  D.~Indumathi  [INO Collaboration],
  Pramana {\bf 63}, 1283 (2004).

\bibitem{Fukushima:2008zz}
  C.~Fukushima {\it et al.},
  Nucl.\ Instrum.\ Meth.\  A {\bf 592} (2008) 56.

\bibitem{Meloni:2008bd}
  D.~Meloni,
  Phys.\ Lett.\  B {\bf 664} (2008) 279
  [arXiv:0802.0086 [hep-ph]].

\bibitem{Migliozzi}
P. Migliozzi, private communication.

\bibitem{ScottoLavina:2008}
L. Scotto Lavina, talk at the NuFact'08 Workshop, Valencia
  PoS {\bf NUFACT08} (2008) 049.

\bibitem{Fuki:2008nufact}
K. Fuki, poster at the NuFact'08 Workshop, Valencia,
  PoS {\bf NUFACT08} (2008) 123.

\bibitem{Meloni:2008ti}
D. Meloni, talk at the NOW'08 Workshop, Otranto,
  arXiv:0812.3555 [hep-ph].

\bibitem{GonzalezGarcia:2004cu}
  M.~C.~Gonzalez-Garcia, M.~Maltoni and A.~Y.~Smirnov,
  Phys.\ Rev.\  D {\bf 70} (2004) 093005
  [arXiv:hep-ph/0408170].

\bibitem{Huber:2002mx}
  P.~Huber, M.~Lindner and W.~Winter,
  Nucl.\ Phys.\  B {\bf 645} (2002) 3
  [arXiv:hep-ph/0204352].

\bibitem{Donini:2003vz}
  A.~Donini, D.~Meloni and S.~Rigolin,
  JHEP {\bf 0406} (2004) 011
  [arXiv:hep-ph/0312072].

\bibitem{Laing:2008zz}
  A.~Laing and F.~J.~P.~Soler,
  AIP Conf.\ Proc.\  {\bf 981} (2008) 166.

\bibitem{Xing:2005gk}
  Z.~z.~Xing and H.~Zhang,
  Phys.\ Lett.\  B {\bf 618} (2005) 131
  [arXiv:hep-ph/0503118].

\bibitem{Yasuda:2007jp}
  O.~Yasuda,
  arXiv:0704.1531 [hep-ph].
  
\bibitem{Zhang:2006yq}
  H.~Zhang,
  Mod.\ Phys.\ Lett.\  A {\bf 22}, 1341 (2007)
  [arXiv:hep-ph/0606040].

\end{thebibliography}
\end{document}